\documentclass [11pt]{article}
\pdfoutput=1

\usepackage{jheppub}

\usepackage[utf8]{inputenc}

\usepackage{amsfonts}

\usepackage{amsmath,amssymb}

\usepackage{mathtools}

\usepackage{graphicx,subfigure}
\usepackage[space]{grffile}
\usepackage{mathtools}

\usepackage{simplewick}

\usepackage{array}

\usepackage{float}

\usepackage{color}

\usepackage{mathrsfs}

\usepackage{ytableau}

\usepackage{tikz}
\usepackage{tikz-feynman}

\usepackage{contour}
\usepackage{slashed}
\def\nn{\nonumber}

\newcommand{\cO}{\mathcal{O}}
\newcommand{\cW}{\mathcal{W}}

\newcommand{\<}{\langle}
\renewcommand{\>}{\rangle}
\newcommand{\p}{\partial}
\newcommand{\be}{\begin{eqnarray}\displaystyle}
\newcommand{\ee}{\end{eqnarray}}
\newcommand{\f}{\frac}

\newcommand{\ve}{\varepsilon}

 \definecolor{verde}{rgb}{0,0.7,0.2}

\newcommand{\myP}{\boldsymbol{p}}

\title{\boldmath  Trace Anomalies and the Graviton-Dilaton Amplitude}

\author[a]{Denis Karateev,}
\author[b]{Zohar Komargodski,}
\author[c]{Jo\~ao Penedones}
\author[c,d]{and Biswajit Sahoo}

\affiliation[a]{
	D\'epartment de Physique Th\'eorique, Universit\'e de Gen\`eve,\\
	24 quai Ernest-Ansermet, 1211 Gen\`eve 4, Switzerland}

\affiliation[b]{
Simons Center for Geometry and Physics, SUNY, Stony Brook, NY 11794, USA
}

\affiliation[c]{Fields and Strings Laboratory, Institute of Physics\\ École Polytechnique Fédéral de Lausanne (EPFL)
	\\ Route de la Sorge, CH-1015 Lausanne, Switzerland}
	
	\affiliation[d]{
	Department of Mathematics, King’s College London,\\ Strand, London WC2R 2LS, United
Kingdom}

\abstract{We consider 3+1 dimensional Quantum Field Theories (QFTs) coupled to the dilaton and the graviton. We show that the graviton-dilaton scattering amplitude receives a universal contribution which is helicity flipping and is proportional to $\Delta c-\Delta a$ along any RG flow, where $\Delta c$ and $\Delta a$ are the differences of the UV and IR $c$- and $a$-trace anomalies respectively. This allows us to relate  $\Delta c-\Delta a$  to spinning massive states in the spectrum of the QFT. We test our predictions in two simple examples: in the theory of a massive free scalar and in the theory of a massive Dirac fermion (a more complicated example is provided in a companion paper \cite{Karateev}). We discuss possible applications.}

\vspace{4cm}

\begin{document}

\maketitle

\section{Introduction}
\label{introduction}

Non-perturbative techniques in Quantum Field Theory (QFT) are of great interest, with applications ranging across many different branches of physics. One striking early example is due to 't Hooft~\cite{hooft1980naturalness} who identified correlation functions which do not evolve under the renormalization group (RG)~\cite{Frishman:1980dq}. 
These special protected correlation functions are related to anomalies of continuous global symmetries. In other words, 't Hooft anomalies of global symmetries are RG independent quantities.
The scale independence of anomalies places interesting constraints on RG flows, commonly referred to as the ``anomaly matching conditions.'' 

In this paper we will view QFT as an RG flow between a UV and an IR fixed points both of which are described by conformal field theories (CFTs). (The IR fixed point can be trivial if the theory is gapped.) The subject of this paper is to consider trace anomalies and explain how they should be matched. Trace anomalies are related to conformal symmetry, which is broken explicitly in theories with a mass scale. Usually, one does not discuss anomaly matching for symmetries which are explicitly broken. 
Yet, for the case of conformal symmetry, there are protected amplitudes which we identify. The protected amplitudes that are determined by the $a$-trace anomaly were discussed in~\cite{Komargodski:2011vj,Komargodski:2011xv}, see also \cite{Luty:2012ww,Baume:2014rla}. In particular, it was emphasized that if the QFT is coupled to a massless dilaton, the dilaton-dilaton scattering amplitude is universally proportional to $\Delta a \equiv a_{UV} - a_{IR}$ at low energies.
 
 In this work we extend this analysis by coupling the theory conformally to a massless graviton.\footnote{In \cite{Gillioz:2019iye} the authors proposed interesting CFT sum-rules using the technology of background gravitational fields.} We identify  protected correlators which are sensitive to the $c$-trace anomaly.  Our analysis is motivated by a discussion in the literature on whether the $c$ trace anomaly should be matched, see e.g.~\cite{Schwimmer:2010za,Nakayama:2017oye,Niarchos:2020nxk,Andriolo:2022lcb,Schwimmer:2023nzk}. We will give a simple argument proving that the answer is positive and demonstrate it explicitly in several examples.

We couple the graviton  to the QFT with strength $\kappa$, which is the square root of Newton constant, and the dilaton couples as usual (canonically) to the trace of the energy momentum tensor with strength $1/f$. 
The dilaton-dilaton scattering amplitude at low energies is given by 
\begin{equation}\label{oldRintro}{\cal T}(s,t,u) = \f{1}{f^{4}}\Delta a (s^2+t^2+u^2)~.\end{equation}
It is completely fixed by the change in the $a$ anomaly and it is insensitive to the precise RG flow taking place, instead, it only depends on the end-points of the RG flow. 

Our new result concerns with the graviton-dilaton amplitude at low energies, in which the graviton flips its helicty: 
\begin{equation}\label{newRintro} {\cal T}{}_{+2}^{-2}(s,t,u) ={\kappa^2\over f^2}(\Delta c-\Delta a) t^2,
\end{equation}
where $\Delta c \equiv c_{UV} - c_{IR}$.\footnote{Sum-rules for $\Delta c $ involving the stress-tensor two-point function and their implications in $d=2$ space-times dimensions were discussed in \cite{Cardy:1988tj,Hartman:2023ccw} and in $d\geq 2$ space-time dimensions were discussed in \cite{Cardy:1988cwa,Cappelli:1990yc,Cappelli:2001pz,Karateev:2020axc}. From these sum-rules one immediately obtains the $c$-theorem in $d=2$ \cite{Zamolodchikov:1986gt} which states that $\Delta c \geq 0$. In $d>2$ instead these sum-rules do not impose any constraint on the sign of $\Delta c$.}
This result should hold in any QFT, regardless of whether it is gapped or not in the infrared. This amplitude is independent of the actual RG flow, as long as the end-points are fixed, as in~\eqref{oldRintro}.

Applying dispersion relations, we can use~\eqref{oldRintro} to rewrite $\Delta a$ in terms of the massive particles in the theory which couple to the trace of the energy-momentum tensor. All the massive particles appear in the corresponding sum rule. 
Similarly, by using \eqref{newRintro}, $\Delta c-\Delta a$ can be associated with spinning massive states (with a partial wave spin of at least 2) that couple to the trace of the energy-momentum tensor, under the assumption that the second derivative in $t$ of the amplitude decays in the Regge limit: 
\begin{equation}
	\label{eq:dispersion_relation}
	\Delta c-\Delta a  = 
	{f^2\over \kappa^2}\int_{m^2}^\infty  \frac{ds}{\pi} \frac{\text{Im}\,\partial_t^2\mathcal{T}{}_{+2}^{-2}(s,0,-s)}{s}.
\end{equation}
The combination $\Delta c-\Delta a$ is not sign definite.

The combination $c-a$ also appears in~\cite{DiPietro:2014bca,Beem:2017ooy,ArabiArdehali:2023bpq} in the context of counting specific operators in supersymmetric theories, as well as in the angular-dependent part of the expectation value of the energy in the state produced by the $U(1)_R$-curent~\cite{Hofman:2008ar}.\footnote{The $c-a$ combination also appears in the computation of the logarithmic term in the entropy of  Schwarszchild black
hole \cite{Solodukhin:2010pk}.} Perhaps more relevant to our discussion here, $c-a$ appeared in~\cite{Camanho:2014apa}, where it was related to spinning primary operators in the large central charge, strong coupling limit of CFTs. Here we are discussing $\Delta c-\Delta a$ in the context of RG flows, and we have related it to spinning massive particles (and our discussion does not require large central charge or strong coupling), so the context is quite different, but there could be a relation. 

A useful reformulation of~\eqref{oldRintro} was recently unveiled in~\cite{Hartman:2023qdn}. There, one considers the state created by acting with the (appropriately smeared) trace of the energy-momentum on the vacuum, $T{}^{\mu}_{\mu} |{\rm VAC}\rangle$ and one studies the null energy radiated off this state. It should be possible to think about~\eqref{newRintro} as 
a helicity flipping process of a graviton propagating through this state. Another interesting quantity where the trace anomalies appear is the entanglement entropy in the vacuum~\cite{Solodukhin:2008dh,Casini:2017vbe}. It would be interesting to see if the combination $\Delta c-\Delta a$ plays a similar central role in that setup.  

The outline of this paper is as follows. In section \ref{sec:review} we review trace anomalies and outline the derivation of \eqref{newRintro}. 
In section \ref{sec:vertices} we compute all the vertices of dilatons and gravitons that will eventually need to be assembled into scattering amplitudes.
In section \ref{sec:examples} we compute the vertices explicitly for a massive fermion and massive boson, which we do to check the general predictions of section \ref{sec:vertices}. In section \ref{sec:scattering_and_bounds} we assemble all the vertices into scattering amplitudes of physical gravitons and dilatons, deriving \eqref{newRintro}. In section \ref{CouplSpac} we discuss less familiar trace anomalies, sometimes called “resonance” or “coupling space” anomalies -- we will derive the low energy vertices protected by these anomalies and test our predictions in some simple RG flows.
In section \ref{sec:discussion} we list a few additional open questions. Three appendices cover technical material.

\section{Review of Trace Anomalies and Outline of the Derivation of~\eqref{newRintro}} 
\label{sec:review}
We will now review trace anomalies and explain the strategy behind deriving~\eqref{newRintro}. 

The discussion of trace anomalies and their matching can be done in the following context
\begin{itemize}
\item Systems with exact conformal symmetry, which is spontaneously broken in the vacuum. This occurs in theories with a moduli space of degenerate vacua (mostly supersymmetric theories or large $N$ theories). The infrared physics in the vacua with spontaneously broken conformal invariance can be nontrivial and the matching of trace anomalies is a powerful constraint on this physics. 
\end{itemize} 
The above setup is conceptually similar to anomaly matching of ordinary spontaneously broken global symmetries.
A more generic setup in which we can discuss the matching of trace anomalies is
\begin{itemize}
\item Ordinary systems undergoing a renormalization group flow. They do not have conformal symmetry but the ideas of trace anomaly matching are still useful. Heuristically, one can see this in the following way. In the very weakly coupled regime of Nambu-Goldstone (i.e. in the limit of very large decay constant) it is impossible to tell apart the case of spontaneously broken symmetry from an explicitly broken one. Therefore, in the case of ordinary RG flows one still expects to be able to define interesting observables that are constrained by the ideas of anomaly matching and this leads to non-trivial results about generic renormalization group flows.
\end{itemize} 

The two setups are quite similar and lead to similar conclusions. Our discussion will be framed in the realm of ordinary relativistic theories undergoing a renormalization group flow, i.e. the second item above, since it is more general.\footnote{One could also consider matching the trace anomaly in nontrivial states, and not just the relativistic invariant vacuum. We do not discuss it here but see~\cite{Eling:2013bj,Benjamin:2023qsc} for a discussion of how the trace anomaly is matched in the (disordered) thermal state.}  
As we will review, trace anomalies can be of ``type A'' or ``type B''~\cite{Deser:1993yx} and we will see that they both need to be matched.

As we have mentioned, theories undergoing a renormalization group flow obviously explicitly break the conformal symmetry since there is an explicit mass scale $M$ or more generally a set of mass scales $M_i$. The conformal symmetry of the ultraviolet fixed point nevertheless imposes constraints on various amplitudes of the theory. To tease out these facts we use  a classical background field  $\Omega(x)$ (the dilaton) which is introduced in the original QFT by replacing all the mass scales $M_i$ as
\begin{equation}
	\label{eq:dilaton_compensation}
	M_i \rightarrow M_i(x) \equiv M_i\, \Omega(x).
\end{equation}
This is sometimes referred to as the Stueckelberg trick.
This includes all mass scales, whether they are generated classically or quantum mechanically.
We also place the QFT on a curved non-dynamical background with metric $g_{\mu\nu}(x)$.
The partition function of the theory $Z[\Omega, g_{\mu\nu}]$ now enjoys Weyl invariance if a Weyl transformation of the metric is accompanied by an appropriate transformation of $\Omega$, namely
\begin{equation}
\label{transrules}
\Omega(x)\to e^{-\sigma(x)}\Omega(x),\qquad g_{\mu\nu}(x)\to e^{2\sigma(x)}g_{\mu\nu}(x),
\end{equation}
where $\sigma(x)$ is the coordinate dependent parameter of the Weyl transformation.
 We can say that $\Omega(x)$ is a ``spurion'' for the conformal symmetry. Also, one can think of $\log \Omega(x)$ as the Nambu-Goldstone boson of spontaneous conformal symmetry breaking. Clearly, the background field $\Omega(x)$ couples to the trace of the energy-momentum tensor (accompanied by additional nonlinear couplings that are necessary to ensure conformal invariance) and hence amplitudes involving $\Omega(x)$ are just a bookkeeping device for correlators of the trace of the energy-momentum tensor.

 In this paper we focus on $d=4$ space-time dimensions and work in mostly plus metric $\eta^{\mu\nu}=\text{diag}\{-1,+1,+1,+1\}$. Instead of the partition function it is more convenient to work with the connected functional $W [\Omega, g_{\mu\nu}]$ related to the partition function as
 \begin{equation}
 	\label{eq:connected_functional}
 	Z[\Omega, g_{\mu\nu}] = e^{i W [\Omega, g_{\mu\nu}]}.
 \end{equation}
When we say that the theory is Weyl invariant this is up to the following variation of the connected functional
   \begin{equation}
 	\label{transrulestwo}
 	\delta_\sigma W [\Omega, g_{\mu\nu}] = \int d^4x \sqrt{-g}\sigma(x)\left(-a_{\text{UV}} E_4+c_{\text{UV}} \cW^2\right),
 \end{equation}
where $E_4$ is the Euler density and $\cW^2$ is the square of the Weyl tensor. For our conventions see appendix \ref{sec:Weyl_anomaly}.
 The coefficients $a_{\text{UV}}$ and $c_{\text{UV}}$ are called the $a$- and $c$-trace anomalies (for a review of the history of trace anomalies, see~\cite{Duff:1993wm}). They also appear as the OPE coefficients in the UV CFT three-point function of the stress-tensor~\cite{Osborn:1993cr,Costa:2011mg}.\footnote{An alternative to~\eqref{transrulestwo} is to express the right hand side as the trace of the stress tensor of the UV CFT:
  \begin{equation}
  	\label{eq:anomaly_via_correlator}
  	\delta_\sigma W[\Omega, g_{\mu\nu}]=
  	\int d^4 x\sqrt{-g}\ \sigma(x)\,\langle 0| T{}^{\mu}_{\mu} (x)|0\rangle_{g}^{\text{UV CFT}} .
  \end{equation}}

 In~\eqref{transrulestwo}, if we choose the infinitesimal Weyl transformation parameter $\sigma(x)$ to be independent of coordinates (i.e. a constant scale transformation), the anomaly contribution associated with the $E_4$ reduces to a boundary term. For this reason, the $a$-trace anomaly is a ``type-A'' anomaly. Conversely, the $\cW^2$ term remains present for a constant $\sigma(x)$, which makes it a ``type-B'' anomaly.
 It is important to keep in mind that there could be additional $c$-number contributions on the right hand side of~\eqref{transrulestwo}, for instance, if there are additional background fields (coupling constants). We will discuss some of these generalizations later. Such additional terms have to be analytic in the background fields, similarly to~\eqref{transrulestwo}.

 The most important property of the Weyl anomaly \eqref{transrulestwo} is that it remains invariant along the whole RG flow (for any energy scale and for arbitrary background fields). In order for this to be true it is crucial that the Weyl invariance is only violated by a $c$-number, i.e. a functional of the background fields. This is why $a_{\text{UV}}$ and $c_{\text{UV}}$ are pure numbers determined by the conformal anomalies at short distances, and independent of the precise trajectory of the RG flow, or the state of the system in the infrared. This universality is the reason that equation~\eqref{transrulestwo} is powerful. 
The connected functional \eqref{eq:connected_functional} can be equivalently written in terms of the low energy degrees of freedom. It will then depend on the low energy effective action $A_\text{EFT}$.
Equation~\eqref{transrulestwo} places constraints on the possible dependence of the IR effective action $A_\text{EFT}$ on the background fields $\Omega(x)$ and $g_{\mu\nu}(x)$. The Weyl anomaly of the connected functional written in terms of the $A_\text{EFT}$  must be the same as \eqref{transrulestwo}. This is the Weyl anomaly matching condition which we will use shortly.

A technically crucial point is that the connected functional $W [\Omega, g_{\mu\nu}] $ is not entirely well defined and neither is its variation $\delta_\sigma W [\Omega, g_{\mu\nu}] $. As usual, the locality of the underlying theory allows us to parameterize the ambiguity in $W [\Omega, g_{\mu\nu}] $ by local terms,
   \begin{equation}\label{ScheDep}
   	A_\text{local}\equiv\int d^4x \sqrt{-g}\,\mathcal{L}_\text{local}(M_i(x) , g_{\mu\nu}(x)),
   \end{equation} 
 where $M_i(x)$ was defined in \eqref{eq:dilaton_compensation}. $A_\text{local}$ cannot contain nonlocal terms such as $\log M_i(x)$ since $\mathcal{L}_{local}$ represents contact terms between local operators to which $M_i(x)$ and $g_{\mu\nu}(x)$ couple. Notice, that $A_\text{local}$ is not required to be Weyl invariant and hence may contribute to the variation 
 $\delta_\sigma W [\Omega, g_{\mu\nu}]$. Therefore, when we write~\eqref{transrulestwo} this must be considered only up to terms of the form $\delta_\sigma A_\text{local}$. Alternatively, one can imagine always working in a scheme where~\eqref{transrulestwo} holds true.

Before addressing general RG flows let us first recall how the variation~\eqref{transrulestwo} is satisfied in a conformal field theory. In this case the partition function does not depend on $\Omega(x)$ since there are no mass scales to start with. The partition function of a CFT as a functional of $g_{\mu\nu}(x)$ is far too complicated to be determined, but~\cite{Deser:1993yx} and~\cite{Deser:1976yx} have constructed particular non-local terms which lead to the right variation for the $a$- and $c$-anomalies:\footnote{In the literature, there are many different constructions of the $a$-anomaly functional, whose variations produce \eqref{transrulestwo}. These constructions are related by some Weyl invariant terms and local terms in the functional, as recently discussed in \cite{Barvinsky:2023exr,Barvinsky:2023aae}.}
\begin{multline}
	\label{eq:connected_functional_CFT_a}
	W_{\text{UV CFT}}^{\text{a-part}}[g_{\mu\nu}] = \f{1}{9}\int d^4x\ \sqrt{-g}\ \square^{-1}\Big[\f{1}{2} R_{\mu\nu\alpha\beta}R^{\mu\nu\alpha\beta}R+10 R_{\mu\nu}R^{\nu\alpha}R_\alpha^{\ \mu}\\
	-13R_{\mu\nu}R^{\mu\nu}R+\f{41}{18}R^3+6R_{\mu\nu\alpha\beta}R^{\mu\alpha}R^{\nu\beta}\Big].
\end{multline}
\begin{equation}
	\label{eq:connected_functional_CFT_c}
	W_{\text{UV CFT}}^{\text{c-part}}[g_{\mu\nu}] =- \f{1}{2}\int d^4x\ \sqrt{-g}\ \cW(x)\ln\left(\frac{\square}{\mu^2}\right)\cW(x)+\cdots .
\end{equation}
The ellipses in \eqref{eq:connected_functional_CFT_c} represent all higher-order curvature corrections to the expression. These corrections are necessary to ensure that the Laplacian in the logarithm's argument becomes homogeneous under a Weyl transformation when it acts on the Weyl tensor. The nonlocal expressions in \eqref{eq:connected_functional_CFT_a} and \eqref{eq:connected_functional_CFT_c}  should be used only for background fields with compact support where the differential operators are invertible. They reproduce the required variation~\eqref{transrulestwo} when the coefficient of \eqref{eq:connected_functional_CFT_a} is $-a_{\text{UV}}$ and the coefficient of \eqref{eq:connected_functional_CFT_c} is $c_{\text{UV}}$. The full effective action could contain additional, generally nonlocal, Weyl invariant terms (as well as scheme dependent local counter-terms which may or may not be Weyl invariant as discussed below~\eqref{ScheDep}). Using \eqref{eq:connected_functional_CFT_a} and \eqref{eq:connected_functional_CFT_c} we can compute correlation functions of the stress-tensor.
The merit of such nonlocal effective actions is that they allow to derive interesting identities satisfied by the energy-momentum tensor correlation functions. 
 For instance, one can see from~\eqref{eq:connected_functional_CFT_c} that the two-point function of stress-tensors in flat spacetime is proportional to $c_\text{UV}$. Instead, $a_\text{UV}$ influences only the three-(and higher-) point functions of  stress-tensors in flat spacetime, simply because the expansion of~\eqref{eq:connected_functional_CFT_a} around flat space starts from three gravitons.

Now let us present the case of a generic renormalization group flow (or, similarly, conformal field theories in a vacuum with spontaneously broken conformal invariance). We will present a local IR effective action that depends on $\Omega(x)\equiv e^{-\tau(x)}$ and $g_{\mu\nu}(x)$ whose variation leads to~\eqref{transrulestwo}. Let us consider three distinct terms of the IR effective action which obey the following constraints
\begin{equation}
	\label{eq:variations_pieces}
	\begin{aligned}
		\delta_\sigma A_a[\tau ,g_{\mu\nu}] &= \int d^4x \sqrt{-g}\sigma(x)E_4,\\
		\delta_\sigma A_c[\tau ,g_{\mu\nu}] &= \int d^4x \sqrt{-g}\sigma(x)\cW^2,\\
		\delta_\sigma A_{\text{invariant}}[\tau ,g_{\mu\nu}]&= 0.
	\end{aligned}
\end{equation}
The solution to these constraints was found in \cite{Fradkin:1983tg,Schwimmer:2010za} using the Wess–Zumino construction \cite{Wess:1971yu}, see also \cite{Komargodski:2011vj}. It reads\footnote{Notice that the IR effective action is local in $\tau(x)$ and $g_{\mu\nu}(x)$, since in the infrared we expand around a non-zero value of $\Omega(x)$ which sets the scale of the RG flow. This situation should be contrasted with the scheme dependent terms~\eqref{ScheDep} which must be local in the UV couplings instead. }
\begin{align}
	\label{eq:Aa}
	A_a[\tau ,g_{\mu\nu}] &=  \int d^4 x \sqrt{-g}\Big( \tau E_{4}+4\left(R^{\mu\nu}-\tfrac{1}{2}g^{\mu\nu}R\right)\partial_\mu\tau \partial_\nu\tau
	+2(\p\tau)^4 -4(\p\tau)^2 \square\tau  \Big),\\
	\label{eq:Ac}
	A_c[\tau ,g_{\mu\nu}] &= \int d^4 x \sqrt{-g}\, \tau\, \cW^2,\\
	\label{eq:Ainvariant}
	A_{\text{invariant}}[ \widehat{g}_{\mu\nu}]&=\int d^4x\ \sqrt{-\widehat{g}}\Bigg(M^4\lambda+M^2r_0\widehat R\ +\ r_1 \widehat R^{2}+r_2 \widehat{\cW}^{2}+r_3 \widehat E_4+\ldots\Bigg).
\end{align}
Here $\widehat g_{\mu\nu}(x) = e^{-2\tau(x)} g_{\mu\nu}(x)$, which is a Weyl invariant combination. 
We have introduced the dimensionless parameters $\lambda$, $r_0$, $r_1$, $r_2$ and $r_3$ whose values depend on a particular theory (and they could depend on the precise RG flow, even if the ultraviolet and infrared fixed points are the same). The parameter $M$ is the mass gap. 
Here, $\widehat R_{\mu\nu}$, $\widehat R$, $\widehat \cW^2$ and $\widehat E_4$ are the Ricci tensor, Ricci scalar,  squared Weyl tensor and the Euler density which are constructed from the metric $\widehat g_{\mu\nu}$.

Let us first suppose for simplicity that our theory is gapped in the IR. Then due to \eqref{eq:variations_pieces} the most general IR effective action which lead to the same Weyl anomaly as \eqref{transrulestwo} is given by
\begin{equation}
	\label{eq:action_IR_free}
	A_\text{EFT}[\Omega,\, g_{\mu\nu}] =  -a_{\text{UV}}\times A_a[\tau,\, g_{\mu\nu}] + c_{\text{UV}} \times A_c[\tau,\, g_{\mu\nu}] \ +\ A_{\text{invariant}}[ \widehat{g}_{\mu\nu}].
\end{equation}
In other words we can say that we have constructed the IR effective action \eqref{eq:action_IR_free} by matching the UV Weyl anomaly \eqref{transrulestwo}. If instead our theory in the IR is described by a non-trivial CFT with trace anomalies $a_\text{IR}$ and $c_\text{IR}$ then the form of the effective action is
\begin{align}
	A_\text{EFT}[\Omega,\, g_{\mu\nu},\Theta] =
	&-\Delta a\times A_a[\tau,\, g_{\mu\nu}] +\Delta c \times A_c[\tau,\, g_{\mu\nu}] 
	+ A_{\text{invariant}}[ \widehat{g}_{\mu\nu}] \nonumber\\ &+
	A_\text{IR CFT}[\Omega,g_{\mu\nu},\Theta]~. \label{eq:EFT_actionSecTwo}
\end{align}
The first line has the same structure as \eqref{eq:action_IR_free}.
The second line describes the IR CFT with (schematically) {\it dynamical} fields denoted by $\Theta$ interacting with the metric and dilaton.
The differences of the UV and IR anomalies are defined as\footnote{In the case of QFTs with of spontaneous conformal symmetry breaking the dilaton is a physical particle. More precisely it is the Goldstone boson of the spontaneous symmetry breaking. Then in the IR the trace anomalies are given by $a_{\text{IR}}=a_{\text{IR CFT}}+ a_\text{free boson}$ and $c_{\text{IR}}=c_{\text{IR CFT}}+ c_\text{free boson}$.\label{f:SSB}}
\begin{equation}
	\Delta a\equiv a_\text{UV}-a_\text{IR},\qquad 
	\Delta c\equiv c_\text{UV}-c_\text{IR}.
\end{equation}
The IR effective action \eqref{eq:EFT_actionSecTwo} is constructed to precisely reproduce the Weyl anomaly \eqref{transrulestwo}.  In the special case when the IR theory is gapped, namely when $a_\text{IR}=0$ and $c_\text{IR}=0$, the effective action \eqref{eq:EFT_actionSecTwo} simply reduces to \eqref{eq:action_IR_free}.

The effective action \eqref{eq:EFT_actionSecTwo} can be used in practice in the following way. One can take variations of this action with respect to background fields in order to compute vertices of the background fields. Some special vertices are universal quantities which are sensitive to $\Delta a$ and $\Delta c$ only. Particular care should be given to the $A_\text{IR CFT}$ term in \eqref{eq:action_IR_free} as was emphasized also in \cite{Nakayama:2011wq,Luty:2012ww}. It turns out that for some special choices of the background fields and vertices, one can safely ignore $A_\text{IR CFT}$. We will see that the garviton-graviton-dilaton three-point vertex and  the graviton-graviton-dilaton-dilaton four-point vertex are sufficient to universally extract $\Delta c,\Delta a$ from the multiple tensor structures in these vertices. We will then combine the vertices into scattering amplitudes and derive  \eqref{newRintro}. 

\section{Vertices of Gravitons and Dilatons}
\label{sec:vertices}

Recall from the previous section, that the effective action describing the probe fields is given by the equation \eqref{eq:EFT_actionSecTwo} together with \eqref{eq:Aa}  - \eqref{eq:Ainvariant}.
Let us now rewrite it using fields $\varphi(x)$ and $h_{\mu\nu}(x)$ defined as
\begin{equation}
	\label{eq:probes_hd}
	\begin{aligned}
	e^{-\tau (x)} &\equiv 1-\f{\varphi(x)}{\sqrt{2}f},\\
	g_{\mu\nu}(x)&\equiv \eta_{\mu\nu}+2\kappa\, h_{\mu\nu}(x).
\end{aligned}
\end{equation}
We will call $\varphi(x)$ and $h_{\mu\nu}(x)$ the dilaton and the graviton fields respectively.
Here we have introduced dimensionful parameters $f$ and $\kappa$ with mass dimensions $[f]=+1$ and $[\kappa]=-1$. We use  $f^{-1}$ and $\kappa$ as expansion parameters, which allow to keep track of the fields $\varphi(x)$ and $h_{\mu\nu}(x)$. The fields $\varphi(x)$ and $h_{\mu\nu}(x)$ at this stage are used simply as bookkeeping devices for correlators of the energy momentum tensor.
Below, we will impose various constraints on $\varphi(x)$ and $h_{\mu\nu}(x)$ which are compatible with the on-shell conditions we will eventually impose when the fields will become propagating and not just background fields.

In subsection \ref{sec:vertices_definitions} we define vertices of probe fields (dilatons and gravitons). In subsection \ref{sec:vertices_computations} we will compute these vertices from the effective action \eqref{eq:EFT_actionSecTwo}. The latter will depend on $\Delta a$ and $\Delta c$ as well as various parameters that are not related to the anomalies -- we show that certain kinematical structures are universal and depend on the anomalies only. We show that this occurs for vertices of three dilatons, four dilatons, two gravitons and one dilaton, one graviton and two dilatons, and finally, two gravitons and two dilatons.

\subsection{Definition of Vertices}
\label{sec:vertices_definitions}

We define the Fourier transforms
\begin{equation}
	\label{eq:Fourier_transform}
	\varphi(x) = \int \f{d^4k}{(2\pi)^4} \exp(i x\cdot k) \varphi(k),\qquad
	h_{\mu\nu}(x) = \int \f{d^4k}{(2\pi)^4} \exp(i x\cdot k) h_{\mu\nu}(k).
\end{equation}
The vertex in Fourier space which contains $m$ gravitons and $n$ dilatons is defined as the following variation
\begin{multline}
	\label{eq:vertices_index_full}
	(2\pi)^4 \delta^{(4)}(p_1+\ldots+p_m+q_{1}+\ldots+q_{n})\times
	V^{\mu_1\nu_1,\ldots, \mu_m\nu_m}_{(h\ldots h\varphi\ldots \varphi)}(p_1,\ldots,p_m,q_{1},\ldots,q_{n})\equiv\\
	(2\pi)^{4(m+n)}\frac{i\,\delta^ {m+n}A_\text{EFT}}{
		\delta h_{\mu_1\nu_1}(p_1)\ldots\delta h_{\mu_m\nu_m}(p_m)
		\delta \varphi(q_{1})\ldots\delta \varphi(q_{n})}\Bigg{|}_{h, \varphi=0}.
\end{multline}
In order to compute variations we use the following formulas
\begin{equation}
	\frac{\delta \varphi (p)}{\delta \varphi (k)} = \delta^{(4)}(p-k),\qquad
	\frac{\delta h_{\mu_1\mu_2} (p)}{\delta h_{\nu_1\nu_2} (k)} = \frac{1}{2}(\delta^{\nu_1}_{\mu_1}\delta^{\nu_2}_{\mu_2}+\delta^{\nu_2}_{\mu_1}\delta^{\nu_1}_{\mu_2})\times\delta^{(4)}(p-k).
\end{equation}
It is very convenient to define the index-free vertices as follows
\begin{equation}
	\label{eq:vertices_index_free}
	V_{(h\ldots h\varphi\ldots \varphi)}	\equiv\\  \varepsilon_{1,\, \mu_1\nu_1}\ldots \varepsilon_{m,\, \mu_m\nu_m} \times
	V^{\mu_1\nu_1,\ldots, \mu_m\nu_m}_{(h\ldots h\varphi\ldots \varphi)}.
\end{equation}
Here $\varepsilon_i$ are auxiliary complex polarization parameters.
Throughout this paper we will work in the harmonic gauge
\begin{equation}
	\label{eq:graviton_def}
	\p^\mu h_{\mu\nu}-\f{1}{2}\p_\nu h=0.
\end{equation}
This allows to significantly simplify all the expressions. Furthermore, we will assume that polarizations are traceless and transverse, namely
\begin{equation}
	\label{eq:transversality_polarizations}
	\eta^{\mu\nu} \varepsilon_{i,\,\mu\nu} = 0,\qquad
	k_i^{\mu} \varepsilon_{i,\,\mu\nu} = 0.
\end{equation}

These restrictions can be chosen without loss of generality for the following reason: diffeomorphism invariance of the effective action $A_\text{EFT}[\Omega,\, g_{\mu\nu}]$ implies that under infinitesimal coordinate transformations $\delta x^\mu=\chi^\mu(x)$, the dilaton and the metric transform as
	\be
	\delta\varphi&=&-\chi^\rho\p_\rho\varphi\\
	\delta h_{\mu\nu}&=& -\f{1}{2\kappa}\left(\p_\mu\chi_\nu+\p_\nu\chi_\mu\right) -\left(h_{\rho\nu} \p_\mu\chi^\rho+h_{\mu\rho} \p_\nu\chi^\rho +\chi^\rho\p_\rho h_{\mu\nu}\right),\label{eq:gaugetr_h}
	\ee
	and together with~\eqref{transrules}, this allows us to remove the restrictions above.

For our purposes we will need to consider the following five vertices 
\begin{equation}
	\label{eq:vertices}
	V_{(\varphi\varphi\varphi)},\qquad
	V^{\mu_1\nu_1}_{(h \varphi\varphi)},\qquad
	V_{(h h\varphi)},\qquad
	V_{(\varphi\varphi\varphi\varphi)},\qquad
	V_{(h h\varphi\varphi)}.
\end{equation}
Notice that all the vertices above, except for the second one, are contracted with polarizations. This stresses the fact that for the computation of $V^{\mu_1\nu_1}_{(h \varphi\varphi)}$ we do not use~\eqref{eq:transversality_polarizations}.
Below we provide the diagrammatic notation for all the vertices \eqref{eq:vertices}.
{\allowdisplaybreaks 
	\begin{align*}
		V_{(\varphi\varphi\varphi)}(k_1,k_2,k_3)&=
		\begin{tikzpicture}[baseline=(a)]
			\begin{feynman}
				\vertex [blob, minimum size=0.75 cm] (a) {};
				\vertex [above left = of a] (i1);
				\vertex [below left = of a] (i2);
				\vertex [right = of a] (o1);
				\diagram* {
					(i1)--[dashed, momentum=$k_1$](a)--[dashed, rmomentum=$k_2$](i2),  
					(o1)--[dashed, momentum=$k_3$](a)
				};
			\end{feynman}
		\end{tikzpicture}\\
V^{\mu_1\nu_1}_{(h\varphi \varphi)} (k_1,k_2,k_3) &= \begin{tikzpicture}[baseline=(a)]
	\begin{feynman}
		\vertex [blob, minimum size=0.75 cm] (a) {};
		\vertex [left = of a , label=left:$\mu_1\nu_1$] (i1);
		\vertex [above right = of a] (o1);
		\vertex [below right = of a] (o2);
		\diagram* {
			(i1)--[photon, momentum'=$k_1$](a)--[dashed, rmomentum=$k_2$](o1), 
			(a)--[dashed, rmomentum=$k_3$](o2),
		};
	\end{feynman}
\end{tikzpicture}\\
	V_{(h h\varphi)}(k_1,k_2,k_3;\varepsilon_1,\varepsilon_2)&=
	\begin{tikzpicture}[baseline=(a)]
		\begin{feynman}
			\vertex [blob, minimum size=0.75 cm] (a) {};
			\vertex [above left = of a , label=left:$\varepsilon_1$] (i1);
			\vertex [below left = of a , label=left:$\varepsilon_2$] (i2);
			\vertex [right = of a] (o1);
			\diagram* {
				(i1)--[photon, momentum=$k_1$](a)--[photon, rmomentum=$k_2$](i2),  
				(o1)--[dashed, momentum'=$k_3$](a)
			};
		\end{feynman}
	\end{tikzpicture}\\
			V_{(\varphi \varphi \varphi \varphi)}(k_1,k_2,k_3,k_4) &=
		\begin{tikzpicture}[baseline=(a)]
			\begin{feynman}
				\vertex [blob, minimum size=0.75 cm] (a) {};
				\vertex [above left = of a ] (i1);
				\vertex [below left = of a ] (i2);
				\vertex [above right = of a] (o1);
				\vertex [below right = of a] (o2);
				\diagram* {
					(i1)--[dashed, momentum'=$k_1$](a)--[dashed, rmomentum=$k_2$](i2), 
					(o1)--[dashed, momentum'=$k_3$ ](a)--[dashed, rmomentum=$k_4$](o2),
				};
			\end{feynman}
		\end{tikzpicture}\\
		V_{( h h \varphi \varphi)}(k_1,k_2,k_3,k_4;\varepsilon_1,\varepsilon_2) &=
	\begin{tikzpicture}[baseline=(a)]
		\begin{feynman}
			\vertex [blob, minimum size=0.75 cm] (a) {};
			\vertex [above left = of a , label=left:$\varepsilon_1$] (i1);
			\vertex [below left = of a , label=left:$\varepsilon_2$] (i2);
			\vertex [above right = of a] (o1);
			\vertex [below right = of a] (o2);
			\diagram* {
				(i1)--[photon, momentum'=$k_1$](a)--[photon, rmomentum=$k_2$](i2), 
				(o1)--[dashed, momentum'=$k_3$ ](a)--[dashed, rmomentum=$k_4$](o2),
			};
		\end{feynman}
	\end{tikzpicture}
	\end{align*}}

\subsection{Computations of Vertices}
\label{sec:vertices_computations}

In this section we will compute the vertices \eqref{eq:vertices} using the effective action~\eqref{eq:EFT_actionSecTwo}. The main results of this section are given by equations \eqref{eq:V3phi}, \eqref{eq:V_hphiphi_EFT_prediction}, \eqref{eq:hhphi_covariant}, \eqref{eq:result_dddd} and \eqref{result_hhdd}.

In what follows we will write some vertices with additional constraints such as $\p^2\varphi(x)=0$ and $\p^2 h_{\mu\nu}(x)=0$. When this is done we carefully emphasise this in the text. At this point these constraints are not the equations of motion but simply a particular choice for the background field configuration which simplifies the expressions.

\subsubsection{Three-Point Vertices}

Let us start by writing the part of the effective action~\eqref{eq:EFT_actionSecTwo} which contains three dilaton fields $\varphi$ and four derivatives
\be
\label{eq:3-dilaton_action}
A_\text{EFT}&=&\Delta a\frac{\sqrt{2}}{f^3}\int d^4x\ (\p\varphi)^2\p^2\varphi +r_1\frac{18 \sqrt{2}}{f^3}\int d^4x\ \varphi (\p^2\varphi)^2 + \ldots.
\ee
We obtain
\begin{multline}
	\label{eq:V3phi}
	V_{(\varphi\varphi\varphi)}(k_1,k_2,k_3)
	=  \frac{i\sqrt{2}}{f^3} \bigg(
	\Delta a\left(\left(k_1^2\right)^2+\left(k_2^2\right)^2+\left(k_3^2\right)^2\right)\\
	+2(18 r_1-\Delta a)\left(k_1^2 k_2^2+k_2^2 k_3^2 +k_3^2 k_1^2\right)+\ldots\bigg),
\end{multline}
where $k_3=-k_1-k_2$. The ellipsis denote terms of different powers of momenta coming from other derivative terms in the effective action. If we impose $\p^2\varphi(x)=0$ the $V_{(\varphi\varphi\varphi)}$ vertex vanishes.

Now consider the four derivative part of the EFT action which contains two dilatons and one graviton. It reads as
\be
A_{\text{EFT}}&=& \f{2\kappa \Delta a}{f^2}\int d^4x\left( \p^2 h^{\mu\nu}-\f{1}{2}\p^2h \eta^{\mu\nu}\right)\p_\mu\varphi\p_\nu\varphi\nn\\
&&+\f{r_1\kappa}{f^2}\int d^4x\left( 18h\p^2\varphi\p^2\varphi -72 h^{\mu\nu}\p_\mu\p_\nu\varphi \p^2\varphi -6\p^2h \varphi\p^2\varphi\right).
\ee
From the above action the 2-dilaton-1-graviton EFT vertex becomes
\begin{equation}
	\label{eq:V_hphiphi_EFT_prediction}
	\begin{aligned}
			V^{\mu\nu}_{(h\varphi \varphi)} (k_1,k_2,k_3) &= 
		\f{i\kappa}{f^2}\left(6M^4\lambda\eta^{\mu\nu}+M^2 r_0\left( \eta^{\mu\nu}(k_1^2+3k_2^2+3k_3^2)-6(k_{2}^{\mu}k_{2}^{\nu}+k_{3}^{\mu}k_{3}^{\nu})\right)\right)\\
		&+\frac{2i\kappa \Delta a}{f^2}k_1^2\,\left(k_{2}^\mu k^\nu_{3}+k^\nu_{2}k^\mu_{3}-\eta^{\mu\nu}k_2.k_3\right)\\\
		&+\f{ir_1\kappa}{f^2}\left( 36\eta^{\mu\nu}k_2^2 k_3^2 -6\eta^{\mu\nu}k_1^2(k_2^2+k_3^2)-72(k^\mu_{2}k^\nu_{2}k_3^2 +k^\mu_{3}k^\nu_{3}k_2^2)\right). 
	\end{aligned}
\end{equation}
Here for completeness we have added terms with zero and two powers of momenta coming from the terms in the Lagrangian with zero and two derivatives. Due to momentum conservation we have $k_3=-k_1-k_2$ as before. Here we do not impose on-shell conditions on the dilaton or the metric since this particular vertex will be useful when these are off-shell. By using momentum conservation and the harmonic gauge the four-momentum part of the equation \eqref{eq:V_hphiphi_EFT_prediction} can be rewritten as 
\be\label{eq:V_hphiphi_EFT_predictiongauge}
V^{\mu\nu}_{(h\varphi \varphi)} (k_1,k_2,k_3)&=&\f{i\kappa}{f^2}\eta^{\mu\nu}\Big[ -12r_1k_1^2 k_1.k_2+(2\Delta a-12r_1)k_1^2k_2^2+36r_1(k_2^2)^2-6r_1(k_1^2)^2\Big]\nn\\
&&-\f{i\kappa}{f^2}k_2^\mu k_2^\nu \Big[ (4\Delta a+72 r_1)k_1^2 +144 r_1(k_2^2 + k_1.k_2)\Big].
\ee
In order to implement harmonic gauge \eqref{eq:graviton_def}, one replaces terms of the form
$k_1^\mu A^\nu$ or $k_1^\nu A^\mu$ by $\f{1}{2}(k_1.A) \eta^{\mu\nu}$ in \eqref{eq:V_hphiphi_EFT_prediction} to bring it to the above form. The above expression of graviton-dilaton-dilaton vertex is closely related to the ANEC construction of~\cite{Hartman:2023qdn}. It turns out that the projection operator $(\p_{k_{2u}}-\p_{k_{3u}})^2 \vec{\p}_{k_2}\cdot \vec{\p}_{k_3}$ of~\cite{Hartman:2023qdn}, when applied to the vertex \eqref{eq:V_hphiphi_EFT_predictiongauge} after setting $k_1=-k_2-k_3$, yields only $\Delta a$, hence it is consistent with the sum rule.

Let us now consider the part of the effective action \eqref{eq:EFT_actionSecTwo} which contains two gravitons and one dilaton, and four derivatives. It reads as
\begin{multline}
\label{eq:4-derivative-phi-h-h}
A_\text{EFT}=  \f{\kappa^2}{\sqrt{2}f} \int d^4x\ \Big(
2(2\Delta a-\Delta c) \varphi\p^2 h_{\mu\nu}\p^2 h^{\mu\nu}\\
+(-\Delta a+\Delta c)\varphi\left( 4\p_\rho\p_\nu h_{\mu\sigma}\p^\rho\p^\nu h^{\mu\sigma}+4\p_\rho\p_\nu h_{\mu\sigma}\p^\mu \p^\sigma h^{\nu\rho}-8\p_\rho\p_\nu h_{\mu\sigma}\p^\mu \p^\rho h^{\nu\sigma}\right) \\
+ 12r_1 \p^2\varphi \big( 4h^{\alpha\beta}\p^2 h_{\alpha\beta}+3\p_\mu h_{\alpha\beta}\p^\mu h^{\alpha\beta}-2\p^\beta h^{\mu\alpha}\p_\alpha h_{\mu\beta}\big) \Big).
\end{multline}
We obtain
\begin{multline}
\label{eq:hhphi_covariant}
V_{(hh \varphi)}(k_1,k_2,k_3;\varepsilon_1,\varepsilon_2) =\\
f_1(k_1,k_2)\times(\varepsilon_1. \ve_2)
+f_2(k_1,k_2)\times(k_1.\varepsilon_2.k_1)(k_2.\ve_1.k_2)+f_3(k_1,k_2)\times(k_1.\ve_2.\ve_1.k_2),
\end{multline}
where the functions $f_i(k_1,k_2)$ read as
\begin{equation}
	\label{eq:vertex_hhd_functions}
	\begin{aligned}
		f_1(k_1,k_2) &= \frac{4i\kappa^2}{\sqrt{2}f}\,\Big(
		2 (-\Delta a+\Delta c+18 r_1)(k_1.k_2)^2
		+(2\Delta a-\Delta c+24 r_1)k_1^2 k_2^2\\
		&+12r_1(k_1^4+k_2^4)+ 42 r_1 (k_1.k_2)(k_1^2+k_2^2)+\ldots\Big),\\
		f_2(k_1,k_2) &= \frac{8i\kappa^2}{\sqrt{2}f}\,\left(-\Delta a+\Delta c+\ldots\right),\\
		f_3(k_1,k_2) &= \frac{8i\kappa^2}{\sqrt{2}f}\,\left(
		2(\Delta a-\Delta c-6r_1)(k_1.k_2)-6r_1(k_1^2+k_2^2)+\ldots\right).
	\end{aligned}
\end{equation}
We used $k_3=-k_1-k_2$. Again the ellipsis denote the terms which are not contributing at four powers of momenta.
In the above we have used the following short-hand notation
\begin{equation}
	\label{eq:shorthand_notation}
(\varepsilon_1. \varepsilon_2)\equiv \varepsilon_{1\mu\nu}\varepsilon_2^{\mu\nu},\quad
(p_i.\varepsilon_j.p_k)\equiv p_{i\mu}\varepsilon_j^{\mu\nu}p_{k\nu},\quad
(p_i.\varepsilon_1.\varepsilon_2.p_j)\equiv p_{i\mu}\varepsilon_1^{\mu\rho}\varepsilon_{2\rho\nu}p_j^\nu.
\end{equation}

The vertex \eqref{eq:hhphi_covariant} is invariant under the $1\leftrightarrow 2$ exchange as expected (this is due to 	$(p_i.\varepsilon_j.p_k) = (p_k.\varepsilon_j.p_i)$, $(p_i.\varepsilon_1.\varepsilon_2.p_j)=(p_j.\varepsilon_2.\varepsilon_1.p_i)$).
Unlike the $V_{(\varphi\varphi\varphi)}$ vertex, we see that~\eqref{eq:4-derivative-phi-h-h} does not trivialize upon imposing  $\p^2\varphi(x)=0$. Only the term proportional to $r_1$ trivializes, so after imposing $\p^2\varphi(x)=0$ we find
\begin{equation}
	\label{eq:vertex_hhd_functionsi}
	\begin{aligned}
		f_1(k_1,k_2) &= \frac{4i\kappa^2}{\sqrt{2}f}\,\Big(
		2 (-\Delta a+\Delta c)(k_1.k_2)^2
		+(2\Delta a-\Delta c)k_1^2 k_2^2+\ldots\Big),\\
		f_2(k_1,k_2) &= \frac{8i\kappa^2}{\sqrt{2}f}\,\left(-\Delta a+\Delta c+\ldots\right),\\
		f_3(k_1,k_2) &= \frac{16i\kappa^2}{\sqrt{2}f}\,\left(
		(\Delta a-\Delta c)(k_1.k_2)+\ldots\right).
	\end{aligned}
\end{equation}

Additionally, imposing  $\p^2\varphi(x)=0$ removes contributions from the infrared CFT as we show in appendix~\ref{DimTwo}. We remained with three distinct kinematical structures that depend on different combinations of the $a$- and $c$-anomalies only.

\subsubsection{Four-point Vertices}
The part of the effective action~\eqref{eq:EFT_actionSecTwo} which contains four dilatons and only four derivatives and with the constraint  $\p^2\varphi(x)=0$ has the following simple form
\begin{equation}
\label{eq:4-dilaton_action}
A_\text{EFT}=\Delta a\f{1}{2f^4}\int d^4x\ (\p\varphi)^4+\ldots.
\end{equation}
We obtain the following four-point vertex
\begin{equation}
\label{eq:result_dddd}
V_{(\varphi\varphi\varphi\varphi)}(k_1,k_2,-k_3,-k_4) = \frac{i\Delta a}{f^4} \left(s^2+t^2+u^2\right) + \ldots.
\end{equation}
Here we have used the Mandelstam variables defined as
\begin{equation}
	\label{eq:mandelstam_variables}
	s\equiv - (k_1+k_2)^2,\qquad
	t\equiv - (k_1-k_3)^2,\qquad
	u\equiv - (k_1-k_4)^2.
\end{equation}
Notice that we have written the vertex \eqref{eq:result_dddd} in such a way that the momenta $k_3$ and $k_4$ can be interpreted as the momenta of outgoing particles. We also have the standard relations that stem from momentum conservation and our choice of background fields $k_i^2=0, s+t+u=0$. Since $V_{(\varphi\varphi\varphi)}(k_1,k_2,k_3)$ vanishes when $\p^2\varphi(x)=0$ is imposed, it is immediate to see that the physical dilaton-dilaton scattering amplitude is then given by~\eqref{oldRintro}. 

The part of the effective action~\eqref{eq:EFT_actionSecTwo} which contains two gravitons and two dilatons and four derivatives has the  form
\begin{multline}\label{comps}
A_\text{EFT}= (-\Delta a +\Delta c)\f{\kappa^2}{4f^2}\int d^4x\ \varphi(x)^2 \Big( 4\p_\rho\p_\nu h_{\mu\sigma}\p^\rho\p^\nu h^{\mu\sigma}+4\p_\rho\p_\nu h_{\mu\sigma}\p^\mu \p^\sigma h^{\nu\rho}-8\p_\rho\p_\nu h_{\mu\sigma}\p^\mu \p^\rho h^{\nu\sigma}\Big)\\
- \Delta a \f{2\kappa^2}{f^2} \int d^4x\ \p_\mu\varphi\p_\nu\varphi \Big( \p^{\mu}h_{\alpha\beta}\p^{\nu}h^{\alpha\beta}+2h^{\alpha\beta}\p^{\mu}\p^{\nu}h_{\alpha\beta}-2h^{\alpha\beta}\p^{\nu}\p_{\beta}h_{\alpha}^{\mu}-2h^{\alpha\beta}\p^{\mu}\p_{\beta}h_{\alpha}^{\nu}\\
+2h^{\alpha\beta}\p_{\alpha}\p_{\beta}h^{\mu\nu} +2\p^{\beta}h^{\nu}_{\alpha}\p_{\beta}h^{\mu \alpha}-2\p_{\beta}h^{\nu}_{\alpha}\p^{\alpha}h^{\mu \beta}-\f{1}{2}\eta^{\mu\nu}\big( 3\p_\rho h_{\alpha\beta}\p^\rho h^{\alpha\beta}-2\p^\beta h^{\rho\alpha}\p_\alpha h_{\rho\beta}\big)\Big)\\
+r_1\f{\kappa^2}{2 f^2}\int d^4x\ \Big( 144\ h^{\mu\nu}h^{\rho\sigma}\p_\mu\p_\nu\varphi \p_\rho\p_\sigma\varphi  \Big)+\ldots
\end{multline}
In writing~\eqref{comps} we have used
\begin{equation}
	\label{eq:condition_2}
\p^2 h_{\mu\nu}(x)=0~,\quad \p^2 \varphi(x)=0~.\end{equation}

We obtain the following form of the vertex
\begin{multline}
	\label{result_hhdd}
	V_{(hh\varphi\varphi)}(k_1,k_2,-k_3,-k_4;\varepsilon_1,\varepsilon_2) =
	g_1(s,t,u)\times (\varepsilon_1. \varepsilon_2)
	+g_2(s,t,u)\times (k_1.\varepsilon_2.\varepsilon_1.k_2)\\
	+g_3(s,t,u)\times(k_3.\varepsilon_2.\varepsilon_1.k_2)
	+g_3(s,u,t)\times(k_1.\varepsilon_2.\varepsilon_1.k_3)
	+g_4(s,t,u)\times(k_3.\varepsilon_2.\varepsilon_1.k_3)\\
	+g_5(s,t,u)\times(k_1.\varepsilon_2.k_1)(k_2.\varepsilon_1.k_2) +g_6(s,t,u)\times(k_3.\varepsilon_2.k_3)(k_2.\varepsilon_1.k_2)\\
	+g_6(s,u,t)\times(k_3.\varepsilon_1.k_3)(k_1.\varepsilon_2.k_1) +g_7(s,t,u)\times(k_2.\varepsilon_1.k_3)(k_1.\varepsilon_2.k_3)\\
	+g_8(s,t,u)\times(k_3.\varepsilon_1.k_3)(k_3.\varepsilon_2.k_3)
	+g_9(s,t,u)\times(k_3.\varepsilon_1.k_3)(k_3.\varepsilon_2.k_1)\\
	+g_9(s,u,t)\times(k_3.\varepsilon_1.k_2)(k_3.\varepsilon_2.k_3),
\end{multline}
where the functions $g_i(s,t,u)$ read as
\begin{equation}
	\label{eq:g_functions}
	\begin{aligned}
		g_1(s,t,u) &= \frac{i\kappa^2}{3f^2}\left(
		-36M^4\lambda- 9M^2 r_0 s +
		3(-\Delta a+\Delta c)s^2 +3\Delta a(s^2+t^2+u^2)+\ldots\right), \\
		g_2(s,t,u) &= \frac{i\kappa^2}{3f^2}\left(
		12M^2 r_0+
		12(-\Delta a+\Delta c)s+\ldots\right),\\
		g_3(s,t,u) &= \frac{i\kappa^2}{3f^2}\left(
		-72M^2 r_0
		-24\Delta a\, u+\ldots\right),\\		
		g_4(s,t,u) &= \frac{i\kappa^2}{3f^2}\left(144M^2 r_0
		-24\Delta a\, s+\ldots\right),\\
		g_5(s,t,u) &= \frac{i\kappa^2}{3f^2}\left(12(-\Delta a +\Delta c)+\ldots\right),\\		
		g_6(s,t,u) &=  \frac{i\kappa^2}{3f^2}\left( 24\Delta a+432r_1+\ldots\right),\qquad
		g_7(s,t,u) = \frac{i\kappa^2}{3f^2}\left(-48\Delta a+\ldots\right),\\
		g_8(s,t,u) &= \frac{i\kappa^2}{3f^2}\left(864r_1+\ldots\right),\qquad\qquad\quad\;\;
        g_9(s,t,u) =\frac{i\kappa^2}{3f^2}\left(-864r_1+\ldots\right).
	\end{aligned}
\end{equation}
Here for completeness we added terms with zero and two powers of momenta.
The ellipsis denote the terms with higher powers of momenta. Using the definitions in~\eqref{eq:shorthand_notation}, it is straightforward to check that the result \eqref{result_hhdd} is symmetric under the $1\leftrightarrow 2$ exchange.

We see that $\Delta a, \Delta c$ appear in multiple kinematical structures of the four-point function. An important property shown in appendix~\ref{DimTwo} is that the contributions from the infrared CFT all have the same kinematical dependence as the term proportional to $r_1$ in~\eqref{comps}. Therefore, our prediction for how $\Delta a, \Delta c$ appear in the four-point vertex $V_{(hh\varphi\varphi)}$ is robust.

\section{Examples of $\Delta a$ and $\Delta c$ from Dilaton-Graviton Vertices}
\label{sec:examples}

Let us consider two simple examples: the theory of a free massive scalar and the theory of a free massive Dirac fermion. In the UV the mass terms are relevant, thus in the UV these theories are described by free conformal field theories. The respective values of the $a$- and $c$-anomalies have been reported in \cite{Osborn:1993cr}. In the IR both theories are empty, which means that both anomalies are zero. Using the results of \cite{Osborn:1993cr} we can write then
\begin{equation}
	\label{eq:answer_anomalies_free}
	\begin{aligned}
		\text{free scalar:}&\qquad
		\Delta a =   \f{1}{5760\pi^2},\qquad 
		\Delta c =   \f{3}{5760\pi^2},\\
		\text{free Dirac fermion:}&\qquad
		\Delta a =  \f{11}{5760\pi^2},\qquad 
		\Delta c =  \f{18}{5760\pi^2}.
	\end{aligned}
\end{equation}

In this section we couple these theories to the dilaton field $\Omega(x)$ and place these theories on a curved background with the metric $g_{\mu\nu}(x)$. We then compute some of the vertices \eqref{eq:vertices}. Comparing the result with the predictions provided in section \ref{sec:vertices_computations} we can extract the values of $\Delta a$ and $\Delta c$. We will see that the obtained values are in a perfect agreement with the known values \eqref{eq:answer_anomalies_free} and the structure of the vertices is in accord with our expectations.

We first report our results in $d$ space-time dimensions. This will be important for regularizing divergent loop integrals in $d=4$ . The final results for the vertices will be given in $d=4$.

In the computations below we will use the following relations
\begin{align}
	\int\f{d^d\ell}{(2\pi)^d} f(\ell^2) \ell^{\mu_1}\ell^{\mu_2} &= \frac{\eta^{\mu_1\mu_2}}{d} \int\f{d^d\ell}{(2\pi)^d} f(\ell^2)
	\ell^2,\\
	\int\f{d^d\ell}{(2\pi)^d} f(\ell^2) \ell^{\mu_1}\ell^{\mu_2} \ell^{\mu_3}\ell^{\mu_4} &= \frac{\eta^{\mu_1\mu_2}\eta^{\mu_3\mu_4}+\eta^{\mu_1\mu_3}\eta^{\mu_2\mu_4}+\eta^{\mu_1\mu_4}\eta^{\mu_2\mu_3}}{d(d+2)}\int\f{d^d\ell}{(2\pi)^d} f(\ell^2) \left(\ell^2\right)^2.\nn
\end{align}
Similar integrals which involve an odd number of $\ell^\mu$ momenta vanish.
Finally, we will need the following scalar integrals
\begin{equation}
	\mathbf{J}\left(a,b;\Delta\right)\equiv  \int\f{d^d\ell}{(2\pi)^d} \f{(\ell^2)^{a}}{\left(\ell^2+\Delta-i\epsilon\right)^b}
	= \f{i}{(4\pi)^{\f{d}{2}}}\f{1}{\Delta^{b-a-\f{d}{2}}}\f{\Gamma\left(a+\f{d}{2}\right)\Gamma\left(b-a-\f{d}{2}\right)}{\Gamma(b)\Gamma\left(\f{d}{2}\right)} \label{eq:master_integral}.
\end{equation}

\subsection{Free Massive Scalar}
\label{sec:example}
Recall the standard action for the free massive scalar
\begin{equation}
	A_\text{free scalar}[\Phi] \equiv \int d^4 x \left(-\f{1}{2}\eta^{\mu\nu}\p_\mu\Phi(x) \p_\nu\Phi(x) -\f{1}{2}m^2\Phi^2(x) \right).
\end{equation}
Adding the dilaton field and placing this theory on a $d$-dimensional curved background we obtain the following action
\begin{multline}
	\label{eq:compensated_free_scalar}
	A^\text{compensated}_\text{free scalar}[\Phi, \varphi, h]\equiv 
	\int d^dx\ \sqrt{-g}\Bigg(-\f{1}{2}g^{\mu\nu}\p_\mu\Phi \p_\nu\Phi -\f{1}{2}m^2 e^{-2\tau}\Phi^2 -\f{d-2}{8(d-1)}R\Phi^2\Bigg).
\end{multline}

Let us now evaluate the Feynman rules for the action \eqref{eq:compensated_free_scalar}. 
We will denote by solid lines  massive scalar particles, by dashed lines dilatons and by wavy lines gravitons. 
We obtain
\begin{equation*}
	\feynmandiagram[horizontal=a to b, inline=(a.base)] {
		a -- [momentum=$p$] b
	};\  = \frac{-i}{p^2 + m^2 -i\epsilon},\qquad
	\begin{tikzpicture}[baseline=(a)]
		\begin{feynman}
			\vertex (a);
			\vertex [above left = of a] (i1);
			\vertex [below left = of a] (i2);
			\vertex [right = of a] (o1);
			\diagram* {
				(i1)--[momentum=$p_1$](a)--[rmomentum=$p_2$](i2),  
				(o1)--[dashed, momentum'=$k$](a),
			};
		\end{feynman}
	\end{tikzpicture} =  \f{i\sqrt{2}}{f}m^2,
\end{equation*}
\begin{equation*}
	\begin{tikzpicture}[baseline=(a)]
	\begin{feynman}
		\vertex (a);
		\vertex [above left = of a] (i1);
		\vertex [below left = of a] (i2);
		\vertex [above right = of a ] (o1);
		\vertex [below right = of a ] (o2);
		\diagram* {
			(i1)--[momentum'=$p_1$](a)--[rmomentum=$p_2$](i2), 
			(o1)--[dashed, momentum'=$k_1$ ](a)--[dashed, rmomentum=$k_2$](o2),
		};
	\end{feynman}
\end{tikzpicture} = -\f{i}{f^2}m^2,\qquad
	\begin{tikzpicture}[baseline=(a)]
		\begin{feynman}
			\vertex (a);
			\vertex [above left = of a] (i1);
			\vertex [below left = of a] (i2);
			\vertex [above right = of a , label=right:$\varepsilon_3$] (o1);
			\vertex [below right = of a , label=right:$\varepsilon_4$] (o2);
			\vertex [right = of a] (o3);
			\diagram* {
				(i1)--[momentum'=$p_1$](a)--[rmomentum=$p_2$](i2), 
				(o1)--[photon ](a)--[photon](o2),
				(a)--[dashed](o3),
			};
		\end{feynman}
	\end{tikzpicture} = -\f{i2\sqrt{2}\kappa^2}{f}m^2 \ \eta_{\mu\alpha}\eta_{\nu\beta} \varepsilon_3^{\mu\nu}\varepsilon_4^{\alpha\beta},
\end{equation*}
\begin{align*}
	\begin{tikzpicture}[baseline=(a)]
		\begin{feynman}
			\vertex (a);
			\vertex [above left = of a] (i1);
			\vertex [below left = of a] (i2);
			\vertex [above right = of a , label=right:$\varepsilon_3$] (o1);
			\vertex [below right = of a , label=right:$\varepsilon_4$] (o2);
			\diagram* {
				(i1)--[momentum'=$p_1$](a)--[rmomentum=$p_2$](i2), 
				(o1)--[photon, momentum'=$k_1$ ](a)--[photon, rmomentum=$k_2$](o2),
			};
		\end{feynman}
	\end{tikzpicture} &=
	\f{i\kappa^2}{3}\varepsilon_3^{\mu\nu}\varepsilon_4^{\alpha\beta}\textbf{Sym}_{(\mu\nu),(\alpha\beta)}\Big( 6\eta_{\mu\alpha}\eta_{\nu\beta}(-p_1.p_2 +m^2)
	+6\eta_{\nu\alpha}(p_{1\mu}p_{2\beta}+p_{1\beta}p_{2\mu})\\ 
	&+6\eta_{\mu\beta}(p_{1\nu}p_{2\alpha}+p_{1\alpha}p_{2\nu})+\f{3(d-2)}{2(d-1)}\left( (2k_1^2+2k_2^2+ 3k_1.k_2) \eta_{\mu\alpha}\eta_{\nu\beta}-2\eta_{\mu\alpha}k_{1\beta}k_{2\nu}\right) \Big).
\end{align*} 
We assumed that the gravitons in these expressions obey traceless-ness and transversality. This is enough since the gravitons in these diagrams will always appear as external and will be contracted with polarizations. In addition we also have the following two diagrams
\begin{equation*}
	\begin{tikzpicture}[baseline=(a)]
		\begin{feynman}
			\vertex (a);
			\vertex [above left = of a] (i1);
			\vertex [below left = of a] (i2);
			\vertex [right = of a , label=right:$\mu\nu$] (o1);
			\diagram* {
				(i1)--[momentum=$p_1$](a)--[rmomentum=$p_2$](i2),  
				(o1)--[photon, momentum'=$k$](a),
			};
		\end{feynman}
	\end{tikzpicture} = - i\kappa\left( p_{1\mu}p_{2\nu}+p_{1\nu}p_{2\mu} +\eta_{\mu\nu}\left(-p_1.p_2+m^2+\f{d-2}{4(d-1)}k^2\right)\right),
\end{equation*}
\begin{equation*}
	\begin{tikzpicture}[baseline=(a)]
		\begin{feynman}
			\vertex (a);
			\vertex [above left = of a] (i1);
			\vertex [below left = of a] (i2);
			\vertex [above right = of a , label=right:$\mu\nu$] (o1);
			\vertex [below right = of a ] (o2);
			\diagram* {
				(i1)--[momentum'=$p_1$](a)--[rmomentum=$p_2$](i2),  
				(o1)--[photon, momentum'=$k_1$](a)--[dashed, rmomentum=$k_2$](o2)
			};
		\end{feynman}
	\end{tikzpicture}= \f{i\sqrt{2}\kappa}{f}m^2\eta_{\mu\nu}.
\end{equation*}
In writing these vertices we did not make any further assumptions. 

The simplest three-point vertex is the three dilaton vertex $V_{(\varphi\varphi\varphi)}$ which has been evaluated in 
\cite{Komargodski:2011vj}. We will not discuss it further here. Instead below we focus on the vertices $V^{\mu\nu}_{(h\varphi\varphi)}$ and $V_{(hh\varphi)}$. We will conclude this section by computing the four-point vertex $V_{(hh\varphi\varphi)}$.

\subsubsection{Graviton-Dilaton-Dilaton Vertex}
The three-point vertex $V^{\mu\nu}_{(h\varphi\varphi)}$ in the theory of massive scalar is given by the sum of the following four diagrams
\begingroup
\allowdisplaybreaks
\be
&&V^{\mu\nu}_{(h\varphi \varphi)} (k_1,k_2,k_3) = \begin{tikzpicture}[baseline=(a)]
	\begin{feynman}
		\vertex [blob, minimum size=0.75 cm] (a) {};
		\vertex [left = of a , label=left:$\mu\nu$] (i1);
		\vertex [above right = of a] (o1);
		\vertex [below right = of a] (o2);
		\diagram* {
			(i1)--[photon, momentum'=$k_1$](a)--[dashed, rmomentum=$k_2$](o1), 
			(a)--[dashed, rmomentum=$k_3$](o2),
		};
	\end{feynman}
\end{tikzpicture}\nn\\
&=& \begin{tikzpicture}[baseline=(a.base)]
	\begin{feynman}
		\vertex (a);
		\vertex [right=2.0 of a] (b);
		\vertex [left=of a, label=left:$k_1$] (i1);
		\vertex [above right=of b, label=right:$k_2$] (i2);
		\vertex [below right=of b, label=right:$k_3$] (o1);
		\diagram* {
			(a) --[half left, momentum={$\ell$}] (b) -- [half left, rmomentum=$k_1-\ell$] (a),
			(i1) --[photon] (a),
			(o1) --[dashed](b)--[dashed]  (i2),
		};
	\end{feynman}
\end{tikzpicture}
\qquad +\qquad
\begin{tikzpicture}[baseline=(a.base)]
	\begin{feynman}
		\vertex (a);
		\vertex [below right=1 and 1.5 of a] (b);
		\vertex [above right=1 and 1.5 of a] (c);
		\vertex [left=of a, label=left:$k_1$] (i1);
		\vertex [right=of b, label=right:$k_2$] (i2);
		\vertex [right=of c, label=right:$k_3$] (i3);
		\diagram* {
			(a) --(b) --(c) --[rmomentum'={$\ell$}] (a),
			(i1) --[photon] (a), (i2)--[dashed] (b),
			(i3) --[dashed] (c),
		};
	\end{feynman}
\end{tikzpicture}\nn\\
&+&
\begin{tikzpicture}[baseline=(a.base)]
	\begin{feynman}
		\vertex (a);
		\vertex [right=2.0 of a] (b);
		\vertex [left=of a, label=left:$k_2$] (i1);
		\vertex [above right=of b, label=right:$k_1$] (i2);
		\vertex [below right=of b, label=right:$k_3$] (o1);
		\diagram* {
			(a) --[half left, momentum={$\ell$}] (b) -- [half left, rmomentum=$k_2-\ell$] (a),
			(i1) --[dashed] (a),
			(o1) --[dashed](b)--[photon]  (i2),
		};
	\end{feynman}
\end{tikzpicture}
\qquad +\qquad
\begin{tikzpicture}[baseline=(a.base)]
	\begin{feynman}
		\vertex (a);
		\vertex [right=2.0 of a] (b);
		\vertex [left=of a, label=left:$k_3$] (i1);
		\vertex [above right=of b, label=right:$k_1$] (i2);
		\vertex [below right=of b, label=right:$k_2$] (o1);
		\diagram* {
			(a) --[half left, momentum={$\ell$}] (b) -- [half left, rmomentum=$k_3-\ell$] (a),
			(i1) --[dashed] (a),
			(o1) --[dashed](b)--[photon]  (i2),
		};
	\end{feynman}
\end{tikzpicture} 
\label{fig:V_hphiphi}
\ee
\endgroup
We can now use the above Feynman diagrams to compute this vertex. Let us stress that we will not make assumptions of transversality and tracelessness in the computation below (but we will employ the harmonic gauge, when said so explicitly).

The first diagram in \eqref{fig:V_hphiphi} reads as
\be
V_{1\mu\nu}&=&\f{\kappa m^2}{2f^2}\int_0^1 dx\Bigg[ \eta_{\mu\nu}\left\lbrace \left(1-\f{2}{d}\right)\mathbf{J}(1,2;\Delta_1)  +k_1^2 \left( \f{d-2}{4(d-1)}-x(1-x) \right)\mathbf{J}(0,2;\Delta_1) \right\rbrace\nn\\
&&+\eta_{\mu\nu}m^2\mathbf{J}(0,2;\Delta_1) +2x(1-x) k_{1\mu}k_{1\nu}\mathbf{J}(0,2;\Delta_1)\Bigg],
\ee
where $\Delta_1 =m^2 +x(1-x)k_1^2$. 
The second diagram in \eqref{fig:V_hphiphi} reads as
\be
V_{2\mu\nu}&=&-\f{4\kappa m^4}{f^2}\int_0^1 dx\int_0^{1-x}dy\ \Bigg[ \eta_{\mu\nu}\left\lbrace \left(1-\f{2}{d}\right)\mathbf{J}(1,3;\Delta_2)+m^2 \mathbf{J}(0,3;\Delta_2)\right\rbrace \nn\\
&& +\eta_{\mu\nu}\left\lbrace -k_1^2 x(1-x)+k_3^3 y^2 +y(1-2x)k_1.k_3 +\f{d-2}{4(d-1)}k_1^2 \right\rbrace \mathbf{J}(0,3;\Delta_2)\nn\\
&& +\left\lbrace (k_1x-k_3y)_\mu \big((1-x)k_1 +k_3y\big)_\nu+(k_1x-k_3y)_\nu \big((1-x)k_1 +k_3y\big)_\mu \right\rbrace\nn\\
&&\times\  \mathbf{J}(0,3;\Delta_2)\Bigg],
\ee
where $\Delta_2 =m^2 +x(1-x)k_1^2 +y(1-y)k_3^2 +2xy(k_1\cdot k_3)$. 
The third diagram in \eqref{fig:V_hphiphi} reads as
\be
V_{3\mu\nu}&=&\f{\kappa m^4}{f^2}\ \eta_{\mu\nu}\int_0^1 dx\ \mathbf{J}(0,2;\Delta_3),
\ee
where $\Delta_3=m^2+x(1-x)k_2^2$. 
The fourth diagram in \eqref{fig:V_hphiphi} reads as
\be
V_{4\mu\nu}&=&\f{\kappa m^4}{f^2}\ \eta_{\mu\nu}\int_0^1 dx\ \mathbf{J}(0,2;\Delta_4),
\ee
where $\Delta_4=m^2+x(1-x)k_3^2$.  The final expression for the one-graviton-two-dilaton vertex is given by the following sum
\be
V^{\mu\nu}_{(h\varphi \varphi)} (k_1,k_2,k_3) = V_1^{\mu\nu} +V_2^{\mu\nu}+ V_3^{\mu\nu} +V_4^{\mu\nu}. \label{eq:Vertex_total}
\ee

The low energy expansion of \eqref{eq:Vertex_total} after substituting $k_3=-k_1-k_2$ and picking up the fourth power of the momenta is given by
\be
V^{\mu\nu}_{(h\varphi \varphi)} (k_1,k_2,k_3)&=&-\frac{i \kappa}{5760 \pi ^2 f^2}\eta^{\mu\nu} \Big[2 k_1^2 \left(4 k_1.k_2+5 k_2^2\right)+7 (k_1^2)^2-2 \big(4 (k_1.k_2)^2+6 k_1.k_2 k_2^2\nn\\
&&+3(k_2^2)^2\big)\Big]+\f{4i\kappa}{5760\pi^2 f^2}k_1^\mu k_1^\nu\left(3 k_1^2 + 6 k_1.k_2 + 5 k_2^2\right) -\f{8i\kappa}{5760\pi^2 f^2}(2k_1^2 +3k_1.k_2\nn\\
&& +3k_2^2)k_2^\mu k_2^\nu -\f{2i\kappa}{5760\pi^2 f^2}(3k_1^2 +4k_1.k_2 +6k_2^2)(k_1^\mu k_2^\nu+k_1^\nu k_2^\mu).
\ee
In harmonic gauge $\p^\mu h_{\mu\nu}-\f{1}{2}\p_\nu h=0$ for the graviton the above expression can be brought to the following form
\be
V^{\mu\nu}_{(h\varphi \varphi)} (k_1,k_2,k_3)&=&-\frac{i \kappa}{5760 \pi ^2 f^2}\eta^{\mu\nu} \Big[ 2k_1^2 k_1.k_2 +(k_1^2)^2 -6(k_2^2)^2\Big]\nn\\
&& -\f{8i\kappa}{5760\pi^2 f^2}(2k_1^2 +3k_1.k_2 +3k_2^2)k_2^\mu k_2^\nu . \label{eq:scalar_model_vertex}
\ee

Now recall that the EFT prediction for the one-graviton-two-dilaton vertex was given in \eqref{eq:V_hphiphi_EFT_predictiongauge}. 
Comparing the above expression with our computation \eqref{eq:scalar_model_vertex} we conclude that
\begin{equation}
	\Delta a=\f{1}{5760\pi^2},\qquad r_1=\f{\Delta a}{6}.
\end{equation}
This is in a perfect agreement with the first line in \eqref{eq:answer_anomalies_free}.

\subsubsection{Graviton-Graviton-Dilaton Vertex}
In this example this vertex is given by the sum of the following two diagrams
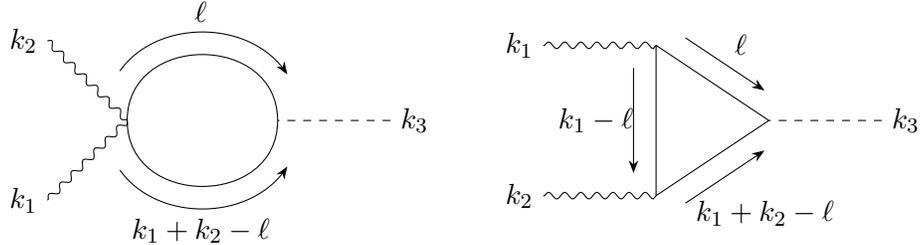
\begingroup
\allowdisplaybreaks
\be
&& \begin{tikzpicture}[baseline=(a.base)]
	\begin{feynman}
		\vertex (a);
		\vertex [right=2.0 of a] (b);
		\vertex [below left=of a, label=left:$k_1$] (i1);
		\vertex [above left=of a, label=left:$k_2$] (i2);
		\vertex [right=of b, label=right:$k_3$] (o1);
		\diagram* {
			(a) --[half left, momentum={$\ell$}] (b) -- [half left, rmomentum=$k_1+k_2-\ell$] (a),
			(i1) --[photon] (a) --[photon] (i2),
			(o1) --[dashed] (b),
		};
	\end{feynman}
\end{tikzpicture}
\qquad
\begin{tikzpicture}[baseline=(c.base)]
	\begin{feynman}
		\vertex (c);
		\vertex [below left=1 and 1.5 of c] (b);
		\vertex [above left=1 and 1.5 of c] (a);
		\vertex [left=of a, label=left:$k_1$] (i1);
		\vertex [left=of b, label=left:$k_2$] (i2);
		\vertex [right=of c, label=right:$k_3$] (o1);
		\diagram* {
			(a) --[momentum'={[label distance=0.5em, style={inner xsep= -0.5em}]$k_1-\ell$}] (b) --[momentum'={[label distance=0.5em, style={inner xsep= -1.5em}]$k_1+k_2-\ell$}] (c) --[rmomentum'={$\ell$}] (a),
			(i1) --[photon] (a), (i2)--[photon] (b),
			(o1) --[dashed] (c),
		};
	\end{feynman}
\end{tikzpicture}\nn
\ee
\endgroup
Let us denote these diagrams by $V_1$ and $V_2$ respectively. Then
\begin{equation}
	\label{eq:sum_vertices}
	V_{(hh\varphi)} (k_1,k_2,k_3;\varepsilon_1,\varepsilon_2) = V_1 (k_1,k_2,k_3;\varepsilon_1,\varepsilon_2) + V_2 (k_1,k_2,k_3;\varepsilon_1,\varepsilon_2).
\end{equation}
The first diagram reads as
\be 
V_1&=& \f{\sqrt{2}\kappa^2 m^2}{6f} \varepsilon^{\mu\nu}(k_1)\varepsilon^{\alpha\beta}(k_2)\int \f{d^d\ell}{(2\pi)^d}\ \f{1}{\ell^2+m^2 -i\epsilon}\f{1}{(k_1+k_2-\ell)^2 +m^2 -i\epsilon}\nn\\
&&\Big[ \eta_{\mu\alpha}\eta_{\nu\beta}\Big\lbrace 6(\ell^2+m^2)-6\ell.(k_1+k_2)+\f{3(d-2)}{2(d-1)}(3k_1.k_2+2k_1^2+2k_2^2)\Big\rbrace\nn\\
&& +\eta_{\nu\alpha}\Big\lbrace -24\ell_\mu \ell_\beta +12(\ell_\mu k_{1\beta}+\ell_\beta k_{2\mu})-\f{3(d-2)}{d-1} k_{1\beta}k_{2\mu}\Big\rbrace\Big].
\ee
The second diagram reads as
\be 
V_2&=& \f{4\kappa^2 \sqrt{2}m^2}{f}\varepsilon^{\mu\nu}(k_1)\varepsilon^{\alpha\beta}(k_2)\int \f{d^d\ell}{(2\pi)^d}\ \f{1}{\ell^2+m^2 -i\epsilon}\f{1}{(k_1+k_2-\ell)^2 +m^2 -i\epsilon}\nn\\
&&\f{1}{(k_1-\ell)^2+m^2 -i\epsilon}\times \ell_\mu \ell_\nu (\ell-k_1)_\alpha (\ell-k_1)_\beta.
\ee

Now we use Feynman's parametrization to combine the denominators in $V_1$ and perform loop momentum shift $\ell\rightarrow \ell+x(k_1+k_2)$ in the above expression. Then the non-vanishing part of $V_1$ under $\ell\leftrightarrow -\ell$ symmetry takes the following form
\be 
V_1 &=& \f{\sqrt{2}\kappa^2 m^2}{6f}\ \int_0^1dx\Bigg[  (\varepsilon_1.\varepsilon_2)\Big\lbrace \Big(6-\f{24}{d}\Big)\mathbf{J}\left(1,2;\Delta\right)\nn\\
&&+\Big(6m^2 -6x(1-x)(k_1+k_2)^2+\f{3(d-2)}{2(d-1)}(3k_1.k_2+2k_1^2+2k_2^2)\Big)\mathbf{J}\left(0,2;\Delta\right) \Big\rbrace\nn\\
&& - (k_2.\varepsilon_1.\varepsilon_2.k_1)\ \Big(\f{3(d-2)}{(d-1)}-24x(1-x)\Big) \mathbf{J}\left(0,2;\Delta\right)\Bigg],
\label{eq:V1}
\ee
where $\Delta=m^2+x(1-x)(k_1+k_2)^2$.

Analogously using Feynman's parametrization to combine the denominators in $V_2$ and performing loop momentum shift $\ell\rightarrow \ell+k_1(y+z)+k_2 y$ the non-vanishing part of $V_2$ under $\ell\leftrightarrow -\ell$ symmetry is
\be 
V_2 &=&
\f{4\kappa^2 \sqrt{2}m^2}{f}\int_0^1 dxdydz\ 2\delta(x+y+z-1) \Big[(\varepsilon_1.\varepsilon_2)\f{2}{d(d+2)}\ \mathbf{J}\left(2,3;\Delta\right)\nn\\
&&-(k_2.\varepsilon_1.\varepsilon_2.k_1)\f{4xy}{d}\ \mathbf{J}\left(1,3;\Delta\right) +(k_2.\varepsilon_1.k_2)(k_1.\varepsilon_2.k_1)y^2x^2\ \mathbf{J}\left(0,3;\Delta\right)\Big]\ , \label{eq:V2}
\ee
where $\Delta=m^2 +x(1-x)k_1^2+y(1-y)k_2^2+2xyk_1.k_2$.

In order to get the final answer for the vertex, we plug \eqref{eq:V1} and \eqref{eq:V2} into \eqref{eq:sum_vertices}. We then set $d=4-\epsilon$ and expand around $\epsilon=0$. Having done that, we expand the expression in powers of momenta keeping only terms containing four powers of momenta (this is the low-energy expansion). Performing the integrals over the Feynman parameters we obtain
\begin{multline}
	V_{(hh \varphi)}(k_1,k_2,k_3;\varepsilon_1,\varepsilon_2) =\\
	f_1(k_1,k_2)\times(\varepsilon_1. \ve_2)
	+f_2(k_1,k_2)\times(k_1.\varepsilon_2.k_1)(k_2.\ve_1.k_2)+f_3(k_1,k_2)\times(k_1.\ve_2.\ve_1.k_2),
\end{multline}
where the functions $f_i(k_1,k_2)$ read as
\begin{equation}
	\label{eq:solution_hhd_scalar}
	\begin{aligned}
		f_1(k_1,k_2) &= \f{i\kappa^2}{1440\sqrt{2}\pi^2 f} \left( 2(k_1^2)^2+2(k_2^2)^2+10(k_1.k_2)^2+7k_1.k_2(k_1^2+k_2^2)+3k_1^2k_2^2 \right),\\
		f_2(k_1,k_2) &= +\frac{i \kappa ^2}{360 \sqrt{2} \pi ^2 f},\\
		f_3(k_1,k_2) &= -\f{i\kappa^2}{720\sqrt{2}\pi^2 f}\left( k_1^2+k_2^2+6(k_1.k_2)\right).
	\end{aligned}
\end{equation}
Comparing \eqref{eq:solution_hhd_scalar} with  \eqref{eq:vertex_hhd_functions} we conclude that
\begin{equation}
	\label{eq:result_anomalies_hhd_fb}
	\Delta a=\f{1}{5760\pi^2},\qquad
	\Delta c=3\Delta a,\qquad
	r_1=\f{\Delta a}{6}.
\end{equation}
This is again in a perfect agreement with the first line in \eqref{eq:answer_anomalies_free}.

\subsubsection{Graviton-Graviton-Dilaton-Dilaton Vertex} 
The simplest four-point vertex is the four dilaton vertex $V_{(\varphi\varphi\varphi\varphi)}$. It was already studied in  \cite{Komargodski:2011vj}, see also \cite{Karateev:2022jdb} for a detailed analysis of this vertex in the free massive theory. We will not consider it here again. Instead, let us focus on the vertex $V_{(hh\varphi\varphi)}$. In our free massive theory this vertex is given by the sum of the following eight diagrams
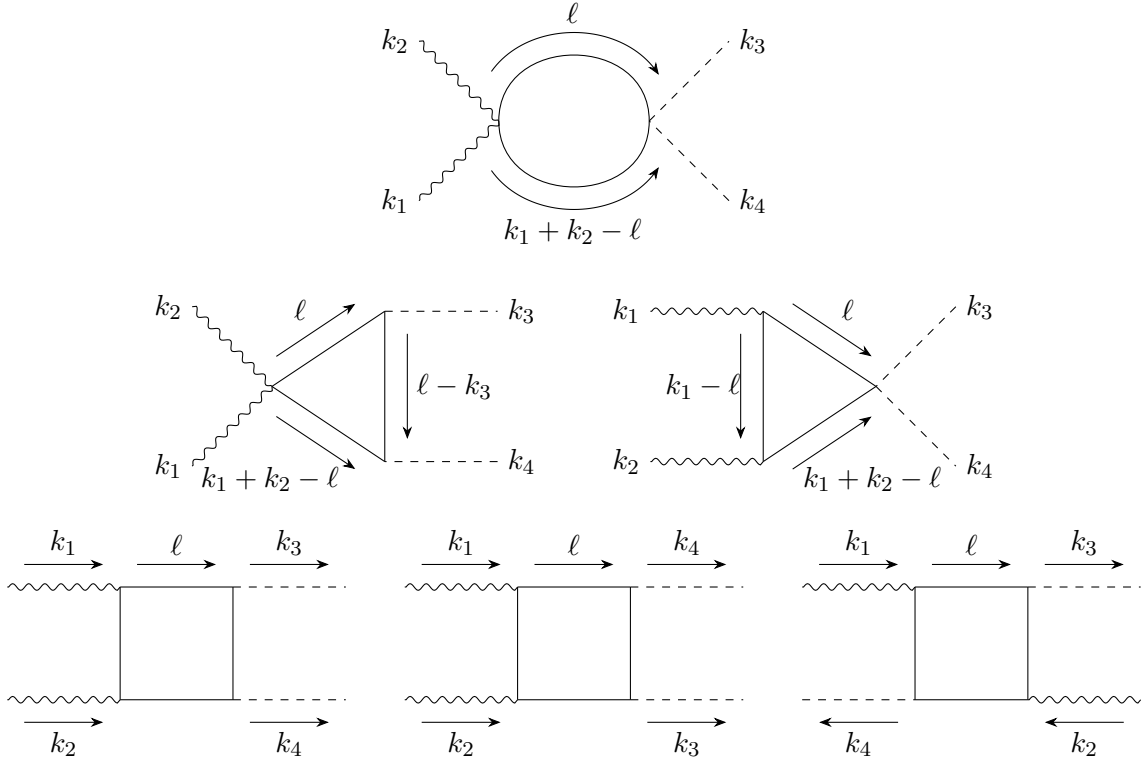
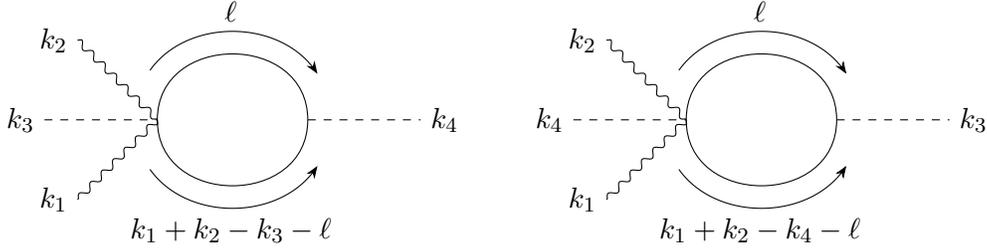
\begin{equation*}
	\begin{tikzpicture}[baseline=(a.base)]
		\begin{feynman}
			\vertex (a);
			\vertex [right=2 of a] (b);
			\vertex [below left=of a, label=left:$k_1$] (i1);
			\vertex [above left=of a, label=left:$k_2$] (i2);
			\vertex [below right=of b, label=right:$k_4$] (o1);
			\vertex [above right=of b, label=right:$k_3$] (o2);
			\diagram* {
				(a) --[half left, momentum={$\ell$}] (b) -- [half left, rmomentum=$k_1+k_2-\ell$] (a),
				(i1) --[photon] (a) --[photon] (i2),
				(o1) --[dashed] (b) --[dashed] (o2),
			};
		\end{feynman}
	\end{tikzpicture}
\end{equation*}
\begin{equation*}
	\begin{tikzpicture}[baseline=(a.base)]
		\begin{feynman}
			\vertex (a);
			\vertex [below right=1 and 1.5 of a] (b);
			\vertex [above right=1 and 1.5 of a] (c);
			\vertex [below left=of a, label=left:$k_1$] (i1);
			\vertex [above left=of a, label=left:$k_2$] (i2);
			\vertex [right=of b, label=right:$k_4$] (o1);
			\vertex [right=of c, label=right:$k_3$] (o2);
			\diagram* {
				(a) --[momentum'={[label distance=0.5em, style={inner xsep= -1.3em}]$k_1+k_2-\ell$}] (b) --[rmomentum'=$\ell-k_3$] (c) --[rmomentum'={$\ell$}] (a),
				(i1) --[photon] (a) --[photon] (i2),
				(o1) --[dashed] (b), (c) --[dashed] (o2),
			};
		\end{feynman}
	\end{tikzpicture}
	\qquad
	\begin{tikzpicture}[baseline=(c.base)]
		\begin{feynman}
			\vertex (c);
			\vertex [below left=1 and 1.5 of c] (b);
			\vertex [above left=1 and 1.5 of c] (a);
			\vertex [left=of a, label=left:$k_1$] (i1);
			\vertex [left=of b, label=left:$k_2$] (i2);
			\vertex [above right=of c, label=right:$k_3$] (o1);
			\vertex [below right=of c, label=right:$k_4$] (o2);
			\diagram* {
				(a) --[momentum'={[label distance=0.5em, style={inner xsep= -0.5em}]$k_1-\ell$}] (b) --[momentum'={[label distance=0.5em, style={inner xsep= -1.5em}]$k_1+k_2-\ell$}] (c) --[rmomentum'={$\ell$}] (a),
				(i1) --[photon] (a), (i2)--[photon] (b),
				(o1) --[dashed] (c), (c) --[dashed] (o2),
			};
		\end{feynman}
	\end{tikzpicture}
\end{equation*}
\begin{equation*}
	\begin{tikzpicture}[baseline=($(a.base)!0.5!(b.base)$)]
		\begin{feynman}
			\vertex (a);
			\vertex [below=of a] (b);
			\vertex [right=of b] (c);
			\vertex [above=of c] (d);
			\vertex [left=of a] (i1);
			\vertex [left=of b] (i2);
			\vertex [right=of c] (o1);
			\vertex [right=of d] (o2);
			\diagram* {
				(a) --(b) -- (c) -- (d) -- [rmomentum'=$\ell$](a),
				(i1) --[photon, momentum=$k_1$] (a),
				(i2) --[photon, momentum'=$k_2$] (b),
				(c) --[dashed, momentum'=$k_4$] (o1),
				(d) --[dashed, momentum=$k_3$] (o2),
			};
		\end{feynman}
	\end{tikzpicture}
	\qquad
	\begin{tikzpicture}[baseline=($(a.base)!0.5!(b.base)$)]
		\begin{feynman}
			\vertex (a);
			\vertex [below=of a] (b);
			\vertex [right=of b] (c);
			\vertex [above=of c] (d);
			\vertex [left=of a] (i1);
			\vertex [left=of b] (i2);
			\vertex [right=of c] (o1);
			\vertex [right=of d] (o2);
			\diagram* {
				(a) --(b) -- (c) -- (d) -- [rmomentum'=$\ell$](a),
				(i1) --[photon, momentum=$k_1$] (a),
				(i2) --[photon, momentum'=$k_2$] (b),
				(c) --[dashed, momentum'=$k_3$] (o1),
				(d) --[dashed, momentum=$k_4$] (o2),
			};
		\end{feynman}
	\end{tikzpicture}
	\qquad
	\begin{tikzpicture}[baseline=($(a.base)!0.5!(b.base)$)]
		\begin{feynman}
			\vertex (a);
			\vertex [below=of a] (b);
			\vertex [right=of b] (c);
			\vertex [above=of c] (d);
			\vertex [left=of a] (i1);
			\vertex [left=of b] (i2);
			\vertex [right=of c] (o1);
			\vertex [right=of d] (o2);
			\diagram* {
				(a) --(b) -- (c) -- (d) -- [rmomentum'=$\ell$](a),
				(i1) --[photon, momentum=$k_1$] (a),
				(i2) --[dashed, rmomentum'=$k_4$] (b),
				(c) --[photon, rmomentum'=$k_2$] (o1),
				(d) --[dashed, momentum=$k_3$] (o2),
			};
		\end{feynman}
	\end{tikzpicture}
\end{equation*}
\begin{equation*}
	\begin{tikzpicture}[baseline=(a.base)]
		\begin{feynman}
			\vertex (a);
			\vertex [right=2.0 of a] (b);
			\vertex [below left=of a, label=left:$k_1$] (i1);
			\vertex [above left=of a, label=left:$k_2$] (i2);
			\vertex [left=of a, label=left:$k_3$] (o1);
			\vertex [right=of b, label=right:$k_4$] (o2);
			\diagram* {
				(a) --[half left, momentum={$\ell$}] (b) -- [half left, rmomentum=$k_1+k_2-k_3-\ell$] (a),
				(i1) --[photon] (a) --[photon] (i2),
				(o1) --[dashed](a),
				(b) --[dashed](o2),
			};
		\end{feynman}
	\end{tikzpicture}
	\qquad
	\begin{tikzpicture}[baseline=(a.base)]
		\begin{feynman}
			\vertex (a);
			\vertex [right=2.0 of a] (b);
			\vertex [below left=of a, label=left:$k_1$] (i1);
			\vertex [above left=of a, label=left:$k_2$] (i2);
			\vertex [left=of a, label=left:$k_4$] (o1);
			\vertex [right=of b, label=right:$k_3$] (o2);
			\diagram* {
				(a) --[half left, momentum={$\ell$}] (b) -- [half left, rmomentum=$k_1+k_2-k_4-\ell$] (a),
				(i1) --[photon] (a) --[photon] (i2),
				(o1) --[dashed](a),
				(b) --[dashed](o2),
			};
		\end{feynman}
	\end{tikzpicture}
\end{equation*}

Let us denote the eight diagrams appearing here by $\mathcal{A}_i$ for $i=1,\cdots,8$. The order is counted from left to right and from top to bottom. Using Feynman's rules the expression of the first diagram becomes
\be 
\mathcal{A}_1 &=& -\f{m^2\kappa^2}{6f^2} \varepsilon_1^{\mu\nu}(k_1)\varepsilon_2^{\alpha\beta}(k_2)\int \f{d^d\ell}{(2\pi)^d}\ \f{1}{\ell^2+m^2 -i\epsilon}\f{1}{(k_1+k_2-\ell)^2 +m^2 -i\epsilon}\nn\\
&&\Big[ \eta_{\mu\alpha}\eta_{\nu\beta}\Big\lbrace 6(\ell^2+m^2)-6\ell.(k_1+k_2)+\f{9(d-2)}{2(d-1)}k_1.k_2\Big\rbrace  \nn\\
&&+\eta_{\nu\alpha}\Big\lbrace-24\ell_\mu \ell_\beta +12(\ell_\mu k_{1\beta}+\ell_\beta k_{2\mu})-\f{3(d-2)}{d-1} k_{1\beta}k_{2\mu}\Big\rbrace\Big].
\ee
Now using Feynman parametrization to combine the denominators and performing  loop momentum shift $\ell\rightarrow \ell+x(k_1+k_2)$ we get
\be
\mathcal{A}_1 &=& -\f{m^2\kappa^2}{6f^2} \int_0^1 dx\ \Big[ (\ve_1\cdot\ve_2)\Big(6-\f{24}{d}\Big) \mathbf{J}\left(1,2;\Delta\right) \nn\\
&&+\Big(6m^2-s\Big(6x^2-6x+\f{9(d-2)}{4(d-1)}\Big)\Big) \mathbf{J}\left(0,2;\Delta\right)\nn\\
&& -(k_2.\ve_1.\ve_2.k_1)\Big(\f{3(d-2)}{(d-1)}-24x+24x^2\Big)\mathbf{J}\left(0,2;\Delta\right) \Big],
\label{eq:A1}
\ee
where $\Delta=m^2-sx(1-x)$.
The expression for  the second diagram is
\be 
\mathcal{A}_2 &=& \f{2m^4\kappa^2}{3f^2} \varepsilon_1^{\mu\nu}(k_1)\varepsilon_2^{\alpha\beta}(k_2)\int \f{d^d\ell}{(2\pi)^d}\ \f{1}{(k_1+k_2-\ell)^2 +m^2 -i\epsilon}\f{1}{(\ell-k_3)^2+m^2 -i\epsilon}\nn\\
&&\f{1}{\ell^2+m^2 -i\epsilon}\Big[ \eta_{\mu\alpha}\eta_{\nu\beta}\Big\lbrace 6(\ell^2+m^2)-6\ell.(k_1+k_2)+\f{9(d-2)}{2(d-1)}k_1.k_2\Big\rbrace\nn\\
&& +\eta_{\nu\alpha}\Big\lbrace -24\ell_\mu \ell_\beta +12(\ell_\mu k_{1\beta}+\ell_\beta k_{2\mu})-\f{3(d-2)}{d-1} k_{1\beta}k_{2\mu}\Big\rbrace\Big].
\ee
Now using Feynman parametrization to combine the denominators and performing  loop momentum shift $\ell\rightarrow \ell+y(k_1+k_2)+zk_3$ we get
\be
\mathcal{A}_2 &=&  \f{2m^4\kappa^2}{3f^2} \int_0^1 dxdydz\ 2\delta(x+y+z-1)\Big[  \nn\\
&&(\ve_1\cdot\ve_2)\Big\lbrace \left(6-\f{24}{d}\right) \mathbf{J}\left( 1,3;\Delta\right) +\Big(6m^2 +6sxy+3sz-\f{9(d-2)}{4(d-1)}s\Big) \mathbf{J}\left( 0,3;\Delta\right)\Big\rbrace \nn\\
&&+\Big\lbrace  (k_2.\ve_1.\ve_2.k_1) \Big( 24y(1-y)-\f{3(d-2)}{(d-1)}\Big)+12(k_2.\ve_1.\ve_2.k_3)z(1-2y)\nn\\
&&+12(k_3.\ve_1.\ve_2.k_1)z(1-2y)-24z^2(k_3.\ve_1.\ve_2.k_3)\Big\rbrace  \mathbf{J}\left( 0,3;\Delta\right)\Big],
\ee
where $\Delta=m^2 -sxy$.
The expression for  the third diagram is
\be 
\mathcal{A}_3 &=& -\f{4m^2\kappa^2}{f^2}\varepsilon_1^{\mu\nu}(k_1)\varepsilon_2^{\alpha\beta}(k_2)\int \f{d^d\ell}{(2\pi)^d}\ \f{1}{(k_1+k_2-\ell)^2 +m^2 -i\epsilon}\f{1}{(k_1-\ell)^2+m^2 -i\epsilon}\nn\\
&&\times \f{1}{\ell^2+m^2 -i\epsilon} \ell_\mu \ell_\nu (\ell-k_1)_\alpha (\ell-k_1)_\beta.
\ee
Now using Feynman parametrization to combine the denominators and performing  loop momentum shift $\ell\rightarrow \ell+k_1(y+z)+k_2 y$ we get
\be
\mathcal{A}_3&=&-\f{4m^2\kappa^2}{f^2} \int_0^1 dxdydz\ 2\delta(x+y+z-1)\Big[\f{2}{d(d+2)}(\ve_1\cdot\ve_2) \mathbf{J}\left( 2,3;\Delta\right)\nn\\
&& -\f{4xy}{d}(k_2.\ve_1.\ve_2.k_1)\mathbf{J}\left( 1,3;\Delta\right)+y^2x^2 (k_2.\ve_1.k_2)(k_1.\ve_2.k_1)\mathbf{J}\left( 0,3;\Delta\right)\Big],
\ee
where $\Delta=m^2 -sxy$.
Using Feynman rules the expression of  the fourth diagram becomes
\be 
\mathcal{A}_4 &=& \f{8m^4\kappa^2}{f^2}\varepsilon_1^{\mu\nu}(k_1)\varepsilon_2^{\alpha\beta}(k_2)\int \f{d^d\ell}{(2\pi)^d}\ \f{1}{\ell^2+m^2 -i\epsilon}\f{1}{(k_1+k_2-\ell)^2 +m^2 -i\epsilon}\nn\\
&&\times\f{1}{(k_1-\ell)^2+m^2 -i\epsilon}\f{1}{(\ell-k_3)^2+m^2 -i\epsilon}\times \ell_\mu \ell_\nu (\ell-k_1)_\alpha (\ell-k_1)_\beta.
\ee
Now using Feynman parametrization to combine the denominators and performing  loop momentum shift $\ell\rightarrow \ell+k_1(y+z)+k_2 y+k_3w$ we get
\be
\mathcal{A}_4&=& \f{8m^4\kappa^2}{f^2}\int_0^1 dxdydzdw \ 6\delta(x+y+z+w-1)  \Big[\f{2}{d(d+2)}(\ve_1\cdot\ve_2)\mathbf{J}\left(2,4; \Delta\right)\nn\\
&& +\f{4}{d}\varepsilon_1^{\mu\nu}(k_1)\varepsilon_{2,\mu}^{\ \beta}(k_2)(k_{2\nu}y+k_{3\nu}w)(k_{1\beta}(y+z-1)+k_{3\beta}w)\mathbf{J}\left(1,4; \Delta\right)\nn\\
&&+\varepsilon_1^{\mu\nu}(k_1)\varepsilon_2^{\alpha\beta}(k_2)(k_{2\mu}y+k_{3\mu}w)(k_{2\nu}y+k_{3\nu}w) (k_{1\alpha}(y+z-1)+k_{3\alpha}w)(k_{1\beta}(y+z-1)\nn\\
&&+k_{3\beta}w)\ \mathbf{J}\left(0,4; \Delta\right)\Big],
\ee
where $\Delta=m^2-sxy-tzw$.
The contribution of the fifth diagram $\mathcal{A}_5$ can be read off from the expression of $\mathcal{A}_4$ above just after exchanging $k_3\leftrightarrow k_4$.

Using Feynman rules the expression of  the sixth diagram becomes
\be 
\mathcal{A}_6 &=& \f{8m^4\kappa^2}{f^2}\varepsilon_1^{\mu\nu}(k_1)\varepsilon_2^{\alpha\beta}(k_2)\int \f{d^d\ell}{(2\pi)^d}\ \f{1}{\ell^2+m^2 -i\epsilon}\f{1}{(k_3-k_2-\ell)^2 +m^2 -i\epsilon}\nn\\
&&\f{1}{(k_1-\ell)^2+m^2 -i\epsilon}\f{1}{(\ell-k_3)^2+m^2 -i\epsilon}\times \ell_\mu \ell_\nu (\ell-k_3)_\alpha (\ell-k_3)_\beta.
\ee
Now using Feynman parametrization to combine the denominators and performing  loop momentum shift $\ell\rightarrow \ell+k_1 z-k_2 y+k_3(y+w)$ we get
\be
\mathcal{A}_6&=& \f{8m^4\kappa^2}{f^2}\int_0^1 dxdydzdw \ 6\delta(x+y+z+w-1)\Big[\f{2}{d(d+2)}(\ve_1\cdot\ve_2)\mathbf{J}\left( 2,4;\Delta\right)\nn\\
&&+\f{4}{d}\varepsilon_1^{\mu\nu}(k_1)\varepsilon_{2,\mu}^{\ \beta}(k_2)(-k_{2\nu}y+k_{3\nu}(y+w))(k_{1\beta}z+k_{3\beta}(y+w-1))\mathbf{J}\left( 1,4;\Delta\right)\nn\\
&&+\varepsilon_1^{\mu\nu}(k_1)\varepsilon_2^{\alpha\beta}(k_2)(-k_{2\mu}y+k_{3\mu}(y+w)) (-k_{2\nu}y+k_{3\nu}(y+w))(k_{1\alpha}z+k_{3\alpha}(y+w-1))\nn\\
&&\times(k_{1\beta}z+k_{3\beta}(y+w-1)) \mathbf{J}\left( 0,4;\Delta \right)\Big],
\ee
where $\Delta=m^2-uxy-tzw$.
The expression for  the seventh diagram is
\be 
\mathcal{A}_7 &=& -\f{2m^4\kappa^2}{f^2} (\ve_1\cdot\ve_2)\int\f{d^d\ell}{(2\pi)^d}\f{1}{\ell^2+m^2-i\epsilon}\f{1}{(k_4-\ell)^2+m^2-i\epsilon}\nn\\
&=& -\f{2m^4\kappa^2}{f^2} (\ve_1\cdot\ve_2)\int_0^1dx\ \mathbf{J}\left(0,2;m^2\right).
\ee
The contribution of the eighth diagram $\mathcal{A}_8$ can be read off from the expression of $\mathcal{A}_7$ above just after exchanging $k_3\leftrightarrow k_4$.

Now we sum the contributions of all eight diagrams and expand around $\epsilon=0$ after setting $d=4-\epsilon$. Having done that, we expand the expression in powers of momenta keeping only terms up to fourth power in momenta. Performing the integrals over the Feynman parameters we obtain
\eqref{result_hhdd} with the functions $g_i(s,t,u)$ given by
\begin{equation}
	\begin{aligned}
		g_1(s,t,u) &=\frac{i\kappa^2}{3f^2}\left(-\frac{3 m^4}{4 \pi ^2 \epsilon }+\frac{15 m^2 s+360 \gamma  m^4+360 m^4 \log \left(\frac{m^2}{4 \pi }\right)}{960 \pi ^2}+\f{2 s^2+s t+t^2}{960 \pi ^2}\right),\\
		g_2(s,t,u) &= \frac{i\kappa^2}{3f^2}\left(-\f{5m^2}{240\pi^2} +\f{s}{240\pi^2}\right), \qquad
		g_3(s,t,u) = \frac{i\kappa^2}{3f^2}\left(\f{m^4}{8\pi^2} -\f{u}{240\pi^2}\right),\\		
		g_4(s,t,u) &= \frac{i\kappa^2}{3f^2}\left(-\f{m^2}{4\pi^2}-\f{s}{240\pi^2}\right), \qquad\quad
		g_5(s,t,u) =  \frac{i\kappa^2}{3f^2}\left(\f{1}{240\pi^2}\right),\\		
		g_6(s,t,u) &= \frac{i\kappa^2}{3f^2}\left(\f{1}{60\pi^2}\right), \qquad	\qquad\qquad\quad
		g_7(s,t,u) = \frac{i\kappa^2}{3f^2}\left(- \f{1}{120\pi^2}\right),\\
		g_8(s,t,u) &= \frac{i\kappa^2}{3f^2}\left(\f{1}{40\pi^2}\right),\qquad\qquad\qquad\quad
		g_9(s,t,u) = \frac{i\kappa^2}{3f^2}\left(-\f{1}{40\pi^2}\right).
	\end{aligned}
\end{equation}
Comparing the above expression with \eqref{eq:g_functions} we get \footnote{
We do not match the cosmological constant $\lambda$ because the free scalar theory has an extra anomaly that affects this term. We discuss this in detail in section \ref{CouplSpac}.
}
\begin{equation}
	\Delta a = \f{1}{5760\pi^2},\qquad
	\Delta c = 3\Delta a,\qquad
	r_0 =-10 \Delta a, \qquad
	r_1= \f{\Delta a}{6}.
\end{equation}
This is also consistent with the result \eqref{eq:result_anomalies_hhd_fb} obtained from the three-point vertex $V_{( h h \varphi)}$.
This example nicely shows how the {\it low-energy} expansion is sensitive to both the $a$ and $c$ ultraviolet trace anomalies, which is the hallmark of anomaly matching.

\subsection{Free Massive Dirac Fermion}\label{S:free_fermion}
In this section we use $a,b,c,...$ as tangent space indices and $\mu ,\nu ,\rho, ...$ as curved space indices in $d$ dimensional curved background. 
With this convention, the massless Dirac field action in a curved background reads
\be 
A_\text{free fermion}[\Psi]&=& \int d^dx\sqrt{-g}\ \overline{\Psi}(x)\big( i\gamma^a E_a^\mu D_\mu \big)\Psi(x),
\label{eq:massless_Dirac_action}
\ee
where $\overline{\Psi}(x)=\Psi^{\dagger}\gamma^0$, $D_\mu\Psi=\p_\mu \Psi +\f{1}{2}\omega_\mu^{bc}\Sigma_{bc}\Psi$ with the following form of Lorentz generator and spin connection
\be 
\Sigma_{ab}&=&-\f{1}{4}[\gamma_a ,\gamma_b],\\
\omega_\mu^{ab}&=& \eta^{bc} e_\nu^{\ a}\p_\mu E_c^{\ \nu}+\eta^{bc} e_\nu^{\ a} \Gamma^{\nu}_{\mu\rho}E_c^{\ \rho}.
\ee
Gamma matrices satisfies the following properties: $\gamma^{a\dagger}=\gamma^0\gamma^a\gamma^0$, $\lbrace \gamma^a ,\gamma^b\rbrace=-2 \eta^{ab}\mathbb{I}$.
The relations between Vielbeins and metric are
\be 
g_{\mu\nu}=e_\mu^{\ a}e_\nu^{\ b}\eta_{ab},\qquad g^{\mu\nu}=E_a^{\ \mu}E_b^{\ \nu}\eta^{ab}.
\ee

\paragraph{Traces of gamma matrices}
For performing computations we will need to evaluate traces of various products of gamma matrices. The basic relations read as
\begin{equation}
	\begin{aligned}
		\textbf{Tr}[\gamma^{a_1}\gamma^{a_2}\cdots\gamma^{a_{2n+1}}]&=0\ \text{for}\ n\in\mathbb{Z}_+,\\
		\textbf{Tr}[\gamma^{a_1}\gamma^{a_2}]&=-d\eta^{a_1 a_2},\\
		\textbf{Tr}[\gamma^{a_1}\gamma^{a_2}\gamma^{a_3}\gamma^{a_4}]&=d(\eta^{a_1 a_2}\eta^{a_3 a_4}-\eta^{a_1 a_3}\eta^{a_2 a_4}+\eta^{a_1 a_4}\eta^{a_2 a_3}).
	\end{aligned}
\end{equation}
In $d=4$ dimensions the last relation would contain the Levi-Civita symbol. This term however will not contribute in parity preserving theories (as the ones considered in this draft). Traces with higher number of gamma matrices can be evaluated using the above relations, anti-commutation relations  and the cyclicity of trace. For example, the trace of six gamma matrices can be written as
\begin{equation}
	\begin{aligned}
		\textbf{Tr}[\gamma^{a_1}\gamma^{a_2}\gamma^{a_3}\gamma^{a_4}\gamma^{a_5}\gamma^{a_6}] =
		&-\eta^{a_1a_2}\textbf{Tr}[\gamma^{a_3}\gamma^{a_4}\gamma^{a_5}\gamma^{a_6}]
		+\eta^{a_1a_3}\textbf{Tr}[\gamma^{a_2}\gamma^{a_4}\gamma^{a_5}\gamma^{a_6}]\\
		&-\eta^{a_1a_4}\textbf{Tr}[\gamma^{a_2}\gamma^{a_3}\gamma^{a_5}\gamma^{a_6}]
		+\eta^{a_1a_5}\textbf{Tr}[\gamma^{a_2}\gamma^{a_3}\gamma^{a_4}\gamma^{a_6}]\\
		&-\eta^{a_1a_6}\textbf{Tr}[\gamma^{a_2}\gamma^{a_3}\gamma^{a_4}\gamma^{a_5}].
	\end{aligned}
\end{equation}

\paragraph{Weyl invariance}
Under a Weyl transformation with parameter $\sigma(x)$
\be 
&& E_a^{\ \mu}(x)\rightarrow e^{-\sigma(x)}E_a^{\ \mu}(x),\\
&& \Psi(x)\rightarrow e^{-\big(\f{d-1}{2}\big)\sigma(x)}\Psi(x),\\
&&\omega_\mu^{ab}\rightarrow \omega_\mu^{ab}+\big( \eta^{bc}e_\mu^{\ a}-\eta^{ac} e_\mu^{\ b}\big)E_c^{\ \rho}\p_\rho\sigma\ .
\ee
Substituting the above transformation relations in the Dirac action \eqref{eq:massless_Dirac_action} we get the following Weyl transformation property
\be 
&&A_\text{free fermion}[\Psi]\rightarrow A_\text{free fermion}[\Psi]\nn\\
&& +\f{i}{2}\int d^d x\sqrt{-g}\ \overline{\Psi}(x)E_a^{\ \mu}\Big[ -(d-1)\gamma^a +(\gamma^c\eta^{ba}-\gamma^b \eta^{ca}) \Sigma_{cb}\Big]\Psi(x) \p_\mu\sigma(x).
\ee
Now using the definition of $\Sigma_{cb}$ given above and anti-commutation relations of gamma matrices we find $(\gamma^c\eta^{ba}-\gamma^b \eta^{ca}) \Sigma_{cb}=(d-1)\gamma^a$.  This implies that the massless Dirac action in general dimension is Weyl invariant, of course.

\paragraph{Compensated action}
Let us now deform the massless free fermion action by a mass term and compensate with dilaton we get the following QFT action of interest
\be 
A^\text{compensated}_\text{free fermion}[\Psi, \varphi, h]=\int d^dx\sqrt{-g}\ \overline{\Psi}(x)\Big( i\gamma^a E_a^\mu D_\mu -m\ e^{-\tau(x)}\Big)\Psi(x) . \label{eq:QFT_fermion_action}
\ee
Let us now evaluate the Feynman rules for the action \eqref{eq:QFT_fermion_action}. 
We will denote massive fermions by solid lines with charge arrows, dilatons by dashed lines  and gravitons by wavy lines.
The Feynman rules for a massive Dirac fermion propagator and non-vanishing interaction vertices are given below
\begingroup
\allowdisplaybreaks
\begin{align}
	\begin{tikzpicture}[baseline=(a)]
		\begin{feynman}
			\vertex (a)[label=left :$\beta$];
			\vertex [right = 3 cm of a, label=right :$\alpha$] (b);
			\diagram* {
				(a) --  [fermion, momentum=$p$] (b),
			};
		\end{feynman}
	\end{tikzpicture}
	&= \frac{i(\slashed{p}-m\mathbb{I})_{\alpha\beta}}{p^2 + m^2 -i\epsilon}\ ,\nn\\		
	\begin{tikzpicture}[baseline=(a)]
		\begin{feynman}
			\vertex (a);
			\vertex [above left = of a] (i1)[label=left :$\beta$];
			\vertex [below left = of a] (i2)[label=left :$\alpha$];
			\vertex [right = of a , label=right:$\varepsilon$] (o1);
			\diagram* {
				(i1)--[fermion, momentum=$p_1$](a)--[fermion, rmomentum=$p_2$](i2),  
				(o1)--[photon, momentum'=$k$](a),
			};
		\end{feynman}
	\end{tikzpicture} &=  i\kappa \varepsilon^{\mu\nu}\ \textbf{Sym}_{(\mu\nu)}\Big[\gamma^\mu p_1^\nu +\gamma^\mu k_\rho\Sigma^{\rho\nu}\Big]_{\alpha\beta}  \ ,        \nn  \\
	\begin{tikzpicture}[baseline=(a)]
		\begin{feynman}
			\vertex (a);
			\vertex [above left = of a] (i1)[label=left :$\beta$];
			\vertex [below left = of a] (i2)[label=left :$\alpha$];
			\vertex [above right = of a , label=right:$\varepsilon_1$] (o1);
			\vertex [below right = of a , label=right:$\varepsilon_2$] (o2);
			\diagram* {
				(i1)--[fermion, momentum'=$p_1$](a)--[fermion, rmomentum=$p_2$](i2), 
				(o1)--[photon, momentum'=$k_1$ ](a)--[photon, rmomentum=$k_2$](o2),
			};
		\end{feynman}
	\end{tikzpicture} &= i\kappa^2\varepsilon_1^{\mu\nu}\varepsilon_2^{\rho\sigma}\ \textbf{Sym}_{(\mu\nu),(\rho\sigma)}\Big[ 2\eta^{\mu\rho}\eta^{\nu\sigma}(\slashed{p}_1+m\mathbb{I})-\f{3}{2}(\eta^{\nu\rho}\gamma^\mu p_1^\sigma+\eta^{\mu\sigma}\gamma^\rho p_{1}^\nu)  \nn \\
	& -\f{1}{2} \gamma^\alpha (k_1-k_2)_\alpha   \eta^{\mu\rho}    \Sigma^{\nu\sigma}      -\gamma^\mu\eta^{\nu\sigma}(k_1+k_2)_\alpha \Sigma^{\alpha\rho}-\gamma^\rho \eta^{\nu\sigma} (k_1+k_2)_\alpha \Sigma^{\alpha\mu}\nn\\
	&-\gamma^\rho k_2^\nu\Sigma^{\mu\sigma} -\gamma^\mu k_1^\sigma\Sigma^{\rho\nu}\Big]_{\alpha\beta}  \nn  \ ,   \\
	\begin{tikzpicture}[baseline=(a)]
		\begin{feynman}
			\vertex (a);
			\vertex [above left = of a] (i1)[label=left :$\beta$];
			\vertex [below left = of a] (i2)[label=left :$\alpha$];
			\vertex [above right = of a , label=right:$\varepsilon_1$] (o1);
			\vertex [below right = of a , label=right:$\varepsilon_2$] (o2);
			\vertex [right = of a] (o3);
			\diagram* {
				(i1)--[fermion, momentum'=$p_1$](a)--[fermion, rmomentum=$p_2$](i2), 
				(o1)--[photon, momentum'=$k_1$ ](a)--[photon, rmomentum=$k_2$](o2),
				(a)--[dashed](o3),
			};
		\end{feynman}
	\end{tikzpicture} &= -\f{2i\kappa^2}{\sqrt{2}f}m\varepsilon_1^{\mu\nu}\varepsilon_2^{\rho\sigma} \ \eta_{\mu\rho}\eta_{\nu\sigma}\mathbb{I}_{\alpha\beta}\ , \hspace{1cm}
	\begin{tikzpicture}[baseline=(a)]
		\begin{feynman}
			\vertex (a);
			\vertex [above left = of a] (i1)[label=left :$\beta$];
			\vertex [below left = of a] (i2)[label=left :$\alpha$];
			\vertex [right = of a] (o1);
			\diagram* {
				(i1)--[fermion, momentum=$p_1$](a)--[fermion, rmomentum=$p_2$](i2),  
				(o1)--[dashed, momentum'=$k$](a),
			};
		\end{feynman}
	\end{tikzpicture} =  \f{i}{\sqrt{2}f}m\ \mathbb{I}_{\alpha\beta}\ .\nn 
\end{align} 
\endgroup

\subsubsection{Dilaton-Dilaton-Dilaton Vertex}  Here we focus on computing the three point vertices $V_{(\varphi\varphi\varphi)}$ and $V_{(hh\varphi)}$ for our free fermion example.
In our convention all the dilatons are incoming and with arbitrary momenta. The vertex $V_{(\varphi\varphi\varphi)}$ is given by the sum of the following two Feynman diagrams
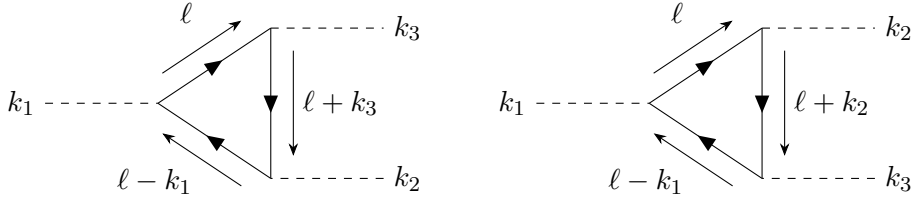
\begin{equation*}
	\begin{tikzpicture}[baseline=(a.base)]
		\begin{feynman}
			\vertex (a);
			\vertex [below right=1 and 1.5 of a] (b);
			\vertex [above right=1 and 1.5 of a] (c);
			\vertex [left=of a, label=left:$k_1$] (i1);
			\vertex [right=of b, label=right:$k_2$] (o1);
			\vertex [right=of c, label=right:$k_3$] (o2);
			\diagram* {
				(a) --[fermion, momentum={[label distance=0.5em, style={inner xsep= -1.3em}]$\ell$}] (c) --[fermion, momentum=$\ell+k_3$] (b) --[fermion, momentum={$\ell-k_1$}] (a),
				(i1) --[dashed] (a),
				(o1) --[dashed] (b), (c) --[dashed] (o2),
			};
		\end{feynman}
	\end{tikzpicture}
	\qquad
	\begin{tikzpicture}[baseline=(a.base)]
		\begin{feynman}
			\vertex (a);
			\vertex [below right=1 and 1.5 of a] (b);
			\vertex [above right=1 and 1.5 of a] (c);
			\vertex [left=of a, label=left:$k_1$] (i1);
			\vertex [right=of b, label=right:$k_3$] (o1);
			\vertex [right=of c, label=right:$k_2$] (o2);
			\diagram* {
				(a) --[fermion, momentum={[label distance=0.5em, style={inner xsep= -1.3em}]$\ell$}] (c) --[fermion, momentum=$\ell+k_2$] (b) --[fermion, momentum={$\ell-k_1$}] (a),
				(i1) --[dashed] (a),
				(o1) --[dashed] (b), (c) --[dashed] (o2),
			};
		\end{feynman}
	\end{tikzpicture}
\end{equation*}
Let us denote these diagrams by $W_1$ and $W_2$ respectively. Then
\begin{equation}
	V_{(\varphi\varphi\varphi)} (k_1,k_2,k_3) = W_1 (k_1,k_2,k_3) + W_2 (k_1,k_2,k_3).
\end{equation}
Using the Feynman rules one can write the above Feynman diagrams in an algebraic form. The first diagram reads as
\be 
W_1&=& \f{ m^3}{2\sqrt{2}f^3}\int \f{d^d\ell}{(2\pi)^d}\f{\textbf{Tr}\big[(\slashed{\ell}-\slashed{k}_1-m\mathbb{I})(\slashed{\ell}+\slashed{k}_3-m\mathbb{I})(\slashed{\ell}-m\mathbb{I})\big]}{[\ell^2+m^2-i\epsilon][(\ell+k_3)^2+m^2-i\epsilon][(\ell-k_1)^2+m^2-i\epsilon]}.
\ee
Evaluating the trace of gamma matrices and replacing $k_3=-k_1-k_2$ the above expression simplifies to
\be 
W_1&=& -\f{d m^4}{2\sqrt{2}f^3}\int \f{d^d\ell}{(2\pi)^d}\f{(-3\ell^2+2k_2.\ell+4k_1.\ell-k_1^2-k_1.k_2+m^2)}{[\ell^2+m^2-i\epsilon][(k_1+k_2-\ell)^2+m^2-i\epsilon][(\ell-k_1)^2+m^2-i\epsilon]}.
\ee
Now using Feynman parametrization to combine the denominators and doing loop momentum shift $\ell\rightarrow \ell+(y+z)k_1+yk_2$ we get
\be 
W_1&=& -\f{d m^4}{2\sqrt{2}f^3}\int_0^1 dxdydz\ 2\delta(x+y+z-1)\Big[ -3\ \mathbf{J}\left(1,3;\Delta\right) +\Big(k_1.k_2(6xy+2z-1)\nn\\
&&+k_1^2 x(3y+3z-1)+k_2^2y(2-3y)+m^2\Big)\ \mathbf{J}\left(0,3;\Delta\right)\Big]\ ,
\ee
where $\Delta=m^2 +x(1-x)k_1^2+y(1-y)k_2^2+2xyk_1.k_2$. The contribution of the second diagram $W_2$ can be read off from the above expression $W_1$ just by exchanging $k_2\leftrightarrow k_3$. 

Now let us sum over $W_1$ and $W_2$ and then set $d=4-\epsilon$ and expand around $\epsilon=0$. Having done that, we expand the expression in powers of momenta keeping only terms containing four powers of momenta. Performing the integrals over the Feynman parameters we obtain
\be
V_{(\varphi\varphi\varphi)}(k_1,k_2,k_3)&=&\frac{i}{2880 \sqrt{2} \pi ^2 f^3}\Big( 11\left((k_1^2)^2+(k_2^2)^2+(k_3^2)^2\right)+14\left(k_1^2k_2^2+k_2^2k_3^2 +k_3^2 k_1^2\right)\Big).\nonumber\\
\ee
Comparing the above expression with \eqref{eq:V3phi} we get,
\be
\Delta a\ =\ \frac{11}{5760 \pi ^2},\qquad r_1\ =\ \frac{1}{5760 \pi ^2}.
\ee
The value of $\Delta a$ perfectly agrees with \eqref{eq:answer_anomalies_free}.

\subsubsection{Graviton-Graviton-Dilaton Vertex}
The vertex $V_{(hh\varphi)}$ is given by the sum of the following three diagrams
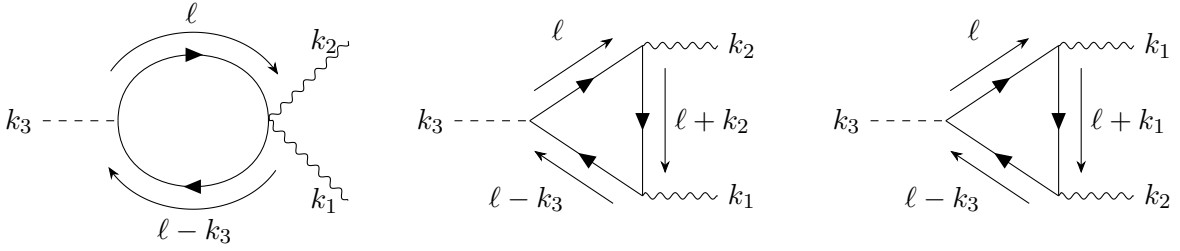
\begin{equation*}
	\begin{tikzpicture}[baseline=(a.base)]
		\begin{feynman}
			\vertex (a);
			\vertex [right=2.0 of a] (b);
			\vertex [below right=1.5 cm of b, label=left:$k_1$] (i1);
			\vertex [above right=1.5 cm of b, label=left:$k_2$] (i2);
			\vertex [left=1 cm of a, label=left:$k_3$] (o1);
			\diagram* {
				(a) --[half left, fermion, momentum={$\ell$}] (b) -- [half left, fermion, momentum=$\ell-k_3$] (a),
				(i1) --[photon] (b) --[photon] (i2),
				(o1) --[dashed](a),
			};
		\end{feynman}
	\end{tikzpicture}
	\qquad
	\begin{tikzpicture}[baseline=(a.base)]
		\begin{feynman}
			\vertex (a);
			\vertex [below right=1 and 1.5 of a] (b);
			\vertex [above right=1 and 1.5 of a] (c);
			\vertex [left=1 cm of a, label=left:$k_3$] (i1);
			\vertex [right=1 cm of b, label=right:$k_1$] (o1);
			\vertex [right=1 cm of c, label=right:$k_2$] (o2);
			\diagram* {
				(a) --[fermion, momentum={[label distance=0.5em, style={inner xsep= -1.3em}]$\ell$}] (c) --[fermion, momentum=$\ell+k_2$] (b) --[fermion, momentum={$\ell-k_3$}] (a),
				(i1) --[dashed] (a),
				(o1) --[photon] (b), (c) --[photon] (o2),
			};
		\end{feynman}
	\end{tikzpicture}
	\qquad
	\begin{tikzpicture}[baseline=(a.base)]
		\begin{feynman}
			\vertex (a);
			\vertex [below right=1 and 1.5 of a] (b);
			\vertex [above right=1 and 1.5 of a] (c);
			\vertex [left=1 cm of a, label=left:$k_3$] (i1);
			\vertex [right=1 cm of b, label=right:$k_2$] (o1);
			\vertex [right=1 cm of c, label=right:$k_1$] (o2);
			\diagram* {
				(a) --[fermion, momentum={[label distance=0.5em, style={inner xsep= -1.3em}]$\ell$}] (c) --[fermion, momentum=$\ell+k_1$] (b) --[fermion, momentum={$\ell-k_3$}] (a),
				(i1) --[dashed] (a),
				(o1) --[photon] (b), (c) --[photon] (o2),
			};
		\end{feynman}
\end{tikzpicture}\end{equation*}
Let us denote these diagrams by $U_1$, $U_2$ and $U_3$ respectively. Then
\begin{equation}
	V_{(hh\varphi)} (k_1,k_2,k_3;\ve_1,\ve_2) = U_1 (k_1,k_2,k_3;\ve_1,\ve_2) + U_2 (k_1,k_2,k_3;\ve_1,\ve_2)+ U_3(k_1,k_2,k_3;\ve_1,\ve_2).
\end{equation}
Using Feynman's rules the algebraic form of the first diagram becomes
\be 
U_1&=&-\f{\kappa^2m}{\sqrt{2}f}\varepsilon_{\mu\nu}(k_1)\varepsilon_{\rho\sigma}(k_2)\int \f{d^d\ell}{(2\pi)^d}\f{1}{\ell^2+m^2-i\epsilon}\f{1}{(\ell-k_3)^2+m^2-i\epsilon}\nn\\
&&\textbf{Tr}\Big[ \Big( 2\eta^{\mu\rho}\eta^{\nu\sigma}(\slashed{\ell}+m\mathbb{I})-\f{3}{2}(\eta^{\nu\rho}\gamma^\mu \ell^\sigma+\eta^{\mu\sigma}\gamma^\rho \ell^\nu)  \nn \\
&& -\f{1}{2} \gamma^\alpha (k_1-k_2)_\alpha   \eta^{\mu\rho}    \Sigma^{\nu\sigma}      -\gamma^\mu\eta^{\nu\sigma}(k_1+k_2)_\alpha \Sigma^{\alpha\rho}-\gamma^\rho \eta^{\nu\sigma} (k_1+k_2)_\alpha \Sigma^{\alpha\mu}\nn\\
&&-\gamma^\rho k_2^\nu\Sigma^{\mu\sigma} -\gamma^\mu k_1^\sigma\Sigma^{\rho\nu}\Big)(\slashed{\ell}-m\mathbb{I})(\slashed{\ell}-\slashed{k}_3-m\mathbb{I})\Big].
\ee
Evaluating the trace of gamma matrices the above expression reduces to
\be 
U_1&=&-\f{dm^2\kappa^2}{\sqrt{2}f}\varepsilon_{\mu\nu}(k_1)\varepsilon_{\rho\sigma}(k_2)\int \f{d^d\ell}{(2\pi)^d}\f{1}{\ell^2+m^2-i\epsilon}\f{1}{(\ell-k_3)^2+m^2-i\epsilon}\nn\\
&&\Big[ \eta^{\mu\rho}\eta^{\nu\sigma}(2(\ell^2+m^2) -2\ell.k_3 +k_3^2)-\f{3}{2}\eta^{\mu\rho}\big\lbrace (2\ell+k_2)^\nu (2\ell+k_1)^\sigma \big\rbrace \Big].
\ee
Now using Feynman parametrization to combine the denominators, doing loop momentum shift $\ell\rightarrow \ell+xk_3$ and keeping the non-vanishing terms under $\ell\leftrightarrow -\ell$ symmetry, we get
\be 
U_1
&=&-\f{dm^2\kappa^2}{\sqrt{2}f}\int_0^1 dx\ \Bigg[ (\varepsilon_1\cdot\varepsilon_2) \Bigg\lbrace \left(2-\f{6}{d}\right) \mathbf{J}\left(1,2;\Delta\right)\nn\\
&& +\left((1-2x+2x^2)(k_1+k_2)^2+2m^2\right)\mathbf{J}\left(0,2;\Delta\right)\Bigg\rbrace\nn\\
&& -\f{3}{2}(k_2.\varepsilon_1.\varepsilon_2.k_1)\ (2x-1)^2 \mathbf{J}\left(0,2;\Delta\right)\Bigg],
\ee
where $\Delta=m^2+x(1-x)(k_1+k_2)^2$.
The second diagram evaluates to
\be 
U_2&=&\f{\kappa^2m}{\sqrt{2}f}\varepsilon_{\mu\nu}(k_1)\varepsilon_{\rho\sigma}(k_2)\int \f{d^d\ell}{(2\pi)^d}\f{1}{\ell^2+m^2-i\epsilon}\f{1}{(\ell-k_3)^2+m^2-i\epsilon}\nn\\
&&\f{1}{(\ell+k_2)^2+m^2-i\epsilon}\textbf{Tr}\Big[  (\slashed{\ell}-\slashed{k}_3-m)\big(\gamma^\mu(\ell+k_2)^\nu+\gamma^\mu k_{1\alpha}\Sigma^{\alpha\nu}\big)\nn\\
&&(\slashed{\ell}+\slashed{k}_2-m)\big(\gamma^\rho \ell^\sigma +\gamma^\rho k_{2\beta}\Sigma^{\beta\sigma}\big)(\slashed{\ell}-m)\Big]
\ee
Now evaluating the gamma matrix traces, combining the denominators using Feynman parametrization and performing loop momentum shift $\ell\rightarrow \ell-xk_1-(x+y)k_2$ the non-vanishing contribution from the above expression becomes
\be
U_2
&=&\f{\kappa^2m^2}{\sqrt{2}f} \int_0^1 dxdydz\ 2\delta(x+y+z-1)\Bigg[ (\varepsilon_1\cdot\varepsilon_2) \Bigg\lbrace \f{d-6}{d+2} \ \mathbf{J}\left(2,3;\Delta\right)\nn\\
&&+\left( m^2+x^2 k_1^2 +(1-2xz)k_1.k_2 +z^2 k_2^2\right) \mathbf{J}\left(1,3;\Delta\right)\Bigg\rbrace\nn\\
&&+(k_2.\varepsilon_1.\varepsilon_2.k_1)\Big\lbrace \left( -1+(12-d)xz\right)\  \mathbf{J}\left(1,3;\Delta\right) -dxz\big( m^2 +x^2 k_1^2\nn\\
&&+\left(1-2xz\right)k_1.k_2 +z^2 k_2^2\big) \  \mathbf{J}\left(0,3;\Delta\right)\Big\rbrace\nn\\
&&+(k_2.\varepsilon_1.k_2)(k_1.\varepsilon_2.k_1)\ dxz (1-4xz)\  \mathbf{J}\left(0,3;\Delta\right)\Bigg]\ ,
\ee
where $\Delta=m^2+x(1-x)k_1^2+z(1-z)k_2^2+2xzk_1.k_2$.
The contribution of the third Feynman diagram $U_3$ can be read off from the above expression of $U_2$ by exchanging $(\varepsilon_1,k_1)\leftrightarrow (\varepsilon_2,k_2)$. 

Now let us sum over the contributions of the three diagrams then set $d=4-\epsilon$ and expand around $\epsilon=0$. Having done that, we expand the expression in powers of momenta keeping only terms containing four powers of momenta. Performing the integrals over the Feynman parameters we obtain
\be 
&& V_{(hh\varphi)}(k_1,k_2,k_3;\varepsilon_1,\varepsilon_2)\nn\\
&=& \f{i\kappa^2}{720 \sqrt{2} \pi ^2 f}\Big[(\varepsilon_1\cdot\varepsilon_2)\left\lbrace 14 k_1^2 k_2^2+6 \left((k_1^2)^2+(k_2^2)^2\right)+25 (k_1.k_2)^2+21 k_1.k_2(k_1^2+ k_2^2)\right\rbrace\nn\\
&&-(k_2.\varepsilon_1.\varepsilon_2.k_1)\  \left\lbrace6(k_1^2+ k_2^2)+26k_1.k_2\right\rbrace+7(k_2.\varepsilon_1.k_2)(k_1.\varepsilon_2.k_1)\Big]+\cdots
\ee
Comparing the above expression with  \eqref{eq:vertex_hhd_functions} we conclude that
\begin{equation}
	\label{eq:result_anomalies_hhd_ff}
	\Delta a=\f{11}{5760\pi^2},\qquad
	\Delta c= \f{18}{5760\pi^2},\qquad
	r_1=\f{1}{5760\pi^2}.
\end{equation}
The values of $\Delta a$ and $\Delta c$ perfectly agree with \eqref{eq:answer_anomalies_free}.

\section{The Dilaton-Graviton Scattering Amplitude}
\label{sec:scattering_and_bounds}
In this section we promote our background fields to dynamical fields by giving them kinetic terms. A convenient choice for these kinetic terms is
\begin{align}
	\label{eq:kin_dil}
	A^\varphi_{\text{kinetic}}&= -\f{\bar f^2}{6}\int d^4x \sqrt{-\widehat{g}}\ \widehat{R},\qquad
	-\f{\bar f^2}{6}\equiv-\f{f^2}{6}-M^2r_0, \\
	\label{eq:kin_gra}
	A^h_\text{kinetic} &= \f{1}{2\bar\kappa^2}\int d^4x \sqrt{-g}\ R,\qquad\;\;\,
	\f{1}{2\bar\kappa^2}\equiv \f{1}{2\kappa^2} +\f{f^2}{6}.
\end{align}
These kinetic terms preserve diffeomorphism invariance. The kinetic term of the dilaton also classically preserves Weyl invariance. Instead, the kinetic term of the graviton breaks Weyl invariance at the Planck scale $\bar\kappa^{-1}$.
The IR action describing our dynamical background fields is now given by\footnote{For a recent discussion on coupling of the dynamical gravity to classically scale invariant QFTs see \cite{Gabadadze:2023quw}.}
\begin{equation}
	\label{eq:action_dynamical}
	A = A_\text{EFT} + A^\varphi_\text{kinetic} + A^h_\text{kinetic},
\end{equation}
where the effective action is given by \eqref{eq:EFT_actionSecTwo}. 

Let us write the dilaton kinetic term in terms of the $\varphi(x)$ fields explicitly
\begin{multline}
	\label{eq:kinetic_dil_explicit}
	A^\varphi_{\text{kinetic}}=
	-M^2r_0\int d^4x\ \sqrt{-\widehat{g}}\widehat R\\
	+ \int d^4x\sqrt{-g}\ \left[ -\f{1}{2}g^{\mu\nu}\p_\mu\varphi \p_\nu\varphi -\f{f^2}{6}R+\f{\sqrt{2}f}{6}R\varphi-\f{1}{12}R\varphi^2\right]
\end{multline}
and make several comments.
The shifted coupling constant $\bar \kappa$ is used in the definition \eqref{eq:kin_gra} in order to effectively remove  the $-\frac{f^2}{6}R$ term in \eqref{eq:kinetic_dil_explicit} once it is summed up with the graviton kinetic term in the combined action \eqref{eq:action_dynamical}. The extra term proportional to $r_0M^2$ is introduced in \eqref{eq:kin_dil} in order to precisely cancel the $r_0M^2$ term in the invariant part of the effective action \eqref{eq:Ainvariant}. This is done in order to slightly simplify the expressions in this section. It allows us to effectively set $r_0=0$ in all the equations.
In order to perform computations from now on we also set the IR cosmological constant in the effective action \eqref{eq:Ainvariant} to zero, namely $\lambda=0$. This can be achieved by appropriately choosing the form of  \eqref{ScheDep}.

In this paper we focus for concreteness on the case of QFTs with explicit UV conformal symmetry breaking which undergo a renormalization group flow. Our results, however, also apply to the case of spontaneous symmetry breaking. Let us briefly explain why. In the case of spontaneous symmetry breaking the dilaton is a physical particle and not simply a probe. It is a Goldstone boson of the spontaneous conformal symmetry breaking. Its kinetic terms is generated automatically and needs not to be added by hand. More precisely the kinetic term for the physical dilaton is given by the $r_0 M^2 \widehat R$ term in the effective action \eqref{eq:Ainvariant}. The $r_0M^2$ is associated to the scale of the spontaneous conformal symmetry breaking $f$ as $r_0M^2 = -\frac{f^2}{6}$. As a result the parameter $r_0$ effectively disappears from the discussion. Also, in the spontaneous symmetry breaking case $\lambda$ is automatically zero. Finally, the difference of the UV and IR trace-anomalies should be interpreted according to the footnote \ref{f:SSB}.

Now using the action \eqref{eq:action_dynamical} we can compute the graviton-dilaton scattering amplitude
\begin{equation}
	\label{eq:scattering_HD}
	\mathcal{T}_{h\varphi\rightarrow h\varphi}(k_1,k_2,k_3,k_4;\varepsilon_1,\varepsilon_3).
\end{equation}
We require that our dilatons and gravitons are weakly coupled -- this is enforced by the ``decoupling'' limit
\begin{equation}
	\label{eq:decoupling_limit}
	\kappa\rightarrow 0,\qquad
	f\rightarrow \infty,\qquad
	\kappa\ll\frac{1}{f}.
\end{equation}
The above limit means in practice that all the expressions should be first expanded in small $\kappa$ keeping $f$ finite and subsequently expanded in small $1/f$. This limit also makes physical sense, as we expect the dilaton decay constant $f$ (if a dynamical dilaton exists in nature) to be not larger than the Planck scale $\bar\kappa^{-1}$.

One can foresee that the scattering amplitude computed using the action \eqref{eq:action_dynamical} at the leading order $\kappa^2$ will contain ``trivial'' terms with no $f$ dependence which come from the kinetic terms \eqref{eq:kin_dil} and \eqref{eq:kin_gra}. The non-trivial dependence on $\Delta a$ and $\Delta c$ will be uncovered instead at the order $\kappa^2f^{-2}$.
Let us also recall the definition of the Mandelstam variables for the scattering amplitude \eqref{eq:scattering_HD}
\begin{equation}
	\label{eq:mandelstam_variables_scattering}
	s\equiv - (k_1+k_2)^2,\qquad
	t\equiv - (k_1-k_3)^2,\qquad
	u\equiv - (k_1-k_4)^2.
\end{equation}
Since both the graviton and the dilaton particles are massless, we have the following constraint
\begin{equation}
	s+t+u=0.
\end{equation}

\subsection{Propagators and Additional Vertices}
In this subsection we summarize propagators and vertices following from the action \eqref{eq:action_dynamical}. As argued above, from now on, we set $r_0=0$ and $\lambda=0$ in \eqref{eq:kin_dil}, \eqref{eq:kin_gra}, and \eqref{eq:action_dynamical} for the computation below.

\paragraph{Two-point vertices and propagators}
Let us begin by examining the propagators and 2-point vertices in the theory described by the action \eqref{eq:action_dynamical}. To do this, we expand the action \eqref{eq:action_dynamical} up to quadratic order in the graviton-dilaton fields. The vertices and propagators resulting from the sum of the kinetic terms \eqref{eq:kin_dil} and \eqref{eq:kin_gra} will be denoted by $U$ and $\Delta$ respectively. The vertices generated from the EFT action \eqref{eq:EFT_actionSecTwo} will be referred to as blob vertices and denoted by $V$. 

 The part of the action \eqref{eq:action_dynamical} which contains two dilatons reads as
\begin{equation}
	A =  - \f{1}{2}\int d^4x\p_\mu\varphi(x)\p^\mu \varphi(x) +\f{1}{f^2} \int d^4x \left(  
	18r_1\p^2\varphi\p^2\varphi + \ldots\right),
\end{equation}
where the ellipses above represents the remaining terms at order $O(f^{-2})$ which contain higher than four power of derivatives of the dilaton field. These terms will always be ignored from now on. The first term of the above action will be used to derive the propagator for the dilaton filed and the order $f^{-2}$ part of the action will be treated as an EFT vertex denoted by the blob. The expressions of the dilaton propagator and the two-dilaton blob vertex read
\begin{align}
\Delta^{(\varphi)}(p)=
\begin{tikzpicture}[baseline=(a)]
		\begin{feynman}
			\vertex (a);
			\vertex [right = of a] (b);
			\diagram* {
				(a)--[dashed, momentum=$p$](b),
			};
		\end{feynman}
	\end{tikzpicture} 
&=\f{-i}{p^2-i\epsilon},\\
V_{(\varphi\varphi)}(p,-p)=\begin{tikzpicture}[baseline=(a)]
	\begin{feynman}
		\vertex [below left = of a ] (i2);
		\vertex [right= of a, blob, minimum size=0.75 cm] (c) {};
		\vertex[right = of c](b);
		\diagram* {
			(a)--[dashed, momentum=$p$](c)--[dashed, rmomentum=$-p$](b)
		};
	\end{feynman}
\end{tikzpicture}&=  \f{i}{f^2}
36r_1(p^2)^2. \label{eq:dd_vertex}
\end{align}

 The part of the action \eqref{eq:action_dynamical} which contains one dilaton and one graviton field reads 
\begin{equation}
	A = -\f{\sqrt{2}\kappa f}{6}\int d^4x\ \varphi \p^2h \ +O(\kappa f^{-1}).
\end{equation}
The graviton-dilaton vertices follow from the above action
\be
\label{eq:dil-grav}
U^{\mu\nu}_{(h\varphi)}(p,-p)&=& \begin{tikzpicture}[baseline=(a)]
	\begin{feynman}
		\vertex (a);
		\vertex [left= of a, label=left:$\mu\nu$] (i1) {};
		\vertex[right = of a](i2);
		\diagram* {
			(i1)--[photon, momentum=$p$](a)--[dashed, rmomentum=$-p$](i2)
		};
	\end{feynman}
\end{tikzpicture}
= \f{i\kappa \sqrt{2}f}{6}p^2\eta_{\mu\nu}, \label{eq:mixed_2vertex}\\
\label{eq:dil-grav_EFT}
V^{\mu\nu}_{(h\varphi)}(p,-p)&=& \begin{tikzpicture}[baseline=(a)]
	\begin{feynman}
		\vertex [blob, minimum size=0.75 cm] (a) {};
		\vertex [left= of a, label=left:$\mu\nu$] (i1);
		\vertex[right = of a](i2);
		\diagram* {
			(i1)--[photon, momentum=$p$](a)--[dashed, rmomentum=$-p$](i2)
		};
	\end{feynman}
\end{tikzpicture}
= O(\kappa f^{-1}).
\ee
Note that since the graviton-dilaton vertex $U_{(h\varphi)}$ is proportional to $\eta_{\mu\nu}$ any Lorentz contraction of this vertex with the on-shell graviton polarization tensor gives zero. As a consequence, any Feynman diagram where the dilaton is converted to the external graviton using $U_{(h\varphi)}$ vanishes on-shell after contraction with the graviton polarization tensor. Therefore, we do not consider such Feynman diagrams. We have also explicitly verified that Feynman diagrams involving vertices $V_{(h\varphi)}$ contribute to a power always higher than $\kappa^2f^{-2}$ in the decoupling limit \eqref{eq:decoupling_limit}. This is the reason we have not explicitly evaluated \eqref{eq:dil-grav_EFT}.

The graviton propagator at the leading order in $\kappa$ simply follows from the sum of the kinetic terms \eqref{eq:kin_dil} and \eqref{eq:kin_gra}. Since it is the standard result we do not derive it here and simply quote the final answer
\begin{align}
	\Delta^{(h)}_{\mu_1\mu_2,\nu_1\nu_2}(p) 	&\equiv \nn
	\begin{tikzpicture}[baseline=(a)]
		\begin{feynman}
			\vertex [label=left:$\mu_1\mu_2$](a);
			\vertex [right = of a, label=right:$\nu_1\nu_2$ ] (b);
			\diagram* {
				(a)--[photon, momentum=$p$](b),
			};
		\end{feynman}
	\end{tikzpicture}\\
	&=\f{1}{2}\,\Big( \eta_{\mu_1\nu_1}\eta_{\mu_2\nu_2}+\eta_{\mu_1\nu_2}\eta_{\mu_2\nu_1}-\eta_{\mu_1\mu_2}\eta_{\nu_1\nu_2}\Big) \f{-i}{p^2-i\epsilon}.
\end{align}
The two-graviton blob vertex that follows from \eqref{eq:EFT_actionSecTwo} turns out to be at order $\kappa^2$, and it does not contribute to the amplitude \eqref{eq:scattering_HD} up to the order $\kappa^2 f^{-2}$. Therefore, we are not presenting it here.

\paragraph{Three and four point vertices} The new vertices following from the sum of the kinetic terms \eqref{eq:kin_dil} and \eqref{eq:kin_gra} will be denoted by $U$, and the vertices generated from the EFT action \eqref{eq:EFT_actionSecTwo} will be referred to as blob vertices and denoted by $V$. The total contribution to the interaction vertices will be the sum of $U$ and $V$. 

The graviton-dilaton-dilaton vertex which follows from \eqref{eq:kin_dil} is given by
\be
U_{(h\varphi \varphi)}^{\mu\nu} (k_1,k_2,k_3) &=& \begin{tikzpicture}[baseline=(a)]
	\begin{feynman}
		\vertex  (a);
		\vertex [left = of a , label=left:$\mu\nu$] (i1);
		\vertex [above right = of a ] (i2);
		\vertex [below right = of a] (i3);
		\diagram* {
			(i1)--[photon, momentum'=$k_1$](a)--[dashed, rmomentum=$k_2$](i2), 
			(a)--[dashed, rmomentum=$k_3$](i3),
		};
	\end{feynman}
\end{tikzpicture}\nn\\
& =& - i\kappa\left( k_{2}^{\mu}k_{3}^{\nu}+k_{2}^{\nu}k_{3}^{\mu} -\eta^{\mu\nu}k_2.k_3\right)-\f{i\kappa}{6}\eta^{\mu\nu}k_1^2.\label{eq:kinetic_3_vertex}
\ee
The graviton-dilaton-dilaton vertex $V_{(h\varphi \varphi)}^{\mu\nu} $ which follows from \eqref{eq:EFT_actionSecTwo} was computed in \eqref{eq:V_hphiphi_EFT_prediction} up to fourth power in momenta. 

The graviton-graviton-dilaton vertex which follows from \eqref{eq:kin_dil} is given by
\be
U_{(hh \varphi)}^{\mu\nu,\rho\sigma} (k_1,k_2,k_3) &=& \begin{tikzpicture}[baseline=(a)]
	\begin{feynman}
		\vertex  (a);
		\vertex [left = of a , label=left:$\mu\nu$] (i1);
		\vertex [above right = of a, , label=right:$\rho\sigma$ ] (i2);
		\vertex [below right = of a] (i3);
		\diagram* {
			(i1)--[photon, momentum'=$k_1$](a)--[photon, rmomentum=$k_2$](i2), 
			(a)--[dashed, rmomentum=$k_3$](i3),
		};
	\end{feynman}
\end{tikzpicture}\nn\\
& =& \f{i\sqrt{2}f \kappa^2}{6}\Big( \eta^{\mu\nu}\eta^{\rho\sigma}(k_1^2 +k_2^2)-4\eta^{\mu\rho}\eta^{\nu\sigma} (k_1^2+k_2^2)-6\eta^{\mu\rho}\eta^{\nu\sigma} k_1.k_2\nn\\
&& +2(k_1^\rho k_2^\mu \eta^{\nu\sigma} +k_1^\sigma k_2^\nu \eta^{\mu\rho})\Big).\label{eq:ggd_vertex}
\ee
The graviton-graviton-dilaton vertex $V_{(hh \varphi)}$ which follows from \eqref{eq:EFT_actionSecTwo} was computed in \eqref{eq:hhphi_covariant}.

The three-graviton vertex which follows from the sum of \eqref{eq:kin_dil} and \eqref{eq:kin_gra} reads
\begingroup
\allowdisplaybreaks
\be
&&U_{(hhh)\lambda\kappa} (k_1,k_2,k_3; \varepsilon_1,\varepsilon_2) = \begin{tikzpicture}[baseline=(a)]
	\begin{feynman}
		\vertex  (a);
		\vertex [left = of a, , label=left:$\varepsilon_1$ ] (i1);
		\vertex [above right = of a, label=right:$\varepsilon_2$ ] (i2);
		\vertex [below right = of a, label=right:$\lambda\kappa$] (i3);
		\diagram* {
			(i1)--[photon, momentum'=$k_1$](a)--[photon, rmomentum=$k_2$](i2), 
			(a)--[photon, rmomentum=$k_3$](i3),
		};
	\end{feynman}
\end{tikzpicture}\nn\\
&=& 2i\kappa\, \Big[ -\f{3}{2} k_1.k_2 (\ve_1.\ve_2)\eta_{\lambda\kappa} +2k_1.k_2 \ve_{1\lambda\alpha}\ve_{2\ \kappa}^{\alpha}+ (\ve_1.\ve_2) k_{1\lambda}k_{1\kappa}+(\ve_1.\ve_2) k_{1\lambda}k_{2\kappa}\nn\\
&& +(\ve_1.\ve_2) k_{2\kappa}k_{2\lambda} +(k_1.\ve_2.k_1) \ve_{1\lambda\kappa}+(k_2.\ve_1.k_2)\ve_{2\lambda\kappa}+(k_2.\ve_1.\ve_2.k_1)\eta_{\lambda\kappa}\nn\\
&&-2k_{1\alpha}\ve_2^{\alpha\beta}\ve_{1\beta\lambda}k_{1\kappa}-2k_{2\alpha}\ve_1^{\alpha\beta}\ve_{2\beta\lambda}k_{2\kappa}-2k_1^{\alpha}\ve_{2\alpha\lambda} k_2^{\beta}\ve_{1\beta\kappa}\Big]. \label{eq:3_graviton_vertex}
\ee
\endgroup
On the other hand the three-graviton vertex $V_{(hhh)} $ following from \eqref{eq:EFT_actionSecTwo} appears at order $\kappa^3$, hence it does not contribute to the amplitude \eqref{eq:scattering_HD} at the $\kappa^2 f^{-2}$ order.

No three-dilaton interaction term is present in \eqref{eq:kin_dil} hence $U_{(\varphi\varphi\varphi)}=0$. On the other hand the three dilaton blob vertex $V_{(\varphi\varphi\varphi)}$ has been computed in \eqref{eq:V3phi} and it is non-vanishing for off-shell dilatons.

The graviton-dilaton-graviton-dilaton vertex which follows from the sum of \eqref{eq:kin_dil} and \eqref{eq:kin_gra} reads
\be
U_{(h\varphi h\varphi)} (k_1,k_2,-k_3,-k_4; \varepsilon_1,\varepsilon_3)&=&
	\begin{tikzpicture}[baseline=(a)]
		\begin{feynman}
			\vertex (a);
			\vertex [above left = of a , label=left:$\varepsilon_1$] (i1);
			\vertex [below left = of a ] (i2);
			\vertex [above right = of a, label=right:$\varepsilon_3$] (o1);
			\vertex [below right = of a] (o2);
			\diagram* {
				(i1)--[photon, momentum'=$k_1$](a)--[dashed, rmomentum=$k_2$](i2), 
				(o1)--[photon, rmomentum'=$k_3$ ](a)--[dashed, momentum=$k_4$](o2),
			};
		\end{feynman}
	\end{tikzpicture}\nn\\
	&=&i\kappa^2\Big(\f{t}{2}(\varepsilon_1.\varepsilon_3)-4(k_2.\varepsilon_1.\varepsilon_3.k_4)-4(k_4.\varepsilon_1.\varepsilon_3.k_2 )\nn\\
	&&+\f{2}{3}(k_1.\varepsilon_3.\varepsilon_1.k_3)\Big). \label{eq:kinetic_4_vertex}
\ee
The graviton-dilaton-graviton-dilaton vertex $V_{(h\varphi h\varphi)}$ which follows from \eqref{eq:EFT_actionSecTwo} was already computed in \eqref{result_hhdd} up to the fourth power in momentum. We write its explicit expression here again. It reads
\begingroup
\allowdisplaybreaks
\be
&&V_{(h\varphi h\varphi)} (k_1,k_2,-k_3,-k_4; \varepsilon_1,\varepsilon_3) = \begin{tikzpicture}[baseline=(a)]
	\begin{feynman}
		\vertex [blob, minimum size=0.75 cm] (a) {};
		\vertex [above left = of a , label=left:$\varepsilon_1$] (i1);
		\vertex [below left = of a ] (i2);
		\vertex [above right = of a, label=right:$\varepsilon_3$] (o1);
		\vertex [below right = of a] (o2);
		\diagram* {
			(i1)--[photon, momentum'=$k_1$](a)--[dashed, rmomentum=$k_2$](i2), 
			(o1)--[photon, rmomentum'=$k_3$ ](a)--[dashed, momentum=$k_4$](o2),
		};
	\end{feynman}
\end{tikzpicture}\nn\\
&=&  \frac{i\kappa^2}{3f^2}\Big[\left(
3\Delta c\,t^2 +3\Delta a(s^2+u^2)\right)\times (\varepsilon_1. \varepsilon_3)
-12(-\Delta a+\Delta c)t\times (k_1.\varepsilon_3.\varepsilon_1.k_3)\nn\\
&&-24\Delta a\, u\times(k_2.\varepsilon_3.\varepsilon_1.k_3)
+24\Delta a\, s\times(k_1.\varepsilon_3.\varepsilon_1.k_2)
-24\Delta a\, t\times(k_2.\varepsilon_3.\varepsilon_1.k_2)\nn\\
&&+12(-\Delta a +\Delta c)\times(k_1.\varepsilon_3.k_1)(k_3.\varepsilon_1.k_3) +\left( 24\Delta a+432r_1\right)\times(k_2.\varepsilon_3.k_2)(k_3.\varepsilon_1.k_3)\nn\\
&&+\left( 24\Delta a+432r_1\right)\times(k_2.\varepsilon_1.k_2)(k_1.\varepsilon_3.k_1) +48\Delta a\times(k_3.\varepsilon_1.k_2)(k_1.\varepsilon_3.k_2)\nn\\
&&+864r_1\times(k_2.\varepsilon_1.k_2)(k_2.\varepsilon_3.k_2)
+864r_1\times(k_2.\varepsilon_1.k_2)(k_2.\varepsilon_3.k_1)\nn\\
&&-864r_1\times(k_2.\varepsilon_1.k_3)(k_2.\varepsilon_3.k_2)\Big] .\label{eq:Feynman_1}
\ee
\endgroup

\subsection{Graviton-Dilaton Amplitude}
\label{sec:amplitude}

The leading order in $1/f$ graviton-dilaton amplitude \eqref{eq:scattering_HD} is given by the sum of the following diagrams which contribute at order $\kappa^2$:
\be
\begin{tikzpicture}[baseline=(a)]
	\begin{feynman}
		\vertex (a);
		\vertex [right = of a] (b);
		\vertex [above left = of a , label=left:$\varepsilon_1$] (i1);
		\vertex [below left = of a ] (i2);
		\vertex [above right = of b, label=right:$\varepsilon_3$] (o1);
		\vertex [below right = of b] (o2);
		\diagram* {
			(i1)--[photon, momentum'=$k_1$](a)--[dashed, rmomentum=$k_2$](i2), 
			(a)--[dashed](b),
			(o1)--[photon, rmomentum'=$k_3$ ](b)--[dashed, momentum=$k_4$](o2),
		};
	\end{feynman}
\end{tikzpicture}
\qquad
\begin{tikzpicture}[baseline=(a)]
	\begin{feynman}
		\vertex (a);
		\vertex [right = of a] (b);
		\vertex [above left = of a , label=left:$\varepsilon_1$] (i1);
		\vertex [below left = of a ] (o2);
		\vertex [above right = of b, label=right:$\varepsilon_3$] (o1);
		\vertex [below right = of b] (i2);
		\diagram* {
			(i1)--[photon, momentum'=$k_1$](a)--[dashed, momentum=$k_4$](o2), 
			(a)--[dashed](b),
			(o1)--[photon, rmomentum'=$k_3$ ](b)--[dashed, rmomentum=$k_2$](i2),
		};
	\end{feynman}
\end{tikzpicture}\nn
\ee
\be
\begin{tikzpicture}[baseline=(a)]
	\begin{feynman}
		\vertex (a);
		\vertex [above left = of a , label=left:$\varepsilon_1$] (i1);
		\vertex [below left = of a ] (i2);
		\vertex [above right = of a, label=right:$\varepsilon_3$] (o1);
		\vertex [below right = of a] (o2);
		\diagram* {
			(i1)--[photon, momentum'=$k_1$](a)--[dashed, rmomentum=$k_2$](i2), 
			(o1)--[photon, rmomentum'=$k_3$ ](a)--[dashed, momentum=$k_4$](o2),
		};
	\end{feynman}
\end{tikzpicture}
\qquad\qquad
\begin{tikzpicture}[baseline=(a)]
	\begin{feynman}
		\vertex (a);
		\vertex [below = of a] (b);
		\vertex [above left = of a , label=left:$\varepsilon_1$] (i1);
		\vertex [above right = of a, label=right:$\varepsilon_3$ ] (o1);
		\vertex [below left = of b ] (i2);
		\vertex [below right = of b] (o2);
		\diagram* {
			(i1)--[photon, momentum'=$k_1$](a)--[photon, momentum=$k_3$](o1), 
			(a)--[photon](b),
			(i2)--[dashed, momentum'=$k_2$ ](b)--[dashed, momentum=$k_4$](o2),
		};
	\end{feynman}
\end{tikzpicture}\nn
\ee
These diagrams arise only from the kinetic terms and contain no information about the QFT to which the graviton and dilaton couple. 
Evaluating these diagrams we obtain the following expression
\begingroup
\allowdisplaybreaks
\be
&&\mathcal{T}^{\text{leading in }1/f}_{h\varphi\rightarrow h\varphi}(k_1,k_2,k_3,k_4;\varepsilon_1,\varepsilon_3)\nn\\
&=& -4\kappa^2\left(\f{(k_2.\varepsilon_1.k_2)(k_4.\varepsilon_3.k_4)}{s}+\f{(k_4.\varepsilon_1.k_4)(k_2.\varepsilon_3.k_2)}{u}\right)\nn\\
&& +\kappa^2\left[t(\varepsilon_1.\varepsilon_3)-4(k_2.\varepsilon_1.\varepsilon_3.k_4+k_4.\varepsilon_1.\varepsilon_3.k_2 )\right]\nn\\
&&-\f{\kappa^2}{t}\Big[ \f{1}{2}(\varepsilon_1.\varepsilon_3)(s^2+t^2+u^2)-4t(k_2.\varepsilon_1.\varepsilon_3.k_2)+4s(k_2.\varepsilon_1.\varepsilon_3.k_1)\nn\\
&&-4u(k_3.\varepsilon_1.\varepsilon_3.k_2) +4(k_3.\varepsilon_1.k_3)(k_2.\varepsilon_3.k_2)+4(k_2.\varepsilon_1.k_2)(k_1.\varepsilon_3.k_1)\nn\\
&&+8(k_3.\varepsilon_1.k_2)(k_1.\varepsilon_3.k_2)\Big].
\label{eq:leading_in_f}
\ee
\endgroup

The sub-leading order in $1/f$ of the amplitude is given by the sum of the following diagrams, which contribute at order $\kappa^2 f^{-2}$ (additionally, we will consider only terms up to fourth order in momentum):
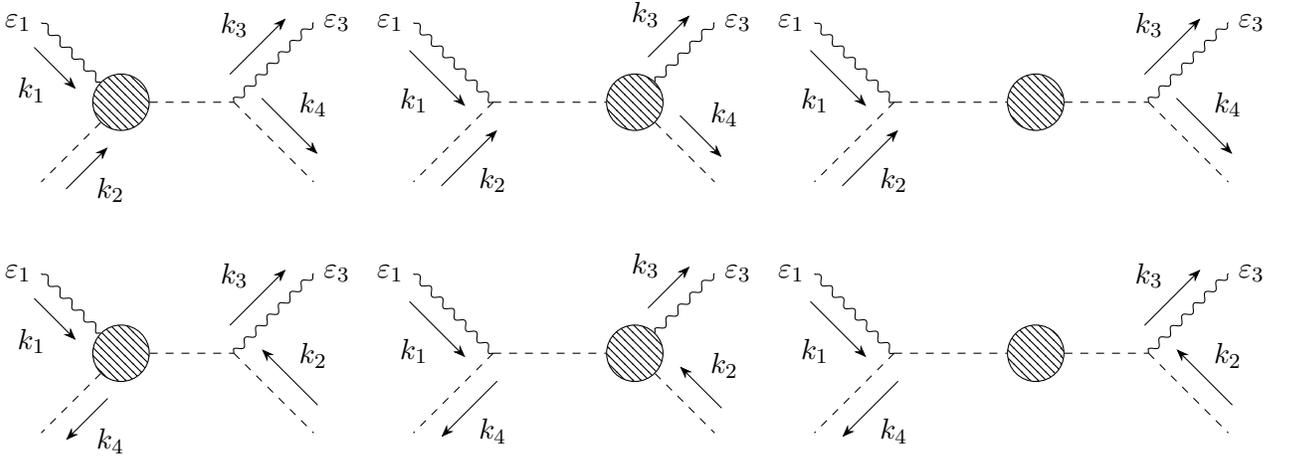
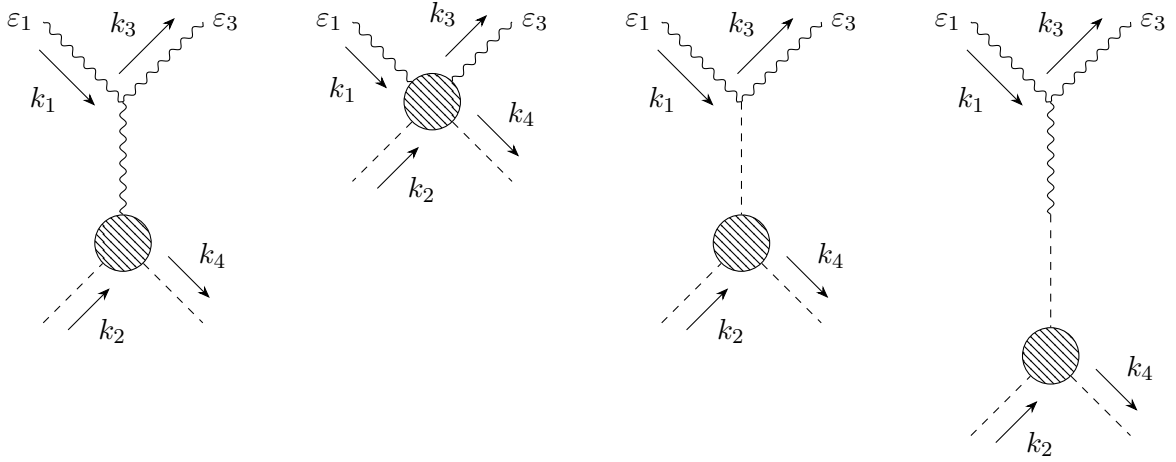
\begin{equation*}
		\begin{tikzpicture}[baseline=(a)]
		\begin{feynman}
			\vertex [blob, minimum size=0.75 cm] (a) {};
			\vertex [above left = of a , label=left:$\varepsilon_1$] (i1);
			\vertex [below left = of a ] (i2);
			\vertex[right = of a](b);
			\vertex [above right = of b, label=right:$\varepsilon_3$] (o1);
			\vertex [below right = of b] (o2);
			\diagram* {
				(i1)--[photon, momentum'=$k_1$](a)--[dashed, rmomentum=$k_2$](i2), 
				(a)--[dashed](b),
				(o1)--[photon, rmomentum'=$k_3$ ](b)--[dashed, momentum=$k_4$](o2),
			};
		\end{feynman}
	\end{tikzpicture}\;
	\begin{tikzpicture}[baseline=(a)]
		\begin{feynman}
			\vertex (a);
			\vertex [right = of a, blob, minimum size=0.75 cm] (b) {};
			\vertex [above left = of a , label=left:$\varepsilon_1$] (i1);
			\vertex [below left = of a ] (i2);
			\vertex [above right = of b, label=right:$\varepsilon_3$] (o1);
			\vertex [below right = of b] (o2);
			\diagram* {
				(i1)--[photon, momentum'=$k_1$](a)--[dashed, rmomentum=$k_2$](i2), 
				(a)--[dashed](b),
				(o1)--[photon, rmomentum'=$k_3$ ](b)--[dashed, momentum=$k_4$](o2),
			};
		\end{feynman}
	\end{tikzpicture}\;
	\begin{tikzpicture}[baseline=(a)]
	\begin{feynman}
		\vertex (a);
		\vertex [above left = of a , label=left:$\varepsilon_1$] (i1);
		\vertex [below left = of a ] (i2);
		\vertex [right= of a, blob, minimum size=0.75 cm] (c) {};
		\vertex[right = of c](b);
		\vertex [above right = of b, label=right:$\varepsilon_3$] (o1);
		\vertex [below right = of b] (o2);
		\diagram* {
			(i1)--[photon, momentum'=$k_1$](a)--[dashed, rmomentum=$k_2$](i2), 
			(a)--[dashed](c)--[dashed](b),
			(o1)--[photon, rmomentum'=$k_3$ ](b)--[dashed, momentum=$k_4$](o2),
		};
	\end{feynman}
\end{tikzpicture}
\end{equation*}
\begin{equation*}
	\begin{tikzpicture}[baseline=(a)]
		\begin{feynman}
			\vertex [blob, minimum size=0.75 cm] (a) {};
			\vertex [above left = of a , label=left:$\varepsilon_1$] (i1);
			\vertex [below left = of a ] (o2);
			\vertex[right = of a](b);
			\vertex [above right = of b, label=right:$\varepsilon_3$] (o1);
			\vertex [below right = of b] (i2);
			\diagram* {
				(i1)--[photon, momentum'=$k_1$](a)--[dashed, momentum=$k_4$](o2), 
				(a)--[dashed](b),
				(o1)--[photon, rmomentum'=$k_3$ ](b)--[dashed, rmomentum=$k_2$](i2),
			};
		\end{feynman}
	\end{tikzpicture}
	\;
	\begin{tikzpicture}[baseline=(a)]
		\begin{feynman}
			\vertex (a);
			\vertex [right = of a, blob, minimum size=0.75 cm] (b) {};
			\vertex [above left = of a , label=left:$\varepsilon_1$] (i1);
			\vertex [below left = of a ] (o2);
			\vertex [above right = of b, label=right:$\varepsilon_3$] (o1);
			\vertex [below right = of b] (i2);
			\diagram* {
				(i1)--[photon, momentum'=$k_1$](a)--[dashed, momentum=$k_4$](o2), 
				(a)--[dashed](b),
				(o1)--[photon, rmomentum'=$k_3$ ](b)--[dashed, rmomentum=$k_2$](i2),
			};
		\end{feynman}
	\end{tikzpicture}
	\;
	\begin{tikzpicture}[baseline=(a)]
	\begin{feynman}
		\vertex (a);
		\vertex [above left = of a , label=left:$\varepsilon_1$] (i1);
		\vertex [below left = of a ] (o2);
		\vertex [right= of a, blob, minimum size=0.75 cm] (c) {};
		\vertex[right = of c](b);
		\vertex [above right = of b, label=right:$\varepsilon_3$] (o1);
		\vertex [below right = of b] (i2);
		\diagram* {
			(i1)--[photon, momentum'=$k_1$](a)--[dashed, momentum=$k_4$](o2), 
			(a)--[dashed](c)--[dashed](b),
			(o1)--[photon, rmomentum'=$k_3$ ](b)--[dashed, rmomentum=$k_2$](i2),
		};
	\end{feynman}
\end{tikzpicture}
\end{equation*}
\begin{equation*}
		\begin{tikzpicture}[baseline=(a)]
		\begin{feynman}
			\vertex (a);
			\vertex [below = of a, blob, minimum size=0.75 cm] (b) {};
			\vertex [above left = of a , label=left:$\varepsilon_1$] (i1);
			\vertex [above right = of a, label=right:$\varepsilon_3$ ] (o1);
			\vertex [below left = of b ] (i2);
			\vertex [below right = of b] (o2);
			\diagram* {
				(i1)--[photon, momentum'=$k_1$](a)--[photon, momentum=$k_3$](o1), 
				(a)--[photon](b),
				(i2)--[dashed, momentum'=$k_2$ ](b)--[dashed, momentum=$k_4$](o2),
			};
		\end{feynman}
	\end{tikzpicture}
	\qquad
	\begin{tikzpicture}[baseline=(a)]
		\begin{feynman}
			\vertex [blob, minimum size=0.75 cm] (a) {};
			\vertex [above left = of a , label=left:$\varepsilon_1$] (i1);
			\vertex [below left = of a ] (i2);
			\vertex [above right = of a, label=right:$\varepsilon_3$] (o1);
			\vertex [below right = of a] (o2);
			\diagram* {
				(i1)--[photon, momentum'=$k_1$](a)--[dashed, rmomentum=$k_2$](i2), 
				(o1)--[photon, rmomentum'=$k_3$ ](a)--[dashed, momentum=$k_4$](o2),
			};
		\end{feynman}
	\end{tikzpicture}
	\qquad
	\begin{tikzpicture}[baseline=(a)]
		\begin{feynman}
			\vertex (a);
			\vertex [below = of a, blob, minimum size=0.75 cm] (b) {};
			\vertex [above left = of a , label=left:$\varepsilon_1$] (i1);
			\vertex [above right = of a, label=right:$\varepsilon_3$ ] (o1);
			\vertex [below left = of b ] (i2);
			\vertex [below right = of b] (o2);
			\diagram* {
				(i1)--[photon, momentum'=$k_1$](a)--[photon, momentum=$k_3$](o1), 
				(a)--[dashed](b),
				(i2)--[dashed, momentum'=$k_2$ ](b)--[dashed, momentum=$k_4$](o2),
			};
		\end{feynman}
	\end{tikzpicture}
	\qquad
		\begin{tikzpicture}[baseline=(a)]
		\begin{feynman}
			\vertex (a);
			\vertex [below = of a] (c);
			\vertex [below = of c, blob, minimum size=0.75 cm] (b) {};
			\vertex [above left = of a , label=left:$\varepsilon_1$] (i1);
			\vertex [above right = of a, label=right:$\varepsilon_3$ ] (o1);
			\vertex [below left = of b ] (i2);
			\vertex [below right = of b] (o2);
			\diagram* {
				(i1)--[photon, momentum'=$k_1$](a)--[photon, momentum=$k_3$](o1), 
				(a)--[photon](c)--[dashed](b),
				(i2)--[dashed, momentum'=$k_2$ ](b)--[dashed, momentum=$k_4$](o2),
			};
		\end{feynman}
	\end{tikzpicture}
	\qquad
	\end{equation*}

We enumerate these diagram from left to right and then from top to bottom. With this convention let us compute them. We find that the first and the second diagrams are given by
\be
i\mathcal{T}_1&=& \f{-i}{(k_1+k_2)^2-i\epsilon}\varepsilon_1^{\mu\nu}V_{(h\varphi\varphi)\mu\nu}(k_1,k_2,-k_1-k_2)U^{\rho\sigma}_{(h\varphi\varphi)}(-k_3,k_1+k_2,-k_4)\ve_{3\rho\sigma}\nn\\
&=&-\f{144r_1i\kappa^2}{f^2}
(k_2.\varepsilon_1.k_2)(k_4.\varepsilon_3.k_4)\nn\\
& =& i\mathcal{T}_2.
\ee
The third diagram is given by 
\be
i\mathcal{T}_3 &=& \left(\f{-i}{(k_1+k_2)^2-i\epsilon}\right)^2 \ve_{1\mu\nu} U^{\mu\nu}_{(h\varphi\varphi)}(k_1,k_2,-k_1-k_2)V_{(\varphi\varphi)}(k_1+k_2,-k_1-k_2)\nn\\
&&\times\ U^{\rho\sigma}_{(h\varphi\varphi)}(-k_3,k_1+k_2,-k_4)\ve_{3\rho\sigma}\nn\\
& =& -i\mathcal{T}_2.
\ee
Analogously for the fourth, fifth and sixth diagrams we find that
\be
	&&i\mathcal{T}_4=i\mathcal{T}_5=-i\mathcal{T}_6\nn\\
	&=&-\f{144 r_1i\kappa^2}{f^2} 
	(k_4.\varepsilon_1.k_4)(k_2.\varepsilon_3.k_2).
\ee
The seventh diagram is given by 
\begingroup
\allowdisplaybreaks
\be
i\mathcal{T}_7 &=& U_{(hhh)}^{\mu\nu}(k_1,-k_3,-k_1+k_3;\varepsilon_1,\varepsilon_3)\Delta^{(h)}_{\mu\nu ,\rho\sigma}(k_1-k_3)V_{(h\varphi\varphi)}^{\rho\sigma}(k_1-k_3,k_2,-k_4)\nn\\
&=&-\f{2i\kappa^2 \Delta a}{f^2}\Big[ \f{1}{2}(\varepsilon_1.\varepsilon_3)(s^2+t^2+u^2)-4t(k_2.\varepsilon_1.\varepsilon_3.k_2)+4s(k_2.\varepsilon_1.\varepsilon_3.k_1)\nn\\
&&-4u(k_3.\varepsilon_1.\varepsilon_3.k_2) +4(k_3.\varepsilon_1.k_3)(k_2.\varepsilon_3.k_2)+4(k_2.\varepsilon_1.k_2)(k_1.\varepsilon_3.k_1)\nn\\
&&+8(k_3.\varepsilon_1.k_2)(k_1.\varepsilon_3.k_2)\Big].
\ee
\endgroup
The eighth diagram is given by
\be
i\mathcal{T}_8= V_{(h\varphi h\varphi)} (k_1,k_2,-k_3,-k_4; \varepsilon_1,\varepsilon_3).
\ee
The ninth diagram is given by
\be
i\mathcal{T}_9= -\f{i\kappa^2}{3f^3}\ \Delta a\ t\left( 3t(\ve_1.\ve_3)-4(k_1.\ve_3.\ve_1.k_3)\right).
\ee
Finally, the tenth diagram is given by
\be
i\mathcal{T}_{10}= \f{i\kappa^2}{3f^3}\ \Delta a\ t\left( 3t(\ve_1.\ve_3)-4(k_1.\ve_3.\ve_1.k_3)\right).
\ee

To summarize, the sub-leading in $1/f$ graviton-dilaton amplitude reads as
\be
&&\mathcal{T}^{\text{sub-leading in }1/f}_{h\varphi\longrightarrow h\varphi} (k_1,k_2,k_3,k_4; \varepsilon_1,\varepsilon_3)\ =\ \sum_{I=1}^{10} \mathcal{T}_I\nn\\
&=& \frac{\kappa^2}{f^2}\left(
\Delta c-\Delta a\right)\times \Big[t^2  (\varepsilon_1. \varepsilon_3)
-4t (k_1.\varepsilon_3.\varepsilon_1.k_3)+4(k_1.\varepsilon_3.k_1)(k_3.\varepsilon_1.k_3)\Big].\label{eq:amplitude}
\ee

Combining \eqref{eq:leading_in_f} and \eqref{eq:amplitude} we obtain the final expression for the graviton-dilaton scattering amplitude
\begin{multline}
	\label{eq:amplitude_final}
	\mathcal{T}_{h\varphi\longrightarrow h\varphi} (k_1,k_2,k_3,k_4; \varepsilon_1,\varepsilon_3) = \mathcal{T}^{\text{leading in }1/f}_{h\varphi\longrightarrow h\varphi} (k_1,k_2,k_3,k_4; \varepsilon_1,\varepsilon_3)+
	\\\mathcal{T}^{\text{sub-leading in }1/f}_{h\varphi\longrightarrow h\varphi} (k_1,k_2,k_3,k_4; \varepsilon_1,\varepsilon_3)+ O(\kappa^3).
\end{multline}
Let us remind that in the computation of the amplitude above, we set the cosmological constant (c.c.) $\lambda$ to zero in the invariant part of the EFT action \eqref{eq:Ainvariant}. This is done for the convenience of having scattering of massless particles and can always be achieved by tuning the UV c.c. appropriately. It is not necessary in general.\footnote{If we do not set $\lambda$ to zero in \eqref{eq:Ainvariant}, it effectively provides mass to the dilaton at order $\frac{M^2}{f} \sqrt{\lambda}$. Now, once we properly amputate the external dilatons in the computation of the scattering amplitude, the c.c. term only contributes to the amplitude $\mathcal{T}{}_{+2}^{+2}(s,t,u)$ in \eqref{eq:linearized transformation_compact} at zeroth power in momenta and at order $\kappa^2f^{-2}$ in the decoupling limit.} However, the physical content of what we are doing can be rephrased as sending a graviton through the state created by the energy-momentum tensor trace, and the massless dilaton is just a convenient choice for this purpose.

The result of the 2-graviton-2-dilaton amplitude in \eqref{eq:amplitude_final} is not affected by the potential presence of interactions between the IR CFT operators and gravitons/dilatons as we show in appendix~\ref{DimTwo}.

\paragraph{Linearized gauge invariance}
The amplitude \eqref{eq:amplitude_final} is  invariant under linearized gauge transformations
\begin{equation}
	\label{eq:linearized transformation}
	\varepsilon^{\mu\nu}_i(k_i) \rightarrow \varepsilon^{\mu\nu}_i(k_i)+\chi^\mu k_i^\nu\ +\ \chi^\nu k_i^\mu,
\end{equation} 
where $\chi^\mu(x)$ is some generic vector field with $k_{i\mu}\chi^{\mu}$=0. Consider the following object
\begin{equation}
	\label{eq:definition_H}
	(H_i)^{\mu\nu,\,\rho\sigma} \equiv 
	k_i^\mu k_i^\rho \varepsilon_i^{\nu\sigma}(k_i)-
	k_i^\mu k_i^\sigma \varepsilon_i^{\nu\rho}(k_i)-
	k_i^\nu k_i^\rho \varepsilon_i^{\mu\sigma}(k_i)+
	k_i^\nu k_i^\sigma \varepsilon_i^{\mu\rho}(k_i).
\end{equation}
It is explicitly invariant under the transformation \eqref{eq:linearized transformation}. Gauge invariant amplitudes should be built only as invariant contractions of  \eqref{eq:definition_H}. We can define the following two tensor structures
\begin{equation}
	\label{eq:3structures}
	\begin{aligned}
		\mathbf{T}_1 &\equiv (H_1)^{\mu\nu,\,\rho\sigma} (H_3)_{\mu\nu,\,\rho\sigma},\\
		\mathbf{T}_2 &\equiv (H_1)^{\mu\nu,\,\rho\sigma} (H_3)_{\mu\nu,\,\rho\sigma}+\left(\frac{16}{s u}\right)\,k_2^{\alpha_1} k_3^{\alpha_2}(H_1)^{\mu\nu}{}_{\alpha_1\alpha_2} k_1^{\beta_1} k_2^{\beta_2} (H_3)_{\mu\nu,\,\beta_1\beta_2}\\
		&+ \left(\frac{8}{s u}\right)^2\,\left(k_2^{\alpha_1} k_3^{\alpha_2}  k_2^{\alpha_3}  k_3^{\alpha_4} (H_1)_{\alpha_1\alpha_2,\,\alpha_3\alpha_4}\right)
		\left(k_1^{\beta_1} k_2^{\beta_2}  k_1^{\beta_3}  k_2^{\beta_4} (H_3)_{\beta_1\beta_2,\,\beta_3\beta_4}\right).
	\end{aligned}
\end{equation}
More generally the construction of tensor structures in $d$ dimensions was recently discussed in \cite{Chowdhury:2019kaq}.
It is straightforward to check that the expression \eqref{eq:amplitude_final} can be rewritten in the following simple form
\begin{equation}
	\label{eq:linearized transformation_compact}
	\mathcal{T}_{h\varphi\longrightarrow h\varphi} (k_1,k_2,k_3,k_4; \varepsilon_1,\varepsilon_3) =
	\kappa^2
	 \frac{su}{t^3}\,\mathbf{T}_2+
	\frac{\kappa^2}{f^2}\left(\Delta c-\Delta a\right)\mathbf{T}_1.
\end{equation}

We use the centre of mass frame approach recently reviewed in \cite{Hebbar:2020ukp} and applied to spin one massless  particles (a.k.a photons) in \cite{Haring:2022sdp}. 
Using the explicit expressions for graviton polarizations given in appendix \ref{app:graviton_polarizations} we can evaluate the graviton-dilaton amplitude in the center of mass frame. It reads
\begin{equation}
	\label{eq:helicity_amplitudes}
	\begin{aligned}
		\mathcal{T}{}_{+2}^{+2} (s,t,u) = \mathcal{T}{}_{-2}^{-2} (s,t,u) &= 
		\kappa^2\,\frac{s u}{t},\\
		\mathcal{T}{}_{+2}^{-2} (s,t,u) = \mathcal{T}{}_{-2}^{+2} (s,t,u) &=  \frac{\kappa^2}{f^2}\left(
		\Delta c-\Delta a\right)t^2.
	\end{aligned}
\end{equation}
We see that, miraculously, the helicity flipping amplitude is universally proportional to $\Delta c-\Delta a$  at low energies.
More formally, we can read off $(\Delta c-\Delta a)$ from taking the second derivative in $t$ and then the following limit 
\be
(\Delta c-\Delta a)=\lim_{s,t\to 0} \lim_{f\rightarrow \infty}\lim_{\kappa\rightarrow 0}\f{f^2}{2\kappa^2}\ \p_t^2 \mathcal{T}{}_{+2}^{-2} (s,t,-s-t).
\ee

In appendix \ref{app:graviton_dilaton_scattering} we discuss some simple properties of $\mathcal{T}{}_{+2}^{-2} (s,t,-s-t)$. First of all, we show that this amplitude vanishes in the forward limit, $ \mathcal{T}{}_{+2}^{-2} (s,0,-s)=0$. This is why the amplitude looks like $t^2$ at low energies and does not have a piece $s^2+u^2$. Second, we write a dispersion relation 
\begin{equation}
	\label{eq:dispersion_relation}
	\Delta c-\Delta a  = 
	{f^2\over \kappa^2}\int_{m^2}^\infty  \frac{ds}{\pi} \frac{\text{Im}\,\partial_t^2\mathcal{T}{}_{+2}^{-2}(s,0,-s)}{s},
\end{equation}
where $m$ is the mass gap which represents the mass of the lowest massive particle integrated out along the RG flow (in gapless theories $m=0$ but $\text{Im}\,\partial_t^2\mathcal{T}{}_{+2}^{-2}(s,0,-s)$ vanishes rapidly at small $s$ since the dilaton is decoupled from the infrared CFT so effectively the sum rule does not probe the infrared CFT degrees of freedom but only the massive particles which are effectively responsible for the $\Delta c-\Delta a$ in the RG flow).

\section{Trace Anomalies in Coupling Space (Resonance Anomalies)}
\label{CouplSpac}

So far we have discussed QFTs coupled to the background fields $\Omega(x),\;g_{\mu\nu}(x)$ and the $a$- and $c$-trace anomalies. These trace anomalies enter through \eqref{transrulestwo} which does not contain $\Omega(x)$ itself.  
A more general class of trace anomalies arises in QFTs where the {\it ultraviolet} CFT has operators with integer (``resonant'') scaling dimension. These anomalies are sometimes called ``resonance'' or ``coupling space'' anomalies.
To be concrete and to avoid cluttered notation, consider a UV CFT which contains a scalar operator $\cO(x)$ whose scaling dimension is $\Delta=2$. We couple this CFT to the background (source) field $J(x)$ in a conformally invariant way as
\begin{equation*}
	 \int d^4x \sqrt{-g} J(x) \cO_{\Delta=2}(x).
\end{equation*}
Notice that $J(x)$ is now required to transform under Weyl transformations like $J(x)\rightarrow e^{-2\sigma(x)}J(x)$. One can show that the partition function of the CFT with the above deformation has the following Weyl anomaly\footnote{Similarly to the discussion after~\eqref{transrulestwo}, the variation~\eqref{transrulesthree} is in a certain scheme. We could add local non-Weyl invariant terms in the UV of the form $\int d^4x \sqrt{-g}  R(x)J(x)$ which would affect the variation~\eqref{transrulesthree} by exact terms of the form $ \int d^4x \sqrt{-g}\sigma(x)\square J(x)$. }${}^,$\footnote{In order to do this one computes the connected functional $W [J,\, g_{\mu\nu}]$ as a series expansion in $J(x)$. The quadratic term in $J(x)$ in this expansion is UV divergent. In order to regularize it we are forced to introduce Weyl symmetry breaking counter terms which introduce the new anomaly.}
  \begin{equation}\label{transrulesthree}
  	\delta_\sigma W [J,\, g_{\mu\nu}] = \int d^4x \sqrt{-g}\sigma(x)\left(-a_{UV} E_4+c_{UV} \cW^2+c^\text{UV}_J J^2\right).
  \end{equation}
Here the real coefficient $c_J^{UV}$ is the new trace anomaly. It is clear why this new trace anomaly arises only for operators of special scaling dimensions: the variation of the connected functional $ W [J,\, g_{\mu\nu}]$ has to be local in the sources of the ultraviolet CFT. We will find that in the free massless scalar CFT with the operator $\cO_{\Delta=2}(x)$ given by the square of the free field $\Phi(x)^2$, the value of the new trace anomaly is $c^\text{UV}_J=\tfrac{1}{8\pi^2}$.

Suppose now that the source field has the following special form $J(x)= \Omega(x)^2M^2$, where $M$ is a mass parameter and  $\Omega(x)\equiv e^{-\tau(x)}$ denotes the dilaton. Such a choice of $J(x)$ triggers an RG flow. Requiring that the IR effective action of this theory  correctly reproduces the Weyl anomaly \eqref{transrulesthree}, we arrive at the following expression
\begin{equation}
	\label{eq:EFT_coupling_space}
	A_\text{EFT}[\Omega,\, g_{\mu\nu}] =  -a_{\text{UV}}\times A_a[\tau,\, g_{\mu\nu}] + c_{\text{UV}} \times A_c[\tau,\, g_{\mu\nu}] + c^\text{UV}_J\times A_{c_J}[\tau ,g_{\mu\nu}] + A_{\text{invariant}}[ \widehat{g}_{\mu\nu}],
\end{equation}
where the new term $A_{c_J}$ in the effective action reads as
\begin{equation}
	\label{eq:AcJ}
	A_{c_J}[\tau ,g_{\mu\nu}] =M^4 \int d^4 x \sqrt{-g}\, \tau(x)\, e^{-4\tau(x)}.
\end{equation}
One can use the effective action \eqref{eq:EFT_coupling_space} to compute vertices of dilatons at the lowest order in momentum in order to extract the value of $c^\text{UV}_J$ and the $\lambda$ coefficient appearing in \eqref{eq:Ainvariant}. We will show that by computing one- and two-point vertices of dilatons one can always disentangle the values of $c^\text{UV}_J$ and $\lambda$.

In the previous paragraph the source $J(x)$ itself triggered the RG flow. There could be another, distinct, situation when the RG flow is triggered by other operators. Assuming the IR theory is gapped, then, in order to match the new trace anomaly the IR effective action must contain instead
 \begin{equation}
	\label{eq:AcJTwo}
	A_{c_J}[\tau ,g_{\mu\nu}] = \int d^4 x \sqrt{-g}\, \tau(x) J^2
\end{equation}
multiplied by $c^\text{UV}_J$. One will also have more Weyl invariant terms compared to \eqref{eq:Ainvariant} such as for example
\begin{equation*}
	\int d^4 x \sqrt{-g}J^2,\qquad
	\int d^4 x \sqrt{-g}e^{-2\tau} J,\qquad
	\int d^4 x \sqrt{-g}\widehat R e^{-2\tau}J,\qquad\ldots
\end{equation*}
In general, $A_{\rm invariant}$ is a local coordinate invariant  functional of $\widehat{g}_{\mu\nu}$ and $\widehat{J}=e^{2\tau} J$. Notice that the new terms involving $\widehat{J}$ in the effective action $A_{\rm invariant}$ do not give anything new if we set $J(x)=M^2\Omega(x)^2$. The only novelty comes from the term \eqref{eq:AcJTwo} as we will discuss below in the  free scalar example.

Finally, if in the IR the theory is described by an IR CFT which contains operators with $\Delta=2$ scaling dimensions which can couple to $J(x)$, it will contain an IR ``coupling space'' anomaly $c^\text{IR}_J$. In order to match the new anomaly one will have to multiply \eqref{eq:AcJTwo} by $\Delta c_J\equiv c_{J}^{ \text{UV}}-c_{J}^{ \text{IR}} $.

In this section we study the ``coupling space'' trace anomaly in a simple example. We start by considering a massless free boson on a curved background coupled to the external background (source) field $J(x)$. We show that this theory has the ``coupling space'' trace anomaly $c_J$ and we compute it explicitly. We then focus on the case when $J(x)=-\frac{1}{2}m^2\Omega(x)^2$, where $\Omega(x)$ is the compensator field and $m$ is the mass parameter. Such choice of $J(x)$ triggers an RG flow. Notice, that this case precisely describes the compensated free massive scalar theory already studied in section \ref{sec:example}. We derive the IR effective action under the requirement that the ``coupling space'' anomaly is invariant along the flow (the anomaly should be matched in IR). We will test our IR effective action by performing an explicit perturbative computation.

\paragraph{Coupling space trace anomaly}
Consider the following action
\begin{equation}
	\label{eq:free_scalar_J}
	A[\Phi, J,g_{\mu\nu}]=
	\int d^4x\sqrt{-g(x)}\Big(-\f{1}{2}g^{\mu\nu}(x)\p_\mu\Phi(x) \p_\nu\Phi(x)  -\f{1}{12}R(x)\Phi^2(x)+J(x)\Phi^2(x)\Big),
\end{equation}
Using the action \eqref{eq:free_scalar_J} let us compute the connected functional defined as
\be
e^{iW[J,g_{\mu\nu}]}\equiv \int [d\Phi]_g\ e^{iA[\Phi, J,g_{\mu\nu}]}.
\ee
One can rewrite this expression as
\be
e^{iW[J,g_{\mu\nu}]}&= &e^{iW_{\text{free}}[g_{\mu\nu}]}\times e^{iW_{\text{source}}[J,g_{\mu\nu}]} + \ldots,
\ee
where the free connected functional is defined as
\be
e^{iW_{\text{free}}[J,g_{\mu\nu}]}\equiv \int [d\Phi]_g\ e^{	\int d^4x\sqrt{-g(x)}\Big(-\f{1}{2}g^{\mu\nu}(x)\p_\mu\Phi(x) \p_\nu\Phi(x)  -\f{1}{12}R(x)\Phi^2(x)\Big)}
\ee
and the source connected functional is given by
\begin{multline}
	\label{eq:source}
W_{\text{source}}[J,g_{\mu\nu}]=  -i\log\Big( 1-\f{1}{2}\int d^4x_1\sqrt{-g(x_1)} \int d^4x_2\sqrt{-g(x_2)} J(x_1)J(x_2)\\
\times \langle \Phi^2(x_1)\Phi^2(x_2)\rangle^g_{\text{free scalar}}
+O(J^3)\Big).
\end{multline}
The time-ordered correlation function is denoted by angular brackets.

Let us now focus on the flat background $g_{\mu\nu}(x)=\eta_{\mu\nu}$. Using the Wick contraction in \eqref{eq:source} and focusing only on the quadratic term in $J(x)$ we obtain
\be
	\label{eq:source_2}
W_{\text{source}}[J,\eta_{\mu\nu}]&=&i\int d^4x_1\int d^4x_2 J(x_1)J(x_2)\left(\langle \Phi(x_1)\Phi(x_2)\rangle_{\text{free scalar}}\right)^2.
\ee
The two-point function in the free massless theory is given by
\begin{equation}
	\langle \Phi(x_1)\Phi(x_2)\rangle_{\text{free scalar}} = \int \f{d^4p}{(2\pi)^4}e^{ip\cdot (x_1-x_2)}\f{-i}{p^2-i\epsilon}.
\end{equation}
Plugging this into \eqref{eq:source_2} and performing the position space integrals we obtain
\be
W_{\text{source}}[J,\eta_{\mu\nu}]&=&-i\mu^{4-d}\int \f{d^dq}{(2\pi)^d}J(q)J(-q)\int \f{d^dp}{(2\pi)^d}\f{1}{p^2-i\epsilon}\f{1}{(p-q)^2-i\epsilon}.
\ee
We wrote the above expression in general $d$-space time dimensions in order to regulate the divergence appearing $d=4$. This allows us to evaluate the above expression and obtain
\be
W_{\text{source}}[J,\eta_{\mu\nu}]&=&\mu^\epsilon \f{\Gamma\left(2-\f{d}{2}\right)}{(4\pi)^{\f{d}{2}}}\frac{\sqrt{\pi } 2^{3-d} \Gamma \left(\frac{d}{2}-1\right)}{\Gamma \left(\frac{d-1}{2}\right)}\int \f{d^d q}{(2\pi)^d}J(q)J(-q) (q^{2}-i\ve)^{\f{d}{2}-2}.
\ee
Focusing on the case of $d=4-\epsilon$ and expanding in the $\epsilon\rightarrow 0$ limit we get
\begin{equation}
	\label{eq:source_3}
	W_{\text{source}}[J,\eta_{\mu\nu}]=\int \f{d^4 q}{(2\pi)^4}J(q)J(-q)\left(\f{1}{8\pi^2 \epsilon} +\f{2-\gamma_E}{16\pi^2} - \f{1}{16\pi^2}\log\left(\f{q^2}{4\pi\mu^2}\right)\right).
\end{equation}
Let us now use the $\overline{MS}$ renormalization scheme. In this scheme we define the counter term in such a way that the pole at $\epsilon=0$ together with the constant term in \eqref{eq:source_3} are cancelled. As a result we get
\be
W_{ \text{source}}[J,\eta_{\mu\nu}]&=&- \f{1}{16\pi^2}\int \f{d^4 q}{(2\pi)^4}J(q)J(-q) \log\left(\f{q^2}{\mu^2}\right).
\ee
In position space the above non-local renormalized connected functional reads as
\be
W_{ \text{source}}[J,\eta_{\mu\nu}]&=&-\f{1}{16\pi^2}\int d^4x \ J(x)\log\left(\f{-\p^2}{\mu^2}\right)J(x).
\ee

Let us now promote the above expression to curved background, namely
\be
\label{eq:source_4}
W_{ \text{source}}[J,g_{\mu\nu}]&=&-\f{1}{16\pi^2}\int d^4x \sqrt{-g}\ J(x)\log\left(-\f{\square+\ldots}{\mu^2}\right)J(x).
\ee
Here the ellipses stand for a term linear in Ricci scalar which is required to obtain the following homogeneous Weyl transformation property\footnote{For instance, in the case of free massless scalar $\Phi(x)$ one can show explicitly that $\left(\square -\f{1}{6}R\right)\Phi(x)\overset{\text{Weyl}}{\longrightarrow} e^{-3\sigma(x)}\left(\square -\f{1}{6}R\right)\Phi(x)$.}
\begin{equation}
	\log\left(-\frac{\square+\ldots}{\mu^2}\right)J(x)
	\overset{\text{Weyl}}{\longrightarrow} 
	e^{-2\sigma(x)}\log\left(-e^{-2\sigma(x)}\frac{\square+\ldots}{\mu^2}\right)J(x).
\end{equation}
It is not important to find this curvature correction term as it vanishes in flat spacetime.
As a result the Weyl variation of the connected functional \eqref{eq:source_4} is given by
\be
\label{eq:coupling_space_anomaly}
\delta_\sigma W_{ \text{source}}[J,g_{\mu\nu}]&=&c_J \int d^4x \sqrt{-g}\ \sigma(x)\ J^2(x), 
\ee
where
\be
c_J=\f{1}{8\pi^2}.\label{eq:c_J_free_scalar}
\ee
As we can see the variation \eqref{eq:coupling_space_anomaly} is non-zero, in other words we have an extra Weyl anomaly described by the coefficient $c_J$ called the ``coupling space'' anomaly.

\paragraph{Compensated massive scalar theory}
Let us now consider the action \eqref{eq:free_scalar_J} in the special case when the source is given by
\begin{equation}
	\label{eq:J_choice}
	J(x)=-\frac{1}{2}m^2\Omega(x)^2,
\end{equation}
where $\Omega(x) = e^{-\tau(x)}$ is the compensator field. Then the action \eqref{eq:free_scalar_J} agrees with the compensated free theory action  \eqref{eq:compensated_free_scalar} in $d=4$.

This action precisely describes the compensated massive scalar theory  \eqref{eq:compensated_free_scalar} studied in section \ref{sec:example}.  Let us consider the following IR effective action
\be
\label{eq:zero_derivative_EFT}
A_{\text{IR}}[\tau ,g_{\mu\nu}]=\f{m^4}{4}\int d^4x\sqrt{-g}e^{-4\tau(x)}\left(\lambda +c_J\  \tau(x)\right).
\ee
It is constructed in such a way that its Weyl transformation has exactly the same form as \eqref{eq:coupling_space_anomaly}. The very first term here is invariant under Weyl transformations. We can expand this action in power of $\tau(x)$, we get then
\begin{multline}
A_{\text{IR}}[\tau ,\eta_{\mu\nu}]=\f{m^4}{4}\int d^4x\ \Bigg(\lambda +\tau  (c_J-4 \lambda )+\tau^2 (8 \lambda -4 c_J)\\+\tau ^3 \left(8 c_J-\frac{32 \lambda }{3}\right)+O\left(\tau ^4\right)\Bigg).
\end{multline}

Let us define the following vertices
\begin{equation}
	(2\pi)^4\delta^{(4)(k_1 + \ldots + k_n)}V_{(\tau\ldots\tau)}(k_1,\ldots k_n) \equiv (2\pi)^{4n} \frac{i\delta^n A_\text{IR}}{\delta\tau(k_1)\ldots\delta\tau(k_n)}.
\end{equation}
Using the above effective action we obtain the following explicit expressions
\be
V_{(\tau)}(k_1)&=& \f{im^4}{4} (c_J-4\lambda), \label{eq:V1phi}\\
V_{(\tau\tau)}(k_1,k_2)&=&\f{im^4}{2} (8 \lambda -4 c_J) ,\label{eq:V2phi}\\
V_{(\tau\tau\tau)}(k_1,k_2,k_3)&=& \f{3im^4}{2}\left(8 c_J-\frac{32 \lambda }{3}\right)\label{eq:V3tau}.
\ee
These vertices can be used to probe the value of $c_J$.

Let us now compute these vertices using the action  \eqref{eq:compensated_free_scalar} in flat background. This will allow us to extract the values of $\lambda$ and the ``coupling space'' anomaly $c_J$. We will get precisely the result  \eqref{eq:c_J_free_scalar} which confirms the consistency of our discussion.

Let us write here the action \eqref{eq:compensated_free_scalar} for convenience (which is equivalent to \eqref{eq:free_scalar_J} combined with \eqref{eq:J_choice}), it reads
\begin{multline}
\label{eq:free_scalar_flat_space}
A_{\text{QFT}}[\Phi, \tau]= \int d^{4-\epsilon}x \Big(-\f{1}{2}\eta^{\mu\nu}\p_\mu\Phi(x) \p_\nu\Phi(x)\\  -\f{1}{2}m^2 \left(1-2 \tau(x) +2 \tau(x)^2
-\f{4}{3}\tau^3+O\left(\tau^4\right)\right)\Phi^2(x)\Big).
\end{multline}
Using this action we can derive the following Feynman rules
\begin{equation*}
	\feynmandiagram[horizontal=a to b, inline=(a.base)] {
		a -- [momentum=$p$] b
	};\  = \frac{-i}{p^2 +m^2 -i\epsilon},\qquad
	\begin{tikzpicture}[baseline=(a)]
		\begin{feynman}
			\vertex (a);
			\vertex [above left = of a] (i1);
			\vertex [below left = of a] (i2);
			\vertex [right = of a] (o1);
			\diagram* {
				(i1)--[momentum=$p_1$](a)--[rmomentum=$p_2$](i2),  
				(o1)--[dashed, momentum'=$k$](a),
			};
		\end{feynman}
	\end{tikzpicture} = 2im^2
\end{equation*}
\begin{equation*}
		\begin{tikzpicture}[baseline=(a)]
		\begin{feynman}
			\vertex (a);
			\vertex [above left = of a] (i1);
			\vertex [below left = of a] (i2);
			\vertex [above right = of a ] (o1);
			\vertex [below right = of a ] (o2);
			\diagram* {
				(i1)--[momentum'=$p_1$](a)--[rmomentum=$p_2$](i2), 
				(o1)--[dashed, momentum'=$k_1$ ](a)--[dashed, rmomentum=$k_2$](o2),
			};
		\end{feynman}
	\end{tikzpicture} = -4im^2,\qquad
\begin{tikzpicture}[baseline=(a)]
	\begin{feynman}
		\vertex (a);
		\vertex [above left = of a] (i1);
		\vertex [below left = of a] (i2);
		\vertex [above right = of a, label=right:$k_1$ ] (o1);
		\vertex [below right = of a, label=right:$k_2$ ] (o2);
		\vertex [right = of a, label=right:$k_3$ ] (o3);
		\diagram* {
			(i1)--[momentum'=$p_1$ ](a)--[rmomentum=$p_2$](i2), 
			(a)--[dashed ](o1),
			(a)--[dashed ](o2),
			(a)--[dashed ](o3),
				};
	\end{feynman}
\end{tikzpicture} = 8im^2,
\end{equation*}
where the solid lines describe the field $\Phi(x)$ and the dashed lines describe the field $\tau(x)$.

The one point vertex reads
\be
V_{(\tau)}(k_1) &\equiv & 
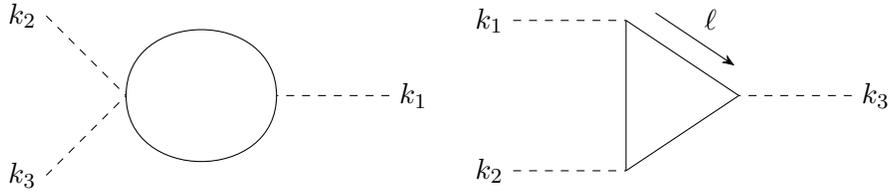
\begin{tikzpicture}[baseline=(a.base)]
	\begin{feynman}
		\vertex (a);
		\vertex [right=2.0 of a] (b);
		\vertex [left=of a, label=left:$k_1$] (i1);
		\diagram* {
			(a) --[half left] (b) -- [half left] (a),
			(i1) --[dashed] (a),
					};
	\end{feynman}
\end{tikzpicture}\nn\\
 &=& m^2\ \bold{J}(0,1;m^2)\nn\\
 &=&- \frac{i m^4}{8 \pi ^2 \epsilon } +\frac{i m^4}{16 \pi^2 }\left( (\gamma_E-1)+\log\left(\f{m^2}{4\pi}\right)\right).
\ee
Recall the the object $\mathbf{J}\left(a,b;\Delta\right)$ was defined and computed in \eqref{eq:master_integral}.
The two-point vertex reads as
\be
V_{(\tau\tau)}(k_1,k_2)&\equiv & 
\begin{tikzpicture}[baseline=(a.base)]
	\begin{feynman}
		\vertex (a);
		\vertex [right=2.0 of a] (b);
		\vertex [above left=of a, label=left:$k_1$] (i1);
		\vertex [below left=of a, label=left:$k_2$] (i2);
		\diagram* {
			(a) --[half left] (b) -- [half left] (a),
			(i1) --[dashed] (a),
			(i2) --[dashed] (a),
		};
	\end{feynman}
\end{tikzpicture} \quad+ \quad
\begin{tikzpicture}[baseline=(a.base)]
	\begin{feynman}
		\vertex (a);
		\vertex [right=2.0 of a] (b);
		\vertex [left=of a, label=left:$k_1$] (i1);
		\vertex [right=of b, label=right:$k_2$] (i2);
		\diagram* {
			(a) --[half left, momentum={$\ell$}] (b) -- [half left, rmomentum={$k_1-\ell$}] (a),
			(i1) --[dashed] (a),
			(i2) --[dashed] (b),
		};
	\end{feynman}
\end{tikzpicture} \nn
\ee
Evaluating these diagrams we obtain
\be
V_{(\tau\tau)}(k_1,k_2)&=&-2m^2\ \bold{J}(0,1;m^2) +2m^4\int_0^1 dx\ \bold{J}(0,2;m^2+x(1-x)k_1^2)\nn\\
&=& \frac{i m^4}{ 2\pi ^2\epsilon} -\f{im^4}{8\pi^2}\left( (2\gamma_E -1)+2\log\left(\f{m^2}{4\pi }\right)\right) +O(k_1^2). 
\ee
Comparing the expressions for the above two vertices with \eqref{eq:V1phi} and \eqref{eq:V2phi} we conclude that
\be
\label{eq:c_J_lambda_sol}
c_J= \frac{1}{8 \pi ^2},\qquad\ \lambda = \f{1}{8 \pi ^2 \epsilon } +\f{1}{32\pi^2}\left( 3-2\gamma_E -2\log\left(\f{m^2}{4\pi}\right)\right).
\ee
The above $c_J$ value agrees with the UV result in \eqref{eq:c_J_free_scalar}. Therefore we see that the ``coupling space'' trace anomaly is matched. Note that the computation above is performed using the bare action \eqref{eq:free_scalar_flat_space}. However, it is important to emphasize that the choice of regularization scheme and counter terms will only impact the constant $\lambda$ and not the result of $c_J$ derived above.

As a consistency check let us also compute the three-point vertex. It is given by the sum of the following diagrams
\begin{equation*}
	 \begin{tikzpicture}[baseline=(a.base)]
		\begin{feynman}
			\vertex (a);
			\vertex [right=2.0 of a] (b);
			\vertex [above left=of a, label=left:$k_1$] (i1);
			\vertex [below left=of a, label=left:$k_2$] (i2);
			\vertex [left=of a, label=left:$k_3$] (i3);
			\diagram* {
				(a) --[half left] (b) -- [half left] (a),
				(i1) --[dashed] (a),
				(i2) --[dashed] (a),
				(i3) --[dashed] (a)
			};
		\end{feynman}
	\end{tikzpicture}\quad
	\begin{tikzpicture}[baseline=(a.base)]
		\begin{feynman}
			\vertex (a);
			\vertex [right=2.0 of a] (b);
			\vertex [above left=of a, label=left:$k_1$] (i1);
			\vertex [below left=of a, label=left:$k_2$] (i2);
			\vertex [right=of b, label=right:$k_3$] (i3);
			\diagram* {
				(a) --[half left] (b) -- [half left] (a),
				(i1) --[dashed] (a),
				(i2) --[dashed] (a),
				(i3) --[dashed] (b)
			};
		\end{feynman}
	\end{tikzpicture}\quad
	\begin{tikzpicture}[baseline=(a.base)]
		\begin{feynman}
			\vertex (a);
			\vertex [right=2.0 of a] (b);
			\vertex [above left=of a, label=left:$k_1$] (i1);
			\vertex [below left=of a, label=left:$k_3$] (i2);
			\vertex [right=of b, label=right:$k_2$] (i3);
			\diagram* {
				(a) --[half left] (b) -- [half left] (a),
				(i1) --[dashed] (a),
				(i2) --[dashed] (a),
				(i3) --[dashed] (b)
			};
		\end{feynman}
	\end{tikzpicture}
\end{equation*}
\begin{equation*}
	\begin{tikzpicture}[baseline=(a.base)]
		\begin{feynman}
			\vertex (a);
			\vertex [right=2.0 of a] (b);
			\vertex [above left=of a, label=left:$k_2$] (i1);
			\vertex [below left=of a, label=left:$k_3$] (i2);
			\vertex [right=of b, label=right:$k_1$] (i3);
			\diagram* {
				(a) --[half left] (b) -- [half left] (a),
				(i1) --[dashed] (a),
				(i2) --[dashed] (a),
				(i3) --[dashed] (b)
			};
		\end{feynman}
	\end{tikzpicture}\quad
	\begin{tikzpicture}[baseline=(c.base)]
		\begin{feynman}
			\vertex (c);
			\vertex [below left=1 and 1.5 of c] (b);
			\vertex [above left=1 and 1.5 of c] (a);
			\vertex [left=of a, label=left:$k_1$] (i1);
			\vertex [left=of b, label=left:$k_2$] (i2);
			\vertex [right=of c, label=right:$k_3$] (i3);
			\diagram* {
				(a) -- (b) --(c) --[rmomentum'={$\ell$}] (a),
				(i1) --[dashed] (a), (i2)--[dashed] (b),
				(i3) --[dashed] (c),
			};
		\end{feynman}
	\end{tikzpicture}
\end{equation*}
Evaluating these diagrams we obtain
\be 
V_{(\tau\tau\tau)}(k_1,k_2,k_3)&=& 4m^2\ \bold{J}(0,1;m^2)-4m^4\int_0^1 dx\ \Big[\bold{J}\left(0,2;m^2+x(1-x)(k_1+k_2)^2\right)\nn\\
&&+\bold{J}\left(0,2;m^2+x(1-x)(k_2+k_3)^2\right)+\bold{J}\left(0,2;m^2+x(1-x)(k_1+k_3)^2\right)\Big] \nn\\
&& +16 m^6\int_0^1 dx\int_0^{1-x}dy\ \bold{J}\left( 0,3;m^2+x(1-x)(k_1+k_2)^2+y(1-y)k_1^2\right)\nn\\
&=& -\frac{2 i m^4}{\pi ^2 \epsilon } +\f{im^4}{\pi^2}\left( \gamma_E +\log\left(\f{m^2}{4\pi}\right)\right)+O(k_i.k_j, k_i^2).
\ee
The above result is consistent with \eqref{eq:V3tau} and \eqref{eq:c_J_lambda_sol}.

\section{A Few Open Questions}
\label{sec:discussion}

\begin{itemize} 
	
         \item There has been a lot of recent progress in the field of the $S$-matrix bootstrap, see \cite{Kruczenski:2022lot} for a review. 
The general direction of extracting central charges from the $S$-matrix bootstrap was initiated in  \cite{Karateev:2019ymz}, see also \cite{Karateev:2020axc}. The special case of 4d was studied in \cite{Karateev:2022jdb}, see also \cite{Marucha:2023vrn} for further numerical explorations.\footnote{In a similar way one can study the chiral anomaly. For the recent applications in large $N_c$ QCD see \cite{Albert:2023jtd,Ma:2023vgc}.}  The idea of \cite{Karateev:2022jdb} is to consider a set of scattering amplitudes which consists not only of physical particles, but also of background dilaton fields. The latter carry information about $a_\text{UV}$. This setup for instance allowed to put lower bounds on $a_\text{UV}$ given some assumptions about the IR spectrum of particles and their interactions.
 The results of our present paper would allow to extend~\cite{Karateev:2022jdb} by adding the graviton particle into the $S$-matrix setup.

	\item In~\cite{Niarchos:2020nxk, Andriolo:2022lcb} the question of type-B trace anomaly matching in some supersymmetric theories was discussed. It would be nice to reconsider those examples in light of our determination of the universal amplitudes that probe the anomalies.
	
	\item It would be nice to test our general predictions for the universal amplitudes in interacting theories. For a class of weakly coupled theories this will appear in  \cite{Karateev}. 
	
	\item The $a$- and $c$-anomalies are matched because they are $c$-number violations of Weyl invariance -- i.e. they cannot depend on expectation values of local operators or on the state of the system. We have discussed how these anomalies are matched in Lorentz invariant vacua of the theory (which may or may not be gapped). The question of how to match these anomalies in other states of the system is far less explored. For the thermal state see~\cite{Eling:2013bj, Benjamin:2023qsc}.
	
	\item The $a$-anomaly cannot depend on exactly marginal coupling constants due to the Wess-Zumino consistency condition but the $c$-anomaly may well depend on such couplings~\cite{Nakayama:2017oye}. This could arise in some large $N$ non-supersymmetric fixed points with exactly marginal couplings, such as~\cite{Chaudhuri:2020xxb} (however, in this model the $c$-anomaly does not appear to depend on the exactly marginal couplings~\cite{Nakayama:2021fgy}). This would be nice to explore further. A more general class of trace anomalies associated to exactly marginal couplings are the Zamolodchikov anomalies~\cite{Osborn:1991gm,Friedan:2012hi,Gomis:2015yaa} that again need to be matched. It would be interesting to explore these further.
	
	\item In renormalization group flows between $\mathcal{N}=4$ supersymmetric fixed points, since $a=c$, we will find that our helicity flipping amplitude vanishes at small $t$. In the AdS dual, this vanishing of the graviton-dilaton helicity-violating scattering should be understood from the coupling of the $D3$-brane to gravity, and with higher derivative terms, it should be possible to activate this amplitude. We leave this to the future.
	
	\item The monotonicity properties of the renormalization group remain elusive in six dimensions~\cite{Elvang:2012st,Heckman:2015axa,Cordova:2015fha,Stergiou:2016uqq,Osborn:2015rna,Grinstein:2014xba}. It would be interesting to throw the graviton into the mix and try to understand the space of amplitudes (four- and six-point functions). Similarly, coupling the dilaton to point-like and extended impurities in space (i.e. line defect in space-time or surface operators and higher dimensional defects in space-time) has led to new results on renormalization group flows on defects~\cite{Jensen:2015swa,Wang:2020xkc,Wang:2021mdq,Cuomo:2021rkm} and it would be interesting to understand what other correlators are protected by the defect monotonic function.
		
\end{itemize}

\section*{Acknowledgements}
We thank Kelian H\"aring, Petr Kravchuk, Marco Meineri, Sasha Monin, Hugh Osborn, Riccardo Rattazzi, Francesco Riva, Slava Rychkov, Adam Schwimmer and Andy Stergiou for very useful conversations. BS also thanks the organizers and participants of the ``Bootstrap 2023'' workshop for fruitful discussions where a part of this work has been presented. We also thank Gregory Gabadadze, Tom Hartman, Grégoire Mathys, Hugh Osborn, Adam Schwimmer, Andy Stergiou and Alexander Zhiboedov for useful comments on the draft.

DK is supported by the SNSF Ambizione grant PZ00P2\_193411. 
ZK is supported in part by the Simons Foundation grant 488657 (Simons Collaboration on the Non-Perturbative Bootstrap), the BSF grant no. 2018204 and NSF award number 2310283.
The work of JP and BS is supported by the Simons Foundation grant 488649 (Simons Collaboration on the Nonperturbative Bootstrap) and by the Swiss National Science Foundation through the project 200020\_197160 and through the National Centre of Competence in Research SwissMAP. BS is also supported by STFC grant number ST/X000753/1.

\appendix
\section{Review of Weyl Anomalies}
\label{sec:Weyl_anomaly}
\subsection{Conventions}The metric is denoted by $g_{\mu\nu}(x)$. The symbol $g^{\mu\nu}$ defines the inverse metric, which is defined through the following relation
\begin{equation}
	g^{\mu\nu} g_{\nu\rho}\equiv \delta^\mu_\rho.
\end{equation}
The object $\sqrt{-g}$ is defined as
\begin{equation}
	\sqrt{-g} \equiv \sqrt{-\det(g_{\mu\nu})}.
\end{equation}

The Christoffel symbol $\Gamma^\mu_{\nu\rho}$ is defined as
\begin{equation}
	\Gamma^{\mu_1}_{\mu_2\mu_3} \equiv \frac{1}{2} g^{\mu_1\mu_4}\left(
	\partial_{\mu_3}g_{\mu_2\mu_4}+\partial_{\mu_2}g_{\mu_3\mu_4}-\partial_{\mu_4}g_{\mu_2\mu_3}
	\right).
\end{equation}
The Riemann tensor is defined as
\begin{equation}
	R^{\mu_1}{}_{\mu_2\mu_3\mu_4} \equiv \partial_{\mu_3}\Gamma^{\mu_1}_{\mu_2\mu_4}  - \partial_{\mu_4}\Gamma^{\mu_1}_{\mu_2\mu_3} 
	+\Gamma^{\mu_1}_{\mu_3\mu_5} \Gamma^{\mu_5}_{\mu_2\mu_4} -
	\Gamma^{\mu_1}_{\mu_4\mu_5} \Gamma^{\mu_5}_{\mu_2\mu_3}.
\end{equation}
The Ricci tensor is defined as
\begin{equation}
	R_{\mu_2\mu_4}\equiv R^{\mu_1}{}_{\mu_2\mu_1\mu_4}.
\end{equation}
The Ricci scalar is defined as
\begin{equation}
	R\equiv g^{\mu_1\mu_2} R_{\mu_1\mu_2}.
\end{equation}
The squared Weyl tensor in 4d can be written as
\begin{equation}
	\cW^2 = R^{\mu_1\mu_2\mu_3\mu_4}R_{\mu_1\mu_2\mu_3\mu_4}-2R^{\mu_1\mu_2}R_{\mu_1\mu_2}+\f{1}{3}R^2.
\end{equation}
The Euler density in 4d is given by 
\begin{equation}
	E_4 = R^{\mu_1\mu_2\mu_3\mu_4}R_{\mu_1\mu_2\mu_3\mu_4}-4R^{\mu_1\mu_2}R_{\mu_1\mu_2}+R^2.
\end{equation}
Let also write explicitly the definition of $\square\tau$ object in curved space-time. It reads as
\begin{equation}
	\square\tau \equiv (g^{\mu\nu}\partial_\nu - g^{\nu_1\nu_2}\Gamma^\mu_{\nu_1\nu_2}) \partial_\mu \tau(x).
\end{equation}

In the $\kappa$ expansion we can write the following main quantities
\begin{align}\nn
	\sqrt{-\text{det}\ g}&=1+\kappa h+\f{\kappa^2}{2}(h^2 -2h^{\mu\nu}h_{\mu\nu})+\f{\kappa^3}{6}(h^3-6hh^{\mu\nu}h_{\mu\nu}+8h^{\mu\rho} h_{\rho\nu} h^\nu_{\ \mu})+O(\kappa^4),\\
	g^{\mu\nu}&=\eta^{\mu\nu}-2\kappa h^{\mu\nu}+4\kappa^2 h^{\mu\rho}h_\rho^{\ \nu} -8\kappa^3 h^{\mu\lambda}h_{\lambda\rho}h^{\rho\nu}+O(\kappa^4),\\
	\Gamma^{\lambda}_{\mu\nu}&= (\kappa\eta^{\lambda\sigma}-2\kappa^2 h^{\lambda\sigma}+4\kappa^3 h^{\lambda\rho}h_\rho^{\ \sigma})\big( \p_\mu h_{\nu\sigma}+\p_\nu h_{\sigma\mu}-\p_\sigma h_{\mu\nu}\big)+\ O(\kappa^4),
	\nn
\end{align}
where $h=\eta^{\mu\nu} h_{\mu\nu}$.
For deriving interaction vertices involving fermions in section \ref{S:free_fermion} we also need to the provide the expansion of vielbeins and spin connection
\be
e_\mu^{\ a} &=&\delta_\mu^a +\kappa h_\mu^{\ a}-\f{\kappa^2}{2}h_\rho^{\ a}h_\mu^{\ \rho}+\f{\kappa^3}{2}h_\rho^{\ a}h_\sigma^{\ \rho}h_\mu^{\ \sigma}+\ O(\kappa^4),\nn\\
E_a^{\ \mu} &=&\delta_a^\mu -\kappa h_a^{\ \mu}+\f{3}{2}\kappa^2 h_a^{\ \rho}h_\rho^{\ \mu}-\f{5}{2}\kappa^3 h_a^{\ \rho}h_{\rho\sigma}h^{\sigma\mu}+\ O(\kappa^4),\nn\\
\omega_\mu^{ab}
&=& \kappa (\p^b h_\mu^{\ a}-\p^a h_\mu^{\ b})+\kappa^2\Big( \f{1}{2}h^{b\sigma}\p_\mu h_\sigma^{\ a}-\f{1}{2}h^{a\sigma}\p_\mu h_\sigma^{\ b}+h^{b\sigma}\p^a h_{\mu\sigma}-h^{a\sigma}\p^b h_{\mu\sigma}\nn\\
&&\ +h^{a\sigma}\p_\sigma h_\mu^{\ b}-h^{b\sigma}\p_\sigma h_\mu^{\ a}\Big)+ \ O(\kappa^3).
\ee

\subsection{Weyl Anomalies}
Define the infinitesimal Weyl transformation of the connected functional of a conformal field theory as
\begin{equation}
	\label{eq:variation_functional}
	\delta_\sigma W_{\text{CFT}}[g_{\mu\nu}] \equiv W_{\text{CFT}}\left[e^{2\sigma}g_{\mu\nu}\right] - W_{\text{CFT}}[g_{\mu\nu}].
\end{equation}
The Weyl transformation \eqref{eq:variation_functional} forms an Abelian group. Thus, Weyl transformations must commute, namely
\begin{equation}
	\label{eq:WZ_CFT}
	\big[ \delta_{\sigma_1},\,\delta_{\sigma_2}\big] W_{\text{CFT}}[g_{\mu\nu}]= 0,
\end{equation}
known as the Wess-Zumino consistency condition. The most general solution has three parameters and is given by
\begin{equation}
	\label{eq:Weyl_anomaly_CFT}
	\delta_\sigma W_{\text{CFT}}[g_{\mu\nu}] = \int d^4 x\sqrt{-g}\ \sigma(x)\big(-a_\text{CFT}\, E_4(x)+c_\text{CFT}\,\cW^2(x)+b \square R\big),
\end{equation}
where $E_4(x)$ is the Euler density and $\cW^2(x)$ is the squared Weyl tensor. 
The constants $a_\text{CFT}$ and $c_\text{CFT}$ are called the $a$- and $c$-anomalies respectively. In a CFT, they appear as the OPE coefficients in the three-point function of the  stress-tensor. For a recent review see section 2.1 in \cite{Karateev:2022jdb}. The coefficient $b$ is scheme dependent and hence is not part of the universal data in the continuum. It is  possible to add the following counterterm
\be 
\f{1}{12}\int d^4x\sqrt{-g}R^2 \label{eq:curvature_counter_term}
\ee
to the original CFT action and remove the coefficient $b$
.\footnote{Under infinitesimal Weyl transformation, $\delta_\sigma \int d^4x\sqrt{-g}R^2=-12\int d^4x\sqrt{-g}(\square \sigma) R$.}

The $c$-anomaly is of ``type-B.'' It was first discovered in \cite{Capper:1974ic}, and the corresponding non-local connected functional was formulated in \cite{Deser:1976yx}. The development of the connected functional associated with the ``type-A'' $a$-anomaly was initiated in \cite{Riegert:1984kt} and eventually resolved in \cite{Deser:1993yx}.

\section{Interaction between the Dilaton and the IR CFT }
\label{DimTwo}
We address some important details about the   situation when there is a nontrivial infrared conformal field theory. The dilaton and metric would couple to these infrared degrees of freedom and it is important to understand the contribution to the various vertices we considered in the main text. The coupling to the metric and dilaton is constrained by requiring Weyl and diffeomorphism invariance. 
Weyl invariance can be made manifest by using the rescaled metric as before, 
$\widehat{g}_{\mu\nu}(x)\equiv e^{-2\tau(x)}g_{\mu\nu}(x)$ and, for primary operators, we rescale them according to their scaling dimension $\widehat{\mathcal{O}}^{\mu\nu\ldots}_{\Delta}(x)\equiv e^{\Delta \tau(x) }\mathcal{O}^{\mu\nu\ldots}_{\Delta}(x)$.

Another very important requirement comes from the fact that we have assumed that, in the absence of background fields, the renormalization group flow leads to the infrared CFT. Therefore, when we set the background fields to be trivial, we should not find relevant operators in flat space.
 
Consider first a scalar irrelevant operator and its simplest coupling to the background fields:
\begin{equation}
	\label{eq:A_interaction_irrelevant}
	A_\text{interaction}[\Omega,\, g_{\mu\nu}] =  \int d^4x \sqrt{-\widehat{g}}\ M^{4-\Delta}  \ \widehat{\mathcal{O}}_{\Delta}(x),
\end{equation}
This will contribute to $V_{\varphi\varphi\varphi\varphi}$ schematically as 
\begin{align}	
&\frac{M^{8-2\Delta}}{f^4}   \int d^4x \int d^4y   \varphi^2(x)  \varphi^2(x)  \ \langle \mathcal{O}_{\Delta}(x) \mathcal{O}_{\Delta}(y) \rangle_{\text{IR CFT}}
	\\
	&\qquad  \to
	\frac{M^{4}}{f^4}  \left[ \left(\frac{s}{M^2}\right)^{\Delta-2} +
	\left(\frac{t}{M^2}\right)^{\Delta-2} +
	\left(\frac{u}{M^2}\right)^{\Delta-2} \right], \nonumber
\end{align}
which is more suppressed than four powers of momentum for $\Delta>4$. Indeed any irrelevant operator dressed by the dilaton is suppressed by the scale $M$ of the RG flow and hence the corresponding amplitudes are subleading in the deep IR and, thus, can be ignored.\footnote{From dimensional arguments, the coupling of irrelevant operators has an IR scale dependence of $M^{-\alpha}$, with $\alpha > 0$. As a consequence, they contribute to our vertices and amplitudes at orders greater than four powers of momenta, as in the example above.} 
It can be quickly seen that CFT operators with spin $\ell\geq 1$ once appropriately contracted with the curvature tensors will only produce irrelevant terms. We conclude that the only terms which we need to worry about read as
\begin{equation}
	\label{eq:A_interaction_relevant}
	A_\text{interaction}[\Omega,\, g_{\mu\nu}] =\sum_{1\leq \Delta\leq 2} \int d^4x \sqrt{-\widehat{g}}\ M^{2-\Delta} R(\widehat{g})\ \widehat{\mathcal{O}}_{\Delta}(x),
\end{equation}
where $\cO$ are scalar primary operators whose scaling dimensions are restricted to the range $1 \leq\Delta\leq 2$. This expression has been originally found in \cite{Luty:2012ww}. Upon setting the metric to be trivial and the dilaton to be a constant, the terms in~\eqref{eq:A_interaction_relevant} all vanish.

Now we substitute $\tau(x)$ in terms of field $\varphi(x)$ and metric in terms of traceless and transverse graviton field $h_{\mu\nu}(x)$ from \eqref{eq:probes_hd} in the above expression. Up to the quadratic order in dilaton and/or graviton fields the action becomes
\be 
&&A_\text{interaction}[\Omega,\, g_{\mu\nu}]\nn\\
&=&\sum_{1\leq \Delta\leq 2} M^{2-\Delta}\int d^4x \Bigg(\f{3 \sqrt{2}}{f} \p^2\varphi +\f{3}{f^2}(\Delta -1)\varphi\p^2\varphi +\kappa^2 (4h^{\alpha\beta}\p^2 h_{\alpha\beta}\nn\\
&& +3\p_\mu h_{\alpha\beta}\p^{\mu}h^{\alpha\beta}-2\p^\beta h^{\mu\alpha} \p_\alpha h_{\mu\beta}) -\f{6\sqrt{2}\kappa}{f} h^{\mu\nu}\p_\mu\p_\nu\varphi +\ \cdots\Bigg)\mathcal{O}_{\Delta}(x). \label{eq:expanded_S}
\ee
Since the terms that involve only the dilaton  contain $\partial^2\varphi$, they do not contribute upon imposing $\partial^2\varphi=0$ as we do in the main body of the text. In particular, the vertices $V_{(\varphi\varphi \varphi)}$, $V_{(\varphi\varphi\varphi \varphi)}$ are not affected by the couplings of the background fields to the gapless infrared modes.
Furthermore,  for a similar reason, the vertex $V_{(hh \varphi)}$ is not affected by these couplings. To see that, remember that the metric also couples to the operators of the infrared CFT via $\kappa\int d^4x\ h_{\mu\nu}(x)T^{\mu\nu}(x)+O(\kappa^2)$, but since the two-point function of the energy momentum tensor and the scalar primaries vanishes, there is no contribution to the vertex $V_{(hh \varphi)}$. The conclusion from this discussion is that with the constraints on the background fields that we have imposed, there are no additional contributions from the infrared CFT to the vertices $V_{(\varphi\varphi \varphi)}$, $V_{(\varphi\varphi\varphi \varphi)}$, and $V_{(hh \varphi)}$.
Finally, it is important to discuss the vertex $V_{(hh \varphi\varphi)}$. Under the condition $\partial^2\varphi=0$ the only  contribution is proportional to 
\begin{align}
&\f{  \kappa^2}{f^2} \sum_{1\leq \Delta\leq 2} M^{4-2\Delta}\int d^4x \int d^4y \  h^{\mu\nu}(x)\p_\mu\p_\nu\varphi(x) h^{\rho\sigma}(y)\p_\rho\p_\sigma \varphi(y) 
\times \langle \mathcal{O}_{\Delta}(x)\mathcal{O}_{\Delta}(y)\rangle_{\text{IR CFT}}\nonumber \\
&\to
\f{  \kappa^2}{f^2} \sum_{1\leq \Delta\leq 2}\left[
 (k_3.\varepsilon_1.k_3) \left(\frac{t}{M^2}\right)^{\Delta-2} (k_4. \varepsilon_2 . k_4  )+
  (k_3. \varepsilon_2 .k_3) \left(\frac{u}{M^2}\right)^{\Delta-2} (k_4. \varepsilon_1.k_4)  \right]
\end{align}
where we use the notation of section \ref{sec:vertices}.
Comparing with \eqref{result_hhdd}, we conclude that each operator of dimension $\Delta$ gives the following contribution  
\begin{align}
g_6(s,t,u) &\to g_6(s,t,u) + i\frac{144 \kappa^2}{f^2} r_\Delta \left(\frac{u}{M^2}\right)^{\Delta-2} \\
g_8(s,t,u) &\to g_8(s,t,u) + i\frac{144 \kappa^2}{f^2} r_\Delta \left[ \left(\frac{t}{M^2}\right)^{\Delta-2} 
+\left(\frac{u}{M^2}\right)^{\Delta-2} \right]\\
g_9(s,t,u) &\to g_9(s,t,u) - i\frac{288 \kappa^2}{f^2}r_\Delta \left(\frac{t}{M^2}\right)^{\Delta-2} \,,
\end{align}
where $r_\Delta$ is a numerical constant that will not be important in what follows. 
In particular, for $\Delta=2$ this  contributes to the $V_{(hh \varphi\varphi)}$ amplitude in the same kinematical structure as the $r_1$ coefficient and it amounts to the shift $r_1 \to r_1 + r_\Delta$.\footnote{More precisely, when $\Delta=2$  the integrals are logarithmically divergent and one needs to introduce a regularization scheme (indeed $r_\Delta \propto \frac{1}{2-\Delta}$).
In practice, one should replace $r_{\Delta} \Big(\f{-p^2}{M^2}\Big)^{\Delta -2} \rightarrow \ln\Big(\f{-p^2}{\mu^2}\Big)$, where $\mu$ is a renormalization scale.
This is related with the coupling space anomaly discussed in section \ref{CouplSpac}.
}
Therefore, it does not affect the extraction of the $a$ and $c$ anomalies from this vertex.

We  also argue that the result of the 2-graviton-2-dilaton amplitude in \eqref{eq:amplitude_final} will not be affected by the presence of any low-dimension scalar operator at the IR fixed point. Note that in the computation of \eqref{eq:amplitude_final}, we used the graviton-dilaton-dilaton EFT vertex $V^{\mu\nu}_{(h\varphi \varphi)}$ in \eqref{eq:V_hphiphi_EFT_prediction} for off-shell dilatons, two dilaton vertex $V_{(\varphi\varphi)}$ for off-shell dilatons in \eqref{eq:dd_vertex}, three dilaton vertex $V_{(\varphi\varphi\varphi)}$ for two on-shell dilatons and one off-shell dilaton, and the vertex $V_{(hh \varphi\varphi)}$ for on-shell dilatons, as evident from the set of diagrams below \eqref{eq:leading_in_f}. 
Above we discussed the effect of an operator of dimension $\Delta<2$ at the IR fixed point, on the on-shell vertex $V_{(hh \varphi\varphi)}$.
The off-shell vertex  $V_{(\varphi\varphi)}$ will also be affected. Namely,  equation \eqref{eq:dd_vertex} will change via $r_1 \to r_1+r_\Delta (-p^2/M^2)^{\Delta-2}$.  The off-shell vertex $V_{(\varphi\varphi\varphi)}$ can receive contribution both from $\langle \mathcal{O}_{\Delta}\mathcal{O}_{\Delta}\rangle$ and $\langle \mathcal{O}_{\Delta}\mathcal{O}_{\Delta}\mathcal{O}_{\Delta}\rangle$, but both the contributions vanish when we set any two of the dilatons on-shell.
The vertex $V^{\mu\nu}_{(h\varphi \varphi)}$ for off-shell dilatons will receive contributions from both $\langle \mathcal{O}_{\Delta}\mathcal{O}_{\Delta}\rangle$ and $\langle T_{\mu\nu}\mathcal{O}_{\Delta}\mathcal{O}_{\Delta}\rangle$. 
The contribution of $\langle T_{\mu\nu}\mathcal{O}_{\Delta}\mathcal{O}_{\Delta}\rangle$ will affect \eqref{eq:V_hphiphi_EFT_prediction} in such a way that it will be non-vanishing only when both dilatons are off-shell. However, as observed from the set of Feynman diagrams below \eqref{eq:leading_in_f}, the only contribution needed from the vertex $V^{\mu\nu}_{(h\varphi \varphi)}$ is when one or both dilatons are on-shell. Therefore, we can ignore the contribution from $\langle T_{\mu\nu}\mathcal{O}_{\Delta}\mathcal{O}_{\Delta}\rangle$.
The two-point function $\langle \mathcal{O}_{\Delta}\mathcal{O}_{\Delta}\rangle$ will lead to an extra contribution of the form
\begin{align}
-i\frac{72 \kappa}{f^2} r_\Delta\left[(k_2.\varepsilon_1.k_2) k_3^2 \left(\frac{-k_3^2}{M^2}\right)^{\Delta-2} +(k_3.\varepsilon_1.k_3) k_2^2  \left(\frac{-k_2^2}{M^2}\right)^{\Delta-2} \right]
\end{align}
to $(\varepsilon_1)_{\mu\nu} V^{\mu\nu}_{(h\varphi \varphi)}$ given in \eqref{eq:V_hphiphi_EFT_prediction}.  
Here we assumed that the graviton is on-shell because this is sufficient to compute the amplitude \eqref{eq:amplitude}. Notice that the only diagram with an off-shell (internal) graviton contributing to the amplitude  \eqref{eq:amplitude} involves 
$V_{(h\varphi \varphi)}$ with two on-shell dilatons. It is easy to see that \eqref{eq:expanded_S} cannot contribute in that case.

We now compute the contribution to the amplitude by summing up the diagrams   below \eqref{eq:leading_in_f}. In order to do that, we first write down the extra contribution from a low dimension scalar operator to the relevant vertices in the conventions used in section \ref{sec:scattering_and_bounds}: 
\be
V_{(\varphi\varphi)}(p,-p)&=& \f{i}{f^2}\times 36r_\Delta (p^2)^2 \left(\f{-p^2}{M^2}\right)^{\Delta-2},\\
V^{\mu\nu}_{(h\varphi\varphi)}(k_1,k_2,k_3)&=& -i\frac{72 \kappa}{f^2} r_\Delta\left[k_2^\mu k_2^\nu\  k_3^2 \left(\frac{-k_3^2}{M^2}\right)^{\Delta-2} +k_3^\mu k_3^\nu \ k_2^2  \left(\frac{-k_2^2}{M^2}\right)^{\Delta-2} \right],
\ee 
\be
V_{(h\varphi h\varphi)}(k_1,k_2,-k_3,-k_4;\varepsilon_1,\varepsilon_3)
 &=&i\frac{144 \kappa^2}{f^2} r_\Delta
 (k_2.\varepsilon_1.k_2)(k_4. \varepsilon_3 . k_4  )  \left(\frac{s}{M^2}\right)^{\Delta-2}  \nn \\
 &&+i\frac{144 \kappa^2}{f^2} r_\Delta
  (k_4. \varepsilon_1 .k_4) (k_2. \varepsilon_3.k_2)   \left(\frac{u}{M^2}\right)^{\Delta-2} 
\ee
The contribution of each diagram below \eqref{eq:leading_in_f} reads
\be
i\mathcal{T}_1&=& -\f{144i\kappa^2}{f^2} r_\Delta (k_4.\ve_3.k_4) (k_2.\ve_1.k_2)\left(\frac{s}{M^2}\right)^{\Delta-2},\nn\\
&=&i\mathcal{T}_2=-i\mathcal{T}_3,\\
i\mathcal{T}_4&=& -\f{144i\kappa^2}{f^2} r_\Delta (k_2.\ve_3.k_2) (k_4.\ve_1.k_4)\left(\frac{u}{M^2}\right)^{\Delta-2},\nn\\
&=&i\mathcal{T}_5=-i\mathcal{T}_6,\\
i\mathcal{T}_7&=&0,\\
i\mathcal{T}_8&=&V_{(h\varphi h\varphi)}(k_1,k_2,-k_3,-k_4;\varepsilon_1,\varepsilon_3),\\
i\mathcal{T}_9&=&i\mathcal{T}_{10}=0.
\ee
Therefore, the contribution to the amplitude vanishes because
\be
\sum_{I=1}^{10}\mathcal{T}_I=0.
\ee

\section{Graviton-Dilaton Scattering in the Center of Mass}
\label{app:graviton_dilaton_scattering}

In this appendix we study the graviton-dilaton scattering amplitude
\begin{equation}
	\label{eq:amplitudes}
	h(p_1,\lambda_1) \varphi(p_2) \longrightarrow h(p_3,\lambda_3) \varphi(p_4)
\end{equation}
in the center of mass frame. Here we denote the gravitons by $h$ and the dilatons  by $\varphi$. The four-momenta are denoted by $p_i$ and the helitcities of the gravitons are denoted by $\lambda_i=\pm 2$. In this appendix we apply the formalism of \cite{Hebbar:2020ukp} in order to study the process \eqref{eq:amplitudes}.\footnote{The problem of studying scattering amplitudes of particles with spin was addressed in the 60s by many authors \cite{Jacob:1959at,  Trueman:1964zzb, Hara:1970gc, Hara:1971kj}, see the older review \cite{Martin:102663} for a more comprehensive list. See also \cite{Bellazzini:2016xrt, deRham:2017zjm} for more recent discussions.}  

Both the graviton and the dilaton are massless. 
The centre of mass frame is defined as
\begin{equation}
	\label{eq:COM_frame}
	\begin{aligned}
		&p_1^{\text{com}}\equiv(\myP,0,0,+\myP),\\
		&p_2^{\text{com}}\equiv(\myP,0,0,-\myP),\\
		&p_3^{\text{com}}\equiv(\myP,+\myP\sin \theta ,0,+\myP\cos \theta),\\
		&p_4^{\text{com}}\equiv(\myP,-\myP\sin \theta ,0,-\myP\cos \theta).
	\end{aligned}
\end{equation}
Here $\theta$ is the scattering angle on the $x-z$ plane.
The sine and cosine of the scattering angle are related to the Mandelstam variables as
\begin{equation}
	\label{eq:t_u_to_scattering_angle}
	\myP= \frac{\sqrt{s}}{2},\qquad
	\sin\theta =\frac{2\sqrt{tu}}{s},\qquad
	\cos\theta =\frac{t-u}{s}.
\end{equation}
These can be inverted and lead to
\begin{equation}
	t = - \frac{s}{2}(1-\cos\theta),\qquad
	u = - \frac{s}{2}(1+\cos\theta).
\end{equation}
The Mandelstam variables are related as
\begin{equation}
	s+t+u=0.
\end{equation}

\subsection{Scattering and Partial Amplitudes}
\label{app:amplitudes}
The scattering process \eqref{eq:amplitudes} is described by the scattering amplitude
\begin{equation}
	\label{eq:amplitude_app}
	(2\pi)^2\delta^{(4)}(p_1+p_2-p_3-p_4)\times
	\mathcal{S}{}_{\lambda_1}^{\lambda_3}(p_1,p_2,p_3,p_4)\equiv {}_\text{out}\<h(p_3,\lambda_3) \varphi(p_4)|h(p_1,\lambda_1) \varphi(p_2)\>_\text{in},
\end{equation}
where the helicity placed downstairs is associated with the incoming graviton and the helicity placed upstairs is associated with  the outgoing graviton. The amplitude \eqref{eq:amplitude_app} can be split into the trivial and the interacting parts as
\begin{multline}
	\label{eq:decomposition}
	\mathcal{S}{}_{\lambda_1}^{\lambda_3}(p_1,p_2,p_3,p_4) = \left[\frac{2p_1^0(2\pi)^3\delta^{(3)}(\vec p_1-\vec p_3) 2p_2^0(2\pi)^3\delta^{(3)}(\vec p_2-\vec p_4)}{(2\pi)^2\delta^{(4)}(p_1+p_2-p_3-p_4)} \right]\delta_{\lambda_1}^{\lambda_3}\\
	+ i \mathcal{T}{}_{\lambda_1}^{\lambda_3}(p_1,p_2,p_3,p_4).
\end{multline}

The centre of mass amplitudes are defined as
\begin{equation}
	\mathcal{T}{}_{\lambda_1}^{\lambda_3}(s,t,u) \equiv \mathcal{T}{}_{\lambda_1}^{\lambda_3}(p^{\text{com}}_1,p^{\text{com}}_2,p^{\text{com}}_3,p^{\text{com}}_4).
\end{equation}
We assume that parity is a symmetry of our theory, then the following relations hold
\begin{equation}
	\label{eq:parity}
	\mathcal{T}{}_{+2}^{+2}(s,t,u) = \mathcal{T}{}_{-2}^{-2}(s,t,u),\qquad
	\mathcal{T}{}_{+2}^{-2}(s,t,u) = \mathcal{T}{}_{-2}^{+2}(s,t,u).
\end{equation}
It  follows from equation (2.64) in \cite{Hebbar:2020ukp}.
As a result there are only two independent amplitudes.
Crossing for massless particles, see equation (2.82) in \cite{Hebbar:2020ukp}, simply implies that both of these amplitude are $s-u$ symmetric, namely
\begin{equation}
	\mathcal{T}{}_{+2}^{+2}(s,t,u) = \mathcal{T}{}_{+2}^{+2}(u,t,s),\qquad
	\mathcal{T}{}_{+2}^{-2}(s,t,u) = \mathcal{T}{}_{+2}^{-2}(u,t,s).
\end{equation}
This is no longer the case if one of the particles is massive. Then, there is mixing between the two amplitudes.

Let us now consider the two-particle state projected to the definite angular momentum
\begin{align}
	|p, \ell, \lambda_1 \>_\text{in} &\equiv \left( |h(p_1,\lambda_1) \varphi(p_2)\>_\text{in}  \right)_\text{projected to $\ell$},\\
	{}_\text{out}\<p', \ell, \lambda_3 | &\equiv \left( {}_\text{out}\<h(p_3,\lambda_3) \varphi(p_4) |  \right)_\text{projected to $\ell$}.
\end{align}
Here $p=p_1+p_2$ and $p'=p_3+p_4$. When both states are in or out projected states they are normalized as
\begin{equation}
	\<p', \ell', \lambda' |p, \ell, \lambda \> = (2\pi)^4\delta^{(4)}(p-p') \delta_{\ell \ell'} \delta_{\lambda \lambda'}.
\end{equation}
The partial amplitudes of the process \eqref{eq:amplitudes} is defined as
\begin{equation}
	[\mathcal{S}_\ell]{}_{\lambda_1}^{\lambda_3}(s) \equiv {}\text{out}\<p', \ell, \lambda_3 | p, \ell, \lambda_1 \>_\text{in}.
\end{equation}
The above definition leads to the following expression
\begin{equation}
	\label{eq:decomposition_partial_amplitudes}
	[\mathcal{S}_\ell]{}_{\lambda_1}^{\lambda_3}(s) = \delta_{\lambda_1}^{\lambda_3} + i[\mathcal{T}_\ell]{}_{\lambda_1}^{\lambda_3}(s),
\end{equation}
where $\ell=2,3,4,\ldots$ is the angular momentum and
\begin{equation}
	\label{eq:projection}
	[\mathcal{T}_\ell]{}_{\lambda_1}^{\lambda_3}(s) = \frac{1}{8\pi} \int_0^\pi d\theta \sin\theta\, d^{(\ell)}_{\lambda_1\lambda_3}(\theta) \mathcal{T}{}_{\lambda_1}^{\lambda_3}\left(s,\,t=-\frac{s}{2}(1-\cos \theta),  \,u=-\frac{s}{2}(1+\cos \theta)\right).
\end{equation}
For details see equations (2.106) and (2.108) in \cite{Hebbar:2020ukp}.
Here $d^{(\ell)}_{\lambda_1\lambda_3}(\theta)$ is the Wigner small $d$-matrix. Due to parity we also have the following relation
\begin{equation}
	[\mathcal{S}_\ell]{}_{\lambda_1}^{\lambda_3}(s) = [\mathcal{S}_\ell]{}_{-\lambda_1}^{-\lambda_3}(s).
\end{equation}

Using the orthogonality of the Wigner's small $d$-matrix we can invert the relation \eqref{eq:projection}. The result reads as
\begin{equation}
	\label{eq:inverted_relation}
	\mathcal{T}{}_{\lambda_1}^{\lambda_3}(s,t,u)=4\pi\sum_{\ell\geq 2}  (2\ell+1)  [\mathcal{T}_\ell]{}_{\lambda_1}^{\lambda_3}(s)d^{(\ell)}_{\lambda_1\lambda_3}(\theta).
\end{equation}
Let us evaluate this relation in the forward limit $\theta=0$. Using the following properties of the Wigner's small $d$-matrix
\begin{equation}
	\ell\geq 2:\qquad d^{(\ell)}_{+2+2}(0)=1,\qquad d^{(\ell)}_{+2-2}(0)=0,
\end{equation}
we conclude that
\begin{equation}
	\label{eq:forward_limit}
	\begin{aligned}
		\mathcal{T}{}_{+2}^{+2}(s,0,-s) &= 4\pi\sum_{\ell\geq 2}  (2\ell+1)  [\mathcal{T}_\ell]{}_{+2}^{+2}(s),\\
		\mathcal{T}{}_{+2}^{-2}(s,0,-s) &=0.
	\end{aligned}
\end{equation}
In writing these relations we only assumed that the forward limit for the amplitudes exist. This type of arguments was also used in appendix A.4 in \cite{Haring:2022sdp} for providing an alternative derivation of unitarity constraints in the forward limit. This is consistent with our main result, which shows that the helicity flipping amplitude vanishes in the forward limit~\eqref{newRintro}.

An alternative derivation of the second equation in \eqref{eq:forward_limit} is done by applying the logic of section 2.7 in \cite{Hebbar:2020ukp} for the forward $t=0$ limit. In the forward limit there is an enhanced $SO(2)$ symmetry around the $z$-axis. Invariance under this symmetry forces the amplitude to vanish in the forward limit unless $\lambda_1=\lambda_3$. For details see equation (2.140) in \cite{Hebbar:2020ukp} and the discussion around it.

To conclude let us derive two useful relations for later. Consider expression \eqref{eq:inverted_relation} with $\lambda_1=+2$ and $\lambda_3=-2$. Then apply one and two derivates in $t$ and evaluate the result in the forward limit. Using the explicit expression of the Wigner $d$-matrix we can write
\begin{equation}
	\label{eq:derivatives_forward_limit}
	\begin{aligned}
	\partial_t\mathcal{T}{}_{+2}^{-2}(s,0,-s) &=0,\\
	\partial_t^2\mathcal{T}{}_{+2}^{-2}(s,0,-s) &= \frac{\pi}{3s^2} \sum_{\ell\geq 2}  
	(2\ell+1)(\ell-1) \ell (\ell+1)(\ell+2)  [\mathcal{T}_\ell]{}_{+2}^{-2}(s).
	\end{aligned}
\end{equation}

\subsection{Graviton Polarizations}
\label{app:graviton_polarizations}
The graviton polarization can simply be written as a product of the photon polarization
\begin{equation}
	\varepsilon_\lambda^{\mu\nu}(p) = \varepsilon_{\lambda/2}^{\mu}(k_i) \varepsilon_{\lambda/2}^{\nu}(k_i),
\end{equation}
where the explicit expressions for $\varepsilon_\lambda^{\mu}(p)$ can be found for example in equation (B.34) in \cite{Haring:2022sdp}. For incoming photons we have
\begin{equation}
	\varepsilon^\mu_{\lambda=\pm 1} (p) = 
	\frac{e^{i \lambda \phi } }{\sqrt{2}}
	\begin{pmatrix}
		0\\
		\cos \theta \cos \phi -i \lambda \sin \phi\\
		\cos \theta \sin\phi  +i \lambda \cos\phi\\
		-\sin \theta 
	\end{pmatrix},
\end{equation}
where $\theta$ and $\phi$ are the spherical angles of the photon incoming 3-momentum $\vec p$. The outgoing polarizations are simply obtained by complex conjugation, namely
\begin{equation}
	\varepsilon^\mu_{\lambda} \left(p\big|_\text{outgoing}\right) = 
	\left(\varepsilon^\mu_{\lambda} \left(p\big|_\text{incoming}\right)\right)^*.
\end{equation}

Suppose the most generic situation of scattering of four photons.
By convention we choose $\phi_i=0$ for all four particles. Instead $\theta_1=\theta_2=0$ for the incoming particles and $\theta_3=\theta$ and $\theta_4=\pi+\theta$. Taking complex conjugate for the outgoing particles we conclude that in the center of mass the photon polarizations reads as
\begin{equation}
	\varepsilon^\mu_{\lambda=\pm 1} (p_1^\text{com}) = 
	\frac{1}{\sqrt{2}}
	\begin{pmatrix}
		0\\
		1\\
		i \lambda\\
		0 
	\end{pmatrix},\qquad
	\varepsilon^\mu_{\lambda=\pm 1} (p_2^\text{com}) = 
\frac{1}{\sqrt{2}}
\begin{pmatrix}
	0\\
	-1\\
	i \lambda\\
	0
\end{pmatrix},
\end{equation}
\begin{equation}
	\varepsilon^\mu_{\lambda=\pm 1} (p_3^\text{com}) = 
	\frac{1}{\sqrt{2}}
	\begin{pmatrix}
		0\\
		\cos \theta\\
		-i \lambda\\
		-\sin \theta 
	\end{pmatrix},\qquad
	\varepsilon^\mu_{\lambda=\pm 1} (p_4^\text{com}) = 
	\frac{1}{\sqrt{2}}
	\begin{pmatrix}
		0\\
		-\cos \theta\\
		-i \lambda\\
		\sin \theta 
	\end{pmatrix},
\end{equation}

\subsection{Unitarity and Positivity Constraints}
\label{app:unitarity}
It is convenient to define the following parity even and parity odd eigenstates
\begin{align}
	|\text{parity }+\> \equiv \frac{1}{\sqrt{2}}\left(|p, \ell, \lambda=+2 \> + (-1)^\ell |p, \ell, \lambda=-2 \>\right),\\
	|\text{parity }-\> \equiv \frac{1}{\sqrt{2}}\left(|p, \ell, \lambda=+2 \> - (-1)^\ell |p, \ell, \lambda=-2 \>\right).
\end{align}
We also define the following short-hand notation
\begin{equation}
	\label{eq:short-hand}
	\begin{aligned}
		\Phi_\ell(s) \equiv [\mathcal{T}_\ell]{}_{+2}^{+2}(s)+(-1)^\ell[\mathcal{T}_\ell]{}_{+2}^{-2}(s),\\
		\chi_\ell(s) \equiv [\mathcal{T}_\ell]{}_{+2}^{+2}(s)-(-1)^\ell[\mathcal{T}_\ell]{}_{+2}^{-2}(s).
	\end{aligned}
\end{equation}

Let us focus on the parity plus eigenstates.
We can then form the following matrix
\begin{center}
	\begin{tabular}{c|cc}
		& $|\psi \>_\text{in}$ & $|\psi \>_\text{out}$ \\
		\hline
		${}_\text{in}\<\psi|$  & $1$ & $1-i\Phi_\ell^*{}(s)$ \\
		${}_\text{out}\<\psi|$ & $1+i\Phi_\ell(s)$ & 1\\
		\hline
	\end{tabular}
\end{center}
Since our theory is unitarity the above matrix is positive semidefinite. Plugging inside the decomposition \eqref{eq:decomposition_partial_amplitudes} we get
\begin{equation}\forall \ell=2,3,4,\ldots\qquad s\geq 0:\qquad
	\begin{pmatrix}
		1 & 1-i\Phi_\ell^*{}(s)\\
		1 +  i\Phi_\ell(s)& 1
	\end{pmatrix}
	\succeq 0.
\end{equation}
An analogous condition holds for the $\Phi_-(s)$ partial amplitude.
These in turn can be rewritten as the following conditions
\begin{equation}\forall \ell=2,3,4,\ldots\qquad s\geq 0:\qquad
	\left| 1 +  i\Phi_\ell(s) \right|^2 \leq 1.
\end{equation}
From this we can read of the positivity condition
\begin{equation}\forall \ell=2,3,4,\ldots\qquad s\geq 0:\qquad
	\text{Im}\Phi_\ell(s) \geq 0.
\end{equation}
Or using \eqref{eq:short-hand} this condition can be equivalently rewritten
\begin{equation}\forall \ell=2,3,4,\ldots\qquad s\geq 0:\qquad
	\text{Im}[\mathcal{T}_\ell]{}_{+2}^{+2}(s)+(-1)^\ell\text{Im}[\mathcal{T}_\ell]{}_{+2}^{-2}(s) \geq 0.
\end{equation}
An analogous condition can be derived using parity odd eigenstates. We have  
\begin{equation}\forall \ell=2,3,4,\ldots\qquad s\geq 0:\qquad
	\text{Im}[\mathcal{T}_\ell]{}_{+2}^{+2}(s)-(-1)^\ell\text{Im}[\mathcal{T}_\ell]{}_{+2}^{-2}(s) \geq 0.
\end{equation}
From the above two conditions we can conclude that
\begin{equation}\forall \ell=2,3,4,\ldots\qquad s\geq 0:\qquad
	\text{Im}[\mathcal{T}_\ell]{}_{+2}^{+2}(s)\geq 0,
\end{equation}
whereas $\text{Im}[\mathcal{T}_\ell]{}_{+2}^{-2}(s)$ does not have a definite sign and it is bounded by 
$\left| \text{Im}[\mathcal{T}_\ell]{}_{+2}^{-2}(s)\right| \le \text{Im}[\mathcal{T}_\ell]{}_{+2}^{+2}(s)$.

\subsection{Dispersion Relation}
\label{app:dispersion_relation}
Let us consider the centre of mass scattering amplitude $\mathcal{T}{}_{+2}^{-2}(s,t,u)$ in the complex $s$ plane for some fixed physical value of $t$. Assuming that this amplitude has a polynomial expansion at low energy we can write the following expression
\begin{equation}
	\label{eq:low_energy_expansion}
	\mathcal{T}{}_{+2}^{-2}(s,t,u)\Big|_\text{low energy} = g_0 + g_1 t + g_2(s^2+u^2) +g_2' t^2
	+g_3(s^3+u^3)+g_3't(s^2+u^2)+O(s^4).
\end{equation}
Due to equation \eqref{eq:forward_limit} we immediately conclude that
\begin{equation}
	\label{eq:EFT_constraints}
	g_0=0,\qquad g_2=0, \qquad g_3=0.
\end{equation}

\begin{figure}
	\centering 
	\begin{tikzpicture}
		
		\draw[->] (-5.2,0)-- (5.2,0);
		\draw[->] (0,-1)--(0,2.8);

		\draw (4,2.5)--(4.5,2.5);
		\draw (4,2.5)--(4,3);
		\draw (4.3,2.8) node{$s$};
		
		\draw[red, decoration = {zigzag,segment length = 2mm, amplitude = 1mm}, decorate] (1,0) -- (5,0);
		\draw[red] (1,0) node{$\bullet$};
		\draw (1, -0.3) node{$4m^2$};
		\draw[red, decoration = {zigzag,segment length = 2mm, amplitude = 1mm}, decorate] (-5,0) -- (-0.5,0);
		\draw[red] (-0.5,0) node{$\bullet$};
		\draw (-0.8, -0.3) node{$-t-4m^2$};

	\end{tikzpicture}
	\caption{The analytic structure of the amplitude $\mathcal{T}{}_{+2}^{-2}(s,t,u)$ on the $s$ complex plane for a fixed physical value of $t$. We draw the cuts associated with multi-particle states of a QFT with mass gap $m$. In general, there are also poles associated with asymptotic single particle states but we do not  draw those to avoid cluttering the figure.
	If the IR theory is gapless then the cuts start from $s=0$ and $s=-t$.}
	\label{fig:analytic_structure}
\end{figure}
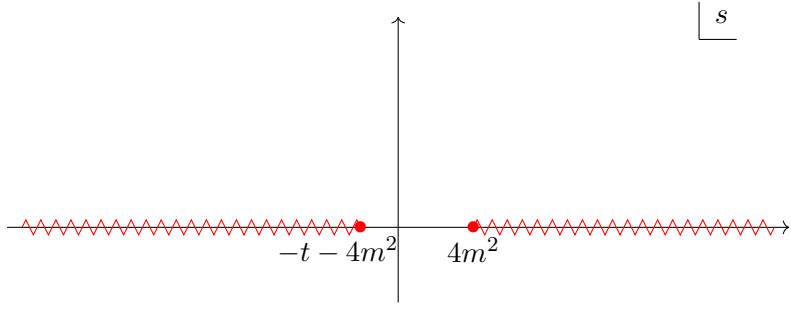

Let us consider a situation when the imaginary part of the amplitude $\mathcal{T}{}_{+2}^{-2}(s,t,u)$ is generated by the presence of some massive degrees of freedom. We denote the mass of the lowest state by $m$. We also assume that neither the dilaton nor the graviton   propagate in the loops. We conclude  that $\text{Im}\mathcal{T}{}_{+2}^{-2}(s,t,u)$ can be non zero for $s\geq m^2$. More precisely, $ \mathcal{T}{}_{+2}^{-2}(s,t,u)$ will have poles associated with single particle states and branch cuts associated with the multi-particle continuum.
The analytic structure of the amplitude $\mathcal{T}{}_{+2}^{-2}(s,t,u)$ is  presented in figure \ref{fig:analytic_structure}. 
The left cut appears at $u=4m^2$ due to the $s-u$ crossing symmetry.

An efficient way to derive dispersion relations for massless particles is by using the technology of \cite{Bellazzini:2020cot}, see appendix F in \cite{Acanfora:2023axz} for its compact review. Similar formalisms were also proposed in \cite{Caron-Huot:2020cmc,Tolley:2020gtv}. (In the case of massive particles the dispersion relations were discussed in section 4 in \cite{Chen:2022nym}.) Consider a scattering amplitude $\Phi(s,t,u)$ which is $s-u$ invariant and obeys $\lim_{|s|\to \infty} \frac{1}{s^2}\Phi(s,t,u) =0$ for some fixed value of $t$.\footnote{This would follow from the Martin-Froissart bound \cite{Martin:1965jj,Jin:1964zza} for a QFT with a mass gap.
It is also expected in gravitational scattering as recently discussed in \cite{Haring:2022cyf}.
Neither of these results applies directly to our case because we have massless particles but the gravitons and dilatons are just probes and therefore the non-perturbative gravitational results of  \cite{Haring:2022cyf} do not apply. Tree-level gravitational interactions should obey the classical Regge growth \cite{Chowdhury:2019kaq} which is weaker, namely 
$\lim_{|s|\to \infty} \frac{1}{s^3}\Phi(s,t,u) =0$.
However, we expect the dilaton scattering amplitudes to be softer at high energies due to the coupling to the trace of the stress tensor that vanishes in the UV limit. 
}
 One can then write the following relation
\begin{equation}
	\label{eq:main_equation}
	\underset{s=0,-t}{\text{Res}}\left(\frac{\Phi(s,t,u)\Big|_\text{low energy}}{s^2(s+t)}\right)=\frac{1}{\pi} \int_{m^2}^\infty ds\left(\frac{1}{s}+\frac{1}{s+t}\right)\frac{\text{Im}\Phi(s,t,u)}{s(s+t)}.
\end{equation}

Let us now set $\Phi(s,t,u) = \mathcal{T}{}_{+2}^{-2}(s,t,u)$. Then using \eqref{eq:low_energy_expansion} we can evaluate the left-hand side in the above equation, we get
\begin{equation}
	\label{eq:residue}
	\underset{s=0,-t}{\text{Res}}\left(\frac{\Phi(s,t,u)\Big|_\text{low energy}}{s^2(s+t)}\right)=
	2g_2+(2g_3'-3g_3)t+\ldots
\end{equation}
We immediately see that $g_2'$ disappears from this relation. 
Plugging this result into \eqref{eq:main_equation} and taking the forward limit we conclude that
\begin{equation}
	g_2=\frac{1}{\pi} \int_{m^2}^\infty ds \frac{\text{Im}\Phi(s,0,-s)}{s^3}.
\end{equation}
The $t=0$ limit is taken in order to remove all the other terms in \eqref{eq:residue} hidden by ellipses.
The left-hand side of this relation is zero due to \eqref{eq:EFT_constraints}. The right-hand side is also zero due to equation \eqref{eq:forward_limit}.
A non-trivial dispersion relation at this level can be derived for the $g_3'$ coefficient.
Taking one derivative in $t$ of both sides in \eqref{eq:main_equation} and using \eqref{eq:residue} we get
\begin{equation}
	2g_3'-3g_3 = \frac{2}{\pi} \int_{m^2}^\infty ds \left(
	-\frac{3\text{Im}\mathcal{T}{}_{+2}^{-2}(s,0,-s)}{2s^4}+
	\frac{\text{Im}\partial_t\mathcal{T}{}_{+2}^{-2}(s,0,-s)}{s^3}
	\right).
\end{equation}
Taking into account equations \eqref{eq:EFT_constraints} and \eqref{eq:forward_limit} we conclude that
\begin{equation}
	g_3' = \frac{1}{\pi} \int_{m^2}^\infty ds
	\frac{\text{Im}\partial_t\mathcal{T}{}_{+2}^{-2}(s,0,-s)}{s^3}.
\end{equation}

Let us now set $\Phi(s,t,u) = (s^2+u^2)\partial_t^2\,\mathcal{T}{}_{+2}^{-2}(s,t,u)$. We will assume that $ \partial_t^2\,\mathcal{T}{}_{+2}^{-2}(s,t,u)$ decays at $s= \infty$ and fixed $t$ so that we can use the relation \eqref{eq:main_equation}.\footnote{For the massive scalar and massive fermionic QFT examples, as described in section~\ref{sec:examples}, the amplitude $\mathcal{T}{}_{+2}^{-2}(s,t,u)$ vanishes in the limit $|s|\rightarrow \infty$ at fixed $t$. 
We expect this in general because the dilaton couples to the trace of the stress tensor which vanishes in the UV limit.
} Due to \eqref{eq:low_energy_expansion} we have
\begin{equation}
	\label{eq:residue_2}
	\underset{s=0,-t}{\text{Res}}\left(\frac{\Phi(s,t,u)\Big|_\text{low energy}}{s^2(s+t)}\right)=
	4\left(g_2+g_2'+(g_3' -g_3) t\right)+\ldots
\end{equation}
Plugging it into \eqref{eq:main_equation}, taking into account \eqref{eq:EFT_constraints}  and evaluating the result in the forward limit we finally conclude that
\begin{equation}
	\label{eq:dispersion_relation_g2}
	g_2' = 
	\int_{m^2}^\infty  \frac{ds}{\pi} \frac{\text{Im}\,\partial_t^2\mathcal{T}{}_{+2}^{-2}(s,0,-s)}{s}.
\end{equation}
Plugging here \eqref{eq:derivatives_forward_limit} we can also rewrite this expression as
\begin{equation}
	\label{eq:dispersion_relation_g2_partial}
	g_2' = \frac{1}{3}\sum_{\ell\geq 2} (2\ell+1)(\ell-1) \ell (\ell+1)(\ell+2) 
	\int_{m^2}^\infty ds \frac{\text{Im}\, [\mathcal{T}_\ell]{}_{+2}^{-2}(s)}{s^3}.
\end{equation}
Due to the discussion in the end of appendix \ref{app:unitarity} the integrand in this dispersion relation is not sign definite.

\bibliographystyle{JHEP}
\bibliography{refs}

\end{document}